\documentclass[
aps,prd,nofootinbib, reprint,preprintnumbers,superscriptaddress
]{revtex4}
\UseRawInputEncoding
\usepackage{supertabular}
\usepackage{lipsum}
\usepackage{blindtext}
\usepackage{amsfonts}
\usepackage{tensor}
\usepackage{graphicx}
\usepackage{pict2e}
\usepackage{balance}
\usepackage{blindtext}
\usepackage{caption}
\usepackage{subfig}
\usepackage{amssymb, amsmath, environ}
\usepackage{soul}
\usepackage[normalem]{ulem}
\usepackage{tcolorbox}
\PassOptionsToPackage{table}{xcolor}
\usepackage{xcolor}
\usepackage{booktabs}              
\usepackage{siunitx}               
\usepackage[colorlinks=true,urlcolor=black,linkcolor=black,citecolor=red]{hyperref}


\newcommand{\bse}{\begin{subequations}}
	\newcommand{\ese}{\end{subequations}}
\newcommand{\be}{\begin{equation}}
\newcommand{\ee}{\end{equation}}

\NewEnviron{eqsplit}{%
	\begin{equation}\begin{split}
	\BODY
	\end{split}\end{equation}
}

\makeatletter
\newcommand*\bigcdot{\mathpalette\bigcdot@{.5}}
\newcommand*\bigcdot@[2]{\mathbin{\vcenter{\hbox{\scalebox{#2}{$\m@th#1\bullet$}}}}}
\makeatother

\newcommand{\bea}{\begin{eqnarray}}
\newcommand{\eea}{\end{eqnarray}}
\newcommand{\ba}{\begin{array}}
	\newcommand{\ea}{\end{array}}

\begin{document}
	\preprint{MITP-23-048}		
	
\title{Non-Perturbative approach to Gravitational Particle Production in the Inflation-Kination Transition:
\\Surveying a wide range of the Parameter Space} 
	\author{Amirsalar Nikandish}
	\email[]{a\_nikandish@sbu.ac.ir}
	\affiliation{Department of Physics, Shahid Beheshti University, 1983969411, Tehran, Iran}
	\author{Shiva Rostam Zadeh}
	\email[]{sh\_rostamzadeh@ipm.ir}
	\affiliation{School of Particles and Accelerators, Institute for Research in Fundamental Sciences (IPM),\\ P.O. Box 19395-5531, Tehran, Iran}
	\author{Nematollah Riazi}
	\email[]{n riazi@sbu.ac.ir}
	\affiliation{Department of Physics, Shahid Beheshti University, 1983969411, Tehran, Iran}
	
\begin{abstract}
	In this study, we investigate the gravitational particle production (GPP) as a natural reheating channel in time-dependent backgrounds. Because conventional reheating scenarios struggle to apply to non-oscillatory inflationary models (NO models) and because of the dark matter abundance puzzle, the GPP has received extensive attention in recent years. We focus on reheating in an inflation-kination background. To maintain generality across inflationary models, we use a continuous scale-factor function that produces this transition and has been proposed in the literature. In this scenario, the gravitational reheating channel arises from the gravitational excitation of a superheavy spectator scalar field with Ricci-matter gravitational coupling. Besides all the perturbative approaches in quantum field theory in curved spacetime, a recently proposed non-perturbative one that uses the notion of the adiabatic vacuum has also been applied to the case of superheavy particles, which overcomes the challenges of the former approaches. Although the effect of the mass of the field on the reheating temperature has been investigated using this approach, the impact of the non-conformal values of the gravitational coupling constant on the GPP rate is not considered. Hence, we modify the non-perturbative approach in such a way that both effects can be traced, and survey a wide range of the parameter space of reheating in the general background. Three parameters affect the reheating temperature in this case: the bare mass of the field, the gravitational coupling constant, and the transition time. In this paper, the effect of each of these parameters alone on the resultant reheating temperature, as well as the combined effects of these parameters, is studied. Since we use the adiabatic vacuum prescription as our physical initial conditions and, we avoid tachyonic instability, the effective frequency of the field should be positive-valued, and hence, gravitational coupling constant must satisfy a stability inequality for each mass. Due to the superheavy values of mass, the negative values for the gravitational coupling constant are interestingly allowed. Using this inequality, we study the effect of non-conformal values of the gravitational coupling on the reheating temperature for each mass and find the highest reheating temperature possible in the parameter space of ``mass-gravitational coupling constant". Then we examine the effect of the reduction of the transition time, which is a characteristic feature of the inflationary model, on the reheating temperature, and find new and interesting results. Taking into account all involved parameters and surveying a wide range of the 3D parameter space of the model, the highest reheating temperature possible is obtained. We will argue that there are generally two scenarios for the decay of generated heavy particles to the thermal bath, where the second scenario yields a higher reheating temperature. All of the aforementioned investigations are done based on numerical calculations, and we will obtain that the highest possible reheating temperature is quite sufficient for the universe to enter the Hot Big-Bang era. For the sake of completeness, we study the GPP in such a transition using an analytical method in an independent, parallel, solvable model. This analysis is presented only as a mathematical tool, and its results should be considered only as a help to obtain a rough estimate for the reheating temperature. Besides, this analysis provides us with a mathematical formula indicating how the values of mass and the gravitational coupling constant manifest themselves in the GPP rate.                                 
\end{abstract}

\maketitle
\section{Introduction} \label{intro}
\indent The paradigm of inflationary cosmology\cite{guth1981inflationary, linde1982new} was first proposed to overcome the puzzles of the standard model of cosmology, such as the spatial flatness and the overall homogeneity (the problem of initial conditions), the absence of topological defects, the size of our universe, etc.\cite{allahverdi2010reheating, riotto2002inflation, baumann2009tasi, mukhanov2005physical}. But the necessity of this paradigm is not limited to the puzzles of the Hot Big-Bang (HBB) cosmology, listed above. Inflation provides us with a very elegant description of the origin of cosmological density perturbations as a consequence of the enhancement of inflaton quantum vacuum fluctuations\cite{allahverdi2010reheating, gorbunov2011introduction, mukhanov1981quantum, hawking1985limits, starobinsky1982dynamics, guth1982fluctuations}. Similarly, inflation justifies how the tensor perturbations, known as the primordial gravitational waves (PGWs), were generated from the vacuum fluctuations of the gravitational fields\cite{fabbri1983effect, abbott1984constraints, grishchuk1974amplification, starobinskii1979spectrum, guzzetti2016gravitational, dos2022gravitational, maggiore2008gravitational, baumann2009tasi, riotto2002inflation}.

Due to the rapidly accelerated expansion caused by inflation, all of the matter content is diluted, and the universe remains cold and empty. Hence, a period of matter creation called reheating is essential for the universe to enter the HBB era\cite{linde2005particle, kofman1994reheating, kofman1997towards, allahverdi2010reheating}. The basic proposal of reheating was first presented in Linde's famous paper on inflationary cosmology \cite{linde1982new}. After that, the reheating proposals studied in the literature were based on perturbation theory\cite{dolgov1982baryon, abbott1982particle, abbott1982particle}\footnote{For a pedagogical review on the perturbative reheating scenario, see also Ref\cite{baumann2009tasi}}. Since, in the first inflationary models, the inflaton field oscillates after the end of inflation (we refer to these models as oscillatory models), it became clear that in such a perturbative approach, the coherent nature of the inflaton is not considered\cite{traschen1990particle, dolgov1990production}. In \cite{traschen1990particle}, the authors discussed the possibility that the reheating phase involves a resonance instability, which means that the reheating phase may have a non-perturbative nature. Linde, Kofman, and Starobinsky studied in detail the parametric resonance during the reheating phase in oscillatory models in \cite{kofman1994reheating} and then in \cite{kofman1997towards}\footnote{See also the important lecture note on reheating during the inflaton's oscillations by Lozanov\cite{lozanov2019lectures}. }. It was shown that after inflation, due to the coherent oscillations of the inflaton, the universe undergoes a matter-dominated phase (MD era) with $W = 0$, where $W$ is the equation of state parameter. At the beginning of this phase, due to the large amplitude of oscillations, the decay of the inflaton to matter is excessive. This phase is known as broad resonance or the preheating stage. After that, by decreasing the amplitude of inflaton oscillations, the resonance band becomes narrow, and the particle production is only efficient for some momenta. At the end, the reheating phase enters the period where the perturbative picture becomes valid. So it seems that the puzzle of reheating in the well-known oscillatory models of Inflation could be explained through parametric resonance, in such a way that, despite the perturbative reheating, particle production occurs in a non-perturbative way. By far, many oscillatory models of inflation have been proposed during the last two decades, and in all of them, the inflaton oscillates after the end of inflation; hence, reheating proceeds through parametric resonance\footnote{Many models of inflation have been proposed in the context of modified gravities. Among them, the $f(R)= R+\frac{R^2}{6m^2}$ or the Starobinsky model is of particular interest. It is conventional to rewrite the Starobinsky model in terms of a scalar field whose potential takes the form of the inflaton and can describe both the slow-roll phase for accelerated expansion and the oscillatory phase for particle production\cite{de2010f}. Hence, these models are also considered oscillatory ones.}.

Besides the oscillatory models of inflation, the non-oscillatory models (NO models) were also proposed to describe both the early (Inflation) and the late (dark energy) accelerated expansion of the universe\cite{felder1999inflation, spokoiny1993deflationary, peebles1999quintessential, ellis2021non, Haro:2019gsv,areste2021quintessential}. See also the review papers \cite{de2021review, dimopoulos2001models} regarding the NO models of inflation. In the inflationary models, the accelerated expansion occurs due to the slow-roll of the inflaton on the plateau of its potential, where its potential energy density is the dominant component of the universe and remains almost constant\footnote{CMB observations strongly confirm the idea that the universe undergoes accelerated expansion when the expectation value of the inflaton corresponds to the plateau and hence, the inflaton's scalar potential is the dominant component of the universe. These observations suggest that its plateau density is around $10^{-10} {M_Pl}^4$ \cite{akrami2020planck, ade2016planck, ade2016improved}.}. The main distinguishing feature between the oscillatory models and NO models is that in NO models, the inflationary potential has no minimum near the inflationary plateau, and hence the inflaton does not oscillate after the end of inflation. Instead, its effective potential decreases gradually at large values of the inflaton field\cite{felder1999inflation, dimopoulos2002modeling, dimopoulos2017quintessential, dimopoulos2021quintessential, sanchez2006trapped, dimopoulos2022modelling, dimopoulos2022palatini, dimopoulos2019quintessential}. As mentioned above, in oscillatory models, the oscillations of the inflaton provide us with an efficient and violent channel of reheating, called parametric resonance. In addition, due to the inflaton's oscillations, the universe experiences a transient period of MD before entering the radiation-dominated (RD) epoch. But in NO models, the situation is completely different. After the end of inflation, the inflaton rolls down the wall of a steep potential, and its kinetic component becomes dominant; hence, the universe undergoes a kinetic component-dominant era, called kination in the literature, with the EoS parameter $W = +1$\cite{joyce1998turning}.

Due to the absence of parametric resonance, the NO models faced a very fundamental challenge, which is the lack of an efficient reheating scenario, and hence, they did not gain much attention for years. In the first attempts to overcome this substantial challenge, the GPP in a curved space-time as a natural channel of particle creation was considered. It is known that in a time-dependent space-time geometry, where some of the background metric coefficients are explicitly a function of time, the GPP occurs due to quantum effects. In fact, in such backgrounds, time translation is not a symmetry of the space-time anymore, or in other words, the metric does not have a time-like Killing vector. As a result, the generator of time translation, or the Hamiltonian, is not a constant of motion, and hence, the physical vacuum in one slice of time is not necessarily the physical vacuum in another one\footnote{In QFT, the vacuum state is the eigenstate of all annihilation operators with the eigenvalue of ${{\hat{a}}^{-}}_k \lvert0> = 0$ for all $k$. But if the background is time-dependent, the Hamiltonian is explicitly a function of time, and hence, it cannot have a time-independent eigenstate that could be considered as the vacuum at all times. Hence, the choice of a vacuum state that could describe the physical particle number is one of the most challenging issues in QFT in curved spacetime.}. This means that an observer at later time, ${\eta}_2$, with a physical vacuum, detects particles in the physical vacuum of another observer at another time, ${\eta}_1$. This phenomenon is interpreted as gravitational particle production. Such matter creation in a time-dependent background is called the GPP process, which, as mentioned above, is based on the change in the definition of the physical vacuum. The Friedmann-Lemaître-Robertson-Walker (FLRW) metric, with the scale factor $a(t)$, is an example of a time-dependent background. As a result, the GPP process occurs in our universe naturally. Especially, this production is more intensive and of particular interest after the end of inflation as a channel of natural reheating, when the universe experiences a transition from de-Sitter (dS) expansion to MD, RD, or kination.

The physics of GPP has been broadly studied in the literature for scalar particles. The observation of the electroweak Higgs boson has assured physicists that fundamental scalar particles exist\cite{dimopoulos2018non}. In addition, the possibility of a matter-gravity (or scalar-gravity) direct coupling of the form $\xi R{\phi}^2$ is also considered for the scalar field in the early universe. In this term, $\xi$ denotes the gravitational coupling constant, and R signifies the Ricci scalar. Such coupling is unique in the way that it is a feasible local scalar interaction with the appropriate dimensions\cite{birrell1984quantum}. One of the first motivations for considering such a term emerges from the theory of self-interacting scalar particles, with a quartic term, $\lambda{\phi}^4$. It is shown in \cite{birrell1984quantum, parker2009quantum} that the term $\xi R{\phi}^2$ is necessary for the theory with $\lambda{\phi}^4$ to be renormalizable. Also, there are other motivations for the term $\xi R{\phi}^2$ from supergravity and superstring\cite{kachru2003towards, kofman2008rapid}. As a consequence, the issue of GPP in curved spacetime is mainly studied for scalar fields with a matter-gravity coupling $\xi R{\phi}^2$. 

Before going further, it is worth mentioning that the value of $\xi$ could be minimal ($\xi = 0$), conformal ($\xi = \frac{1}{6}$), or otherwise. In general, like other coupling constants, $\xi$ may behave as a running coupling constant through different energy scales, and hence can not be set universally to zero\cite{Figueroa:2021iwm}. Among other values of $\xi$, the two aforementioned minimal and conformal cases are of particular interest. It is known that the conformal coupling for massless fields leads to the conformal invariance of the action and the field equation of motion (EoM)\cite{birrell1984quantum, ford2021cosmological}. The GPP can not occur for fields with conformal EoM, even in a time-dependent background\cite{ford1987gravitational, birrell1984quantum, mukhanov2007introduction}. But we must notice that there are no compelling reasons for $\xi$ to be minimal or conformal in the real world, because no symmetry is involved\cite{carroll2019spacetime}. From the viewpoint of our paper, $\xi$ has another important feature: it is dimensionless. Hence, we have no constraint to limit ourselves to its small value; i.e., it can take non-minimal and non-conformal values of the order of unity or more\cite{carroll2019spacetime}\footnote{In some recent studies, the great non-minimal values for $\xi$ are considered\cite{Figueroa:2021iwm, dimopoulos2018non, opferkuch2019ricci, dorsch2024gravitational, jin2020production}}. Indeed, there is only a weak constraint from current experiments on the value of $\xi$ ($\xi < 10^{15}$)\cite{Figueroa:2021iwm, atkins2013bounds}. Although it is due to the feeble gravitational interaction with SM fields\footnote{The authors in \cite{atkins2013bounds} show that if the scalar field is supposed to be the SM Higgs, its non-minimal coupling to Ricci should be smaller than $2 \times 10^{15}$.}, recent studies indicate that the primordial gravitational waves (PGWs) might provide us with the opportunity to examine the existence and strength of such coupling\cite{nikandish2024signature, chakraborty2025probing}.

According to the above explanations, the issue of GPP has been considered in the theory of the scalar field with a $\xi R {\phi}^2$ coupling. As mentioned earlier, despite the conventional QFT in Minkowski space-time, the vacuum state is not unique in curved space-time at each specific time. That is, if one vacuum state is chosen as a preferred vacuum state at one time, it is not necessarily the preferred vacuum state at a later time. The choice of vacuum is equivalent to the choice of creation and annihilation operators, and it, in turn, is equivalent to the choice of the mode functions of the EoM. The two sets of mode functions (corresponding to the two vacuum states of two distinct times) are related to each other through the Bogoliubov transformation. The value of the Bogoliubov coefficient, $\beta$, determines the particle production rate as ${\lvert \beta \lvert}^2$. Hence, to compute the GPP between two different times, it is essential to determine the physical vacuum that describes the true particle occupation number in the final state. This is because the GPP is based on the change in the physical vacuum between the initial and final times.

There are many approaches in the literature proposed by many physicists to determine the physical vacuum at a specific time. These approaches have been applied to the theory of a gravitationally coupled scalar field with the EoM of the form Eq. \ref{eomphi4} and Eq. \ref{eomx}. In the first attempts, Bernard and Duncan consider the GPP for a hypothetical time-dependent background where the scale factor tends to constant values at the initial and final times (statically bounded expansion) \cite{bernard1977regularization, birrell1984quantum, ford2021cosmological}. In the case of conformal coupling, they studied the effect of mass on the GPP. It is essential to note that they solved the EoM of the mode function exactly for both the initial and final times, finding that the exact form of the mode functions corresponds to the in and out vacua by applying the appropriate initial conditions. In another attempt, Parker and Toms studied the GPP in the same background, but for a massless scalar minimally coupled to gravity\cite{parker2009quantum, parker1976thermal, ford2021cosmological}. In such a background, they had the chance of solving the mode function EoM exactly and defining the physical vacuum for the in and out regimes without any ambiguity. Hence, in these two models, the mode functions defining the two different vacuum states are found exactly by using meaningful initial conditions, and hence, the physical particle number can be found and interpreted without any challenge or ambiguity. But, in general, it is not the case for the FLRW background, where the scale factor is not bounded and, on the contrary, is always growing. As a consequence, in the FLRW background, it is usually (but not always) impossible to find the exact solution of the mode function for different cosmological eras, and hence, the physicists' attention turned to other methods of defining the physical vacuum state for different cosmological eras. It is important to keep in mind that their main goal was the finding of the reheating temperature caused by the GPP that happened in the inflation-post-inflation transition.

After the approaches discussed above, and due to their incompatibility with the cosmological expansion, another approach was proposed by the other authors, usually known as the instantaneous lowest-energy state\footnote{This method is also named as the instantaneous diagonalization of the Hamiltonian in the literature\cite{winitzki2005cosmological}.}. See \cite{mukhanov2007introduction} for a comprehensive discussion and detailed mathematical explanation for this approach. Although this method is based on a very physical intuition of the vacuum state and the Hamiltonian, and has a very beautiful mathematical description that explains how the physical vacuum should be chosen to minimize the expectation value of the instantaneous time-dependent Hamiltonian, there are situations where such a description fails. It is discussed by Fulling \cite{fulling1989aspects, fulling1979remarks} that in some generic cases (anisotropic space-times) the total density of particles (the apparent particle production rate) becomes infinite, even in cases where the geometry is slowly changing with time and the time interval between the initial and the final state is small\cite{winitzki2005cosmological,mukhanov2007introduction}. Due to this challenge, another approach was proposed in \cite{parker1969quantized}, where Parker proposed that the physical vacuum state should be defined in such a way to minimize the apparent particle number observed at a certain time. This proposal, like the instantaneous lowest energy state, has a very strong intuition and is physically reasonable, but there is also another challenge: the resulting vacuum state depends on the choice of that specific time. So the particle number observed in $\lvert 0>_{\eta}$ is not in general an adequate particle number at the time $\eta$. Hence, these proposals seem to be disputable, and the physicists continued to struggle to define the physical vacuum state.

The next effort was based on invoking perturbation theory to find the physical vacuum and the physical particle number. This perturbation theory was based on the fact that finding an exact solution for the mode function EoM is too difficult in general (except for some special cases). The perturbative approach was initially proposed by Zeldovich and Starobinsky in \cite{Zeldovich:1971mw, Zeldovich:1977vgo} and also by Birrell and Davis in \cite{birrell1984quantum}. After that, Ford in \cite{ford1987gravitational} (see also \cite{ford2021cosmological} for review) invoked this approach to study the GPP in an inflation-radiation background and to find the reheating temperature initiated from the GPP. It was, in fact, the first attempt to study the GPP as a channel to reheat the universe after inflation. This analysis was done for the case of a massless scalar field whose gravitational coupling is close to the conformal case, $\lvert \xi - \frac{1}{6}  \lvert \ll 1$, since it is the case where the perturbative approach could be applied. But as discussed in \cite{dimopoulos2018non, felder1999instant, ellis2021non}, this perturbative analysis is not an efficient reheating scenario at all since the $T_{reh}$ obtained in this scenario is of the order of $T_{reh} \simeq 10^6 GeV$, which is much less than the $T_{reh}$ obtained in oscillatory models of inflation through the parametric resonance. Such a conclusion was, in turn, to the detriment of the NO models again \footnote{For a more recent study on the perturbative approach to QFT in curved space-time, see \cite{huhtala2022perturbative}.}. 

Although the reheating process proposed through the perturbative analysis was not efficient at all, the attempts to find a more adequate description of the physical vacuum and the correct physical number were not stopped. In fact, the path was continued by the physicists by proposing the``adiabatic vacuum"\cite{mukhanov2007introduction, winitzki2005cosmological}. From the point of view of our study in this paper, this notion of physical vacuum, known as ``adiabatic vacuum", is of particular interest. If the effective frequency in the EoM in Eq. \ref{eomx} is a sufficiently slow function of time and the spatial part of the universe is sufficiently flat, the vacuum state in such a regime is called the adiabatic vacuum, and the particle states can be defined through this vacuum. It is essential to keep in mind that to invoke the notion of adiabatic vacuum and to compute the particle number by the Bogoliubov transformation, the effective frequency of the field, ${{{\omega}_k}^2(\eta)}$ must be positive-valued within the whole adiabatic regime. The adiabatic vacuum is described through the Wentzel-Kramers-Brillouin (WKB) approximate solution to Eq. \ref{eomx}. Since the notion of this vacuum is one of the bases of our calculations as our physical initial conditions in this paper, we explain its definition and the calculation of GPP through this vacuum in App.~\ref{wkbadi} in detail. It is worth noting that although this method is applicable in the cosmological eras when the background is slowly changing, the precision of the WKB approximation is insufficient for some calculations. This is due to the fact that the WKB approximation is, in fact, a divergent asymptotic series, and hence, the calculation of particle number using the WKB adiabatic vacuum has a fundamental accuracy limit\cite{mukhanov2007introduction, winitzki2005cosmological}. In \cite{winitzki2005cosmological}, Winitzki tried to explain this uncertainty in detail, and also tried to improve the WKB approximation using the perturbative method discussed above, which was initially developed by \cite{Zeldovich:1971mw, Zeldovich:1977vgo} and also \cite{grib1994vacuum}\footnote{For a more recent study regarding the issue of the WKB solution for GPP see also\cite{enomoto2021exact}.}.

Due to the limits and inaccuracy of the WKB solution and also the inefficiency of perturbative GPP, the issue of reheating in NO models remained an essential challenge. To resolve this, Felder, Kofman, and Linde proposed the novel ``instant preheating" scenario \cite{felder1999instant, deHaro:2023xcc}. This is a non-perturbative scenario that can happen in NO models where the inflationary effective potential decreases slowly at large values of the inflaton field, for instance, in the theory of quintessence. Recently, this theory has been revisited by Ellis et. al. in \cite{ellis2021non}. Although this scenario is very efficient and overcomes the problem of reheating in NO models, in \cite{dimopoulos2018non}, the author argued that the model building would be restrictive in this proposal. So, the scenario of ``Ricci reheating" based on the tachyonic instability of a spectator field was proposed \cite{figueroa2017standard, dimopoulos2018non, opferkuch2019ricci, laverda2024ricci, Figueroa:2021iwm}. In this scenario, the role of the term $\xi R{\phi}^2$ is essential for the tachyonic instability of the spectator field to occur. In their study, the GPP is not calculated using the notion of the adiabatic vacuum and the Bogoliubov transformations. Instead, their primary goal was to propose a channel of reheating that occurred because of the changing of the sign of the Ricci scalar during the inflation-kination transition. Due to this sign changing, the effective mass of the field becomes negative-valued and so, the tachyonic instability occurs, which in turn, leads to the excitation of the field. At the end, the excited spectator field oscillates and transfer its energy to the thermal bath. It turned out that the ``Ricci reheating" scenario, just like the ``instant preheating", yields an efficient reheating mechanism which results in a very high reheating temperature.

Although the last two aforementioned proposals were very successful solutions to the issue of reheating in the NO models, progress in the numerical methods of calculation brought attention back to the GPP calculations. In many studies, Haro et al. considered the issue of GPP in several famous NO models of inflation. By developing the numerical analysis and invoking different methods of the WKB solution for the mode function, the GPP was studied in the quintessential inflationary models by them. They have also developed the analytical calculation necessary to compute the $T_{reh}$ \cite{deHaro:2023gho, deHaro:2022ukj, deHaro:2022vxc, Salo:2021piz, Salo:2021vdv, deHaro:2019oki, Haro:2018zdb, Haro:2018jtb, deHaro:2017ale, AresteSalo:2017lkv, DeHaro:2017abf, deHaro:2015hdp, Haro:2015zda, salo2021gravitational2, salo2021gravitational, hashiba2021particle, hashiba2019dark}. After that, the path of study of GPP, not only as a natural channel of reheating in the FLRW background, but also as a way of explaining the dark matter (DM) abundance, was continued. Such studies were mainly based on invoking numerical calculations to find the solution of the mode EoM. In fact, the GPP rate $\lvert\beta\lvert^2$ was calculated in the post-inflationary era between the physical mode function of the inflationary era - whose evolution is calculated numerically from an ${\eta}_{in}$ in the inflation phase to the ${\eta}_{f}$ in the post-inflationary phase, invoking the well-known Bunch-Davis initial conditions - and the physical vacuum of the post-inflationary era. Then, the ${\beta}_{\kappa}$ coefficient can be obtained using a limiting formula in the limit $\eta \rightarrow \infty$. In many important studies conducted over recent years, the role of the GPP in explaining the dark matter abundance has been considered \cite {pierre2023scalar, kaneta2023gravitational, cembranos2020gravitational, capanelli2024gravitational, fairbairn2019despicable, racco2024gravitational, verner2025nonminimal, bertuzzo2024gravitationally, cembranos2024vector, haro2019gravitational, haro2020gravitational, de2025note, chung2003classical, li2019gravitational, li2020cosmological, hashiba2019gravitational, babichev2020gravitational}.

In recent years, the reheating process due to the GPP in NO models of inflation has also been considered in various scenarios other than the inflation-kination transition. For instance, Vilja et. al in \cite{lankinen2017gravitational, lankinen2020reheating, lankinen2021reheating} studied the GPP during the kination itself or the stiff matter-dominated era. Their results demonstrate that the $T_{reh}$ in such a scenario could take values between $10^6 - 10^{12}  GeV$ or $10^{13} GeV$ for some specific channels of decay. In addition, the role of GPP in explaining the baryon asymmetry of the universe (Baryogenesis) is also studied in \cite{flores2024role, fujikura2023generation, kamada2020gravitational, kamada2020renormalization}. There are also interesting studies regarding the issue of GPP done during the last decade, which are worth paying attention to. For instance, see \cite{clery2023gravitational, enomoto2021exact, belfiglio2025comparing}. 

As mentioned above, it is accepted that the reheating process in oscillatory models of inflation happens due to the coupling of the inflaton to the Standard Model (SM) fields and through the process of parametric resonance. But as discussed in \cite{peebles1999quintessential}, there are constraints on the value of such couplings\footnote{Such couplings must be weak enough to preserve the flat shape of the inflaton potential and, on the other hand, must be strong enough for a successful thermalization.}, and models satisfying these constraints do not naturally appear in particle physics. As a consequence, reheating through the gravitational production of spectator field particles might be considered as an interesting possible channel of reheating even in oscillatory models. Besides, there is another important motivation, discussed earlier, that the channel of GPP reheating naturally happens in any space-time varying with time, including the background in oscillatory models. For these reasons, during the last five years, this channel of reheating has also been studied for the oscillatory models of inflation, especially in the context of the Starobinsky model\cite{cembranos2023late, dorsch2024gravitational, jin2020production, feiteira2025cosmological}. The Starobinsky model is one of the most successful models of inflation from the viewpoint of the PLANCK results, which is compatible with observations\cite{akrami2020planck}. The result of such studies is that even in the case of oscillatory models, and in backgrounds where the Ricci scalar is oscillating with time\cite{de2010f}, such as the Starobinsky model, the GPP is definitely a possible channel of reheating after the inflationary era\cite{dorsch2024gravitational}. For the sake of completeness, see also \cite{haro2020note} and a new study in \cite{Kolb:2026xfp}\footnote{For detailed reviews on the topic of GPP and its applications, especially including reheating and the DM abundance, see also the important review papers\cite{ford2021cosmological, kolb2024cosmological}. Also, for a detailed and comprehensive discussion on the notion and the basis of the particle creation and particle number in an expanding universe, see \cite{parker2012particle}.}.

As a consequence of all of the discussions mentioned above, it is time to explain our motivations regarding the use of a non-perturbative approach to the issue of GPP. First of all, this procedure was first proposed in the famous paper by Damour and Vilenkin in 1995\cite{damour1996string}, as a way to study the spectrum of created dilatons. After that, this approach was invoked in \cite{hashiba2019gravitational} to study the gravitational reheating for superheavy conformally coupled particles. This paper is, in fact, the main reference of our study, which we will refer to throughout this paper. Recently, this approach was invoked again by Yann Mambrini et. al. to study the Bogoliubov vs Boltzmann approach as well as the scalar fluctuations and the production of DM \cite{chakraborty2025generalizing, garcia2025scalar}. Hence, we found this approach to be very up-to-date and applicable, as attention has been brought to it in recent years. This approach is built on two sturdy pillars: 
\begin{itemize}
\item On the one hand, in the case of the inflation-kination transition, the exact form of the mode function could be obtained for both the inflation and the kination eras. This is just like the case in the Bernard-Duncan\cite{bernard1977regularization, birrell1984quantum, ford2021cosmological} and Parker-Toms\cite{parker2009quantum, parker1976thermal, ford2021cosmological} models discussed above, and is one of the key advantages of this approach. This is because in the previous studies, the physical mode function defining the physical vacuum for the kination era was defined solely using an approximation formula and hence must be considered just as an approximate physical mode function corresponding to the physical vacuum of the kination era. But in the non-perturbative approach, the physical mode function defining the physical vacuum of the kination era (K-vacuum) could be found exactly, and hence the physical vacuum and the particle numbers could be defined using this non-perturbative mode functions. As we will explain in the following sections, it is especially vital in the case of numerical analysis, where the physical mode function of the inflationary phase is found using numerical analysis and its evolution is followed from inflation until an arbitrary time in the kination era. Then the Bogoliubov coefficient $\beta$ is calculated between the numerical mode function and the K-vacuum. In addition, in our attempts to propose a parallel toy model for which an analytic expression for the particle production rate can be found, the coefficient $\beta$ is calculated between the K-vacuum and the well-known physical mode function corresponding to the physical vacuum of the inflationary era, the Bunch-Davies vacuum (BD-vacuum). Hence, the first bargaining chip of the non-perturbative approach is that it is built up by defining the exact form of the K-vacuum.
\item On the other hand, in this approach, the choice of the physical mode function or, equivalently, the physical vacuum is based on the notion of the adiabatic vacuum discussed above as physically relevant initial values\cite{mukhanov2007introduction}. After solving the EoM of the mode function for the inflationary era, the initial conditions are invoked from the fact that in the limit of $\eta \rightarrow -\infty$ (deep inside the horizon), all of the modes with any momentum $k$ are inside the horizon and hence, the curvature of the space-time becomes irrelevant for them. As a result, they behave just like the modes in Minkowski (flat) spacetime (notion of adiabatic vacuum)\cite{mukhanov2007introduction}. Besides, for the physical modes in the kination era, the initial conditions arise from the fact that in the limit of $k \rightarrow \infty$ (high-frequency modes), the mode lies well inside the horizon and hence does not feel the curvature of spacetime again (the notion of vacuum for short distances)\cite{mukhanov2007introduction}. So, although the form of the mode function could be obtained in an exact form, its initial conditions are invoked from the notion of adiabatic vacuum.
\end{itemize}
\textbf{Henceforth, in this paper, our main goal is to study a wide range of parameter space of the GPP channel of reheating using the adiabatic notion of vacuum as initial conditions and, in a non-perturbative manner. During the survey of the parameter space, we will consider the non-minimal as well as the conformal value for the gravitational coupling constant, $\xi$. But it is worth emphasising that although the non-minimal values of $\xi$ were considered in studies such as \cite{opferkuch2019ricci, dimopoulos2018non, figueroa2017standard, Figueroa:2021iwm} to study the tachyonic instability, in our study, the GPP is formulated through the notion of adiabatic vacuum (as meaningful and physical initial conditions) and the Bogoliubov transformation, so that the effective frequency of the spectator field is considered to be always positive and hence, the tachyonic instability can not happen.}

This paper is organized as follows: In Sec. \ref{theo} we provide all of the theoretical tools that will be needed for our calculations throughout this paper. This includes the total action and equations of motion, a comprehensive review of gravitational particle production and Bogoliubov transformation, the second quantization, and the adiabatic vacuum notion, the functional form of the scale factor and derivation of the K-vacuum, the stability relation, and the scenarios of decay and their corresponding reheating temperature after the particle production. Sec. \ref{numresults} is dedicated to the results of our numerical calculations. These results are presented in three classes to see the effect of mass, the effect of the non-conformal coupling, and the effect of the transition time. These results are depicted in relevant plots, and we provide this section with comprehensive explanations, physical interpretations, and mathematical justifications for the sake of completeness and also to build a very strong physical description of the gravitational particle production issue. In Sec. \ref{anaresults}, we derive an analytical formula for the GPP rate in an independent toy model which describes the inflation-kination transition too. With this formula, we recalculate the GPP rate, as well as the energy density and the reheating temperature, to compare these results with the numerical ones, and also to show that this analytical formula can be used as a rough estimation of the GPP rate and $T_{reh}$ in the modified non-perturbative approach. At the end, we present our conclusion and further arguments as a comprehensive summary of all of our calculations. In the appendices, the details of all of our calculations are presented. These details can provide our readers with a deeper physical understanding of the nature of this channel of reheating.

Throughout this paper, all of our calculations are done in natural units, where we have set $\hbar=k_B=c=1$. Also, the signature of the metric is chosen to be space-positive $(-,+,+,+)$.

\section{Theoretical Framework}\label{theo}
In this section, our primary purpose is to provide the theoretical foundations necessary for calculating GPP. The first step in our study must be the procedure of second quantization. To this end, we have to write down the total action of our theory, including the Einstein-Hilbert part, the inflationary background part, the scalar field with gravitational coupling, and the action of the ordinary matter (into which the scalar field will decay and reheat the universe). Then we will derive the EoM of the scalar field, $\phi$. By introducing an auxiliary field, $\chi$, the second quantization procedure can be applied in a conventional manner, and then we upgrade the field $\chi$ to an operator and write down the EoM for each mode. To calculate the GPP rate, we will review the Bogoliubov transformations. Then the functional form of the scale factor for our numerical analysis will be introduced. The notion of adiabaticity will also be discussed, and the relevant adiabatic initial conditions for both eras will be presented. Then we will calculate the physical vacuum of the Kination era and present the equations of $T_{reh}$ for both scenarios of decay. To prevent tachyonic instability, we will derive a stability-constrained relation for each value of mass that also guarantees the validity of adiabatic initial conditions, i.e. ${{\omega}_{\kappa}}^2>0$. 

\subsection{Equation of Motion and the Procedure of the Second Quantization}\label{eomsec}
Let us start with the well-known action which includes the spectator scalar field $\phi$ with scalar-gravity coupling and also with appropriate coupling to the SM fields, in a background initially dominated by the inflaton field:
\begin{equation}
\mathcal{S} = \int d^4x \, \sqrt{-g} \left(
\frac{1}{16\pi} M_{Pl}^2 R 
+ \mathcal{L}_{inf} 
- \frac{1}{2} g^{\mu\nu} \partial_\mu \phi \, \partial_\nu \phi
- \frac{1}{2} m_\phi^2 \phi^2
- \frac{1}{2} \xi R \phi^2
+ \mathcal{L}_M
\right).
\label{action}
\end{equation}
In this equation, $R$ is the Ricci curvature scalar, and $\xi$ stands for the coupling strength of the $\phi$-gravity interaction, which could be minimal, conformal, or non-minimal. $m_{\phi}$ describes the bare mass of the spectator scalar field, which in our study we consider to be superheavy (of the order of $H_{inf}$). Also, $M_{pl}$ is the Planck mass with the value of $1.22 * 10^{19} GeV$, ${{\mathcal{L}}}_{inf}$ is the Lagrangian density of the dominant inflaton, containing the kinetic term as well as the inflationary potential, and ${{\mathcal{L}}}_{M}$ describes the Lagrangian density of other matter fields (SM matter fields). All of the couplings between the spectator field $\phi$ and the SM fields, which provide us with the decay channels of $\phi$ to SM particles, are coded in ${{\mathcal{L}}}_{M}$. These couplings are essential from the reheating point of view. As a result, the Lagrangian density of the free $\phi$ could be written as :
\begin{equation}
\mathcal{L}_\phi^{free} = \sqrt{-g} \left(
-\frac{1}{2} g^{\mu\nu} \partial_\mu \phi \, \partial_\nu \phi
-\frac{1}{2} \xi R \phi^2
-\frac{1}{2} m_\phi^2 \phi^2
\right)
\label{freephi}
\end{equation}
Varying the ${{\mathcal{L}}_{\phi}^{free}}$ with respect to $\phi$, the EoM of the sepctator field is obtained as:
\begin{equation}
\square \phi - \left(\xi R + m_\phi^2\right) \phi = 0
\label{eomphi1}
\end{equation}
where the mathematical symbol ``$\Box$" stands for the D'alembertian in curved spacetime, $$\Box {\phi}=\frac{1}{\sqrt{-g}}\partial_\mu(\sqrt{-g}\partial^\mu {\phi}).$$ Hence, the EoM of $\phi$ can be re-written in the following form:
\begin{equation}
\frac{1}{\sqrt{-g}} \, \partial_\mu \left( \sqrt{-g} \, g^{\mu\nu} \, \partial_\nu \phi \right)
- \left(\xi R + m_\phi^2\right)\phi = 0.
\label{eomphi2}
\end{equation}
The metric of the background employed in our calculations is the one for a homogeneous and isotropic universe, i.e., the flat Friedmann-Lemaître-Robertson-Walker (FLRW) metric, which could be described using either the cosmic time ($t$) or the conformal time ($\eta$), as:
\begin{equation}
ds^2 = -dt^2 + a^2(t)\,\delta_{ij}\,dx^i dx^j
= a^2(\eta)\left(-d\eta^2 + \delta_{ij}\,dx^i dx^j\right)
\label{metric}
\end{equation}
where $d\eta = \frac{dt}{a}$ and $i,j = 1-3$.
Here, the spectator field $\phi$ is not a dominant field, and hence, there is no constraint on it to be a homogeneous and isotropic field. In fact, the field $\phi$ is a function of space also, and hence, has a Fourier expansion as: 
\begin{equation}
{\phi}(\mathbf{x},t) =
\int \frac{d^3k}{(2\pi)^{3/2}} 
e^{i\mathbf{k}\cdot\mathbf{x}} \phi_k(t).
\label{phimode}
\end{equation}
Here, ${{\phi}_k}(t)$ is the mode function with momentum $k$ of the field $\phi$\footnote{The FLRW metric components are solely functions of time. Hence, the $\sqrt{-g}$ entering the EoM of $\phi$ is just a function of time. As a result, the EoM of $\phi$, Eq. \ref{eomphi1}, in the FLRW background could be solved using the separation of variables method. The spatial part of the equation is just the Laplace equation that yields the plane wave solution. Due to the homogeneity or the spatial translation symmetry of the FLRW background, each mode of the Fourier expansion evolves separately. Hence, the field expansion could be written as the sum of the terms in the form $e^{i\bold{k}.\bold{x}}{{\phi}_k}(\eta)$. Where the $e^{i\bold{k}.\bold{x}}$ are known as the Fourier basis and ${{\phi}_k}(\eta)$ the field mode function. In fact, because of the symmetries of the FLRW background, the issue of studying the field evolution reduces to studying the evolution of separated and independent harmonic oscillators. These independent harmonic oscillators are known as the mode functions.}. 
Now, substituting the mode expansion, Eq. \ref{phimode}, in the EoM, Eq. \ref{eomphi2}, and invoking the fact that the Fourier modes evolve independently, the EoM for each mode function of $\phi$ is obtained as:
\begin{equation}
\ddot{\phi}_k(t) + 3H \dot{\phi}_k(t)
+ \left( \frac{k^2}{a^2} + \xi R + m_\phi^2 \right)\phi_k(t) = 0.
\label{eomphi3}
\end{equation}
To obtain the above equation, the FLRW metric in terms of cosmic time is used, where the determinant of the metric becomes $\sqrt{-g} = {a^3}(t)$. It is obvious that due to the presence of the damping or dissipative term, $3H\dot{\phi}$, the general form of the EoM is not of the form of the simple harmonic oscillator (SHO), $\Ddot{\phi}(t) + {{\omega}^2}(t)\phi(t) = 0$. To apply the procedure of second quantization, it is usually conventional to write down the EoM in the form of that of the harmonic oscillator. Hence, the procedure of second quantization could not be applied, as is conventionally done, to the quantum field $\phi$ with such an EoM\footnote{The presence of this term in the EoM is a manifestation of the expansion of the universe. The amplitude of any field that lives in the FLRW background experiences dissipation or damping behavior because of the expansion of the universe.}. As a consequence, there remain two more steps to convert the form of Eq. \ref{eomphi3} to the form of the SHO EoM. First, it is convenient to rewrite the EoM in terms of the conformal time, $d\eta = \frac{dt}{a(t)}$ or $\eta = \int \frac{dt}{a(t)}$, as:
\begin{equation}
\phi_k''(\eta) + 2aH \phi_k'(\eta)
+ \left(k^2 + a^2 \xi R + a^2 m_\phi^2 \right)\phi_k(\eta) = 0.
\label{eomphi4}
\end{equation}
Throughout this paper, ``prime" denotes the derivative with respect to the conformal time, $\eta$, while ``dot" represents the derivative with respect to the cosmic time, $t$. It is also convenient to introduce a new auxiliary field, $\chi(\bold{x},\eta) = a(\eta) \phi(\bold{x},\eta)$. This field is also a time- and coordinate-dependent field and has a Fourier expansion just like the $\phi$ field, as $\chi(\bold{x},\eta) = \int \frac{{d^3}k}{(2\pi)^{\frac{3}{2}}} e^{i\bold{k}.\bold{x}} {{\chi}_k}(\eta)$. Hence, by substituting its expansion in Eq. \ref{eomphi4} and keeping in mind the fact that each Fourier mode evolves independently, the EoM of each mode function is obtained as:
\begin{align}
\chi_k''(\eta)
+\Big[\,k^2 + \big(\xi-\tfrac{1}{6}\big)a^2(\eta)R(\eta) + a^2(\eta)m_\phi^2\,\Big]\,
\chi_k(\eta)=0,\\
\chi_k''(\eta)+\omega^2(\eta)\,\chi_k(\eta)=0,
\qquad
\label{eomx}
\end{align}
where
\begin{equation}
\omega^2(\eta)=\,k^2 + \big(\xi-\tfrac{1}{6}\big)a^2(\eta)R(\eta) + a^2(\eta)m_\phi^2.
\label{omegi}
\end{equation}
For the details of the calculations, see App. \ref{EoMder}. As it is obvious from the above equation, the general form of the EoM becomes exactly like the form of the SHO equation, and hence, now, the procedure of second quantization could be applied to the auxiliary field, $\chi$, as is conventional. To this end, we upgrade the field $\chi$ (or equivalently $\phi$) to an operator and write down its operator expansion as:
\begin{equation}
\hat{\phi}(\mathbf{x},\eta)=
\int \frac{d^3k}{(2\pi)^{3/2}\,a(\eta)}
\left(
e^{i\mathbf{k}\ \cdot\ \mathbf{x}}\;\hat{a}_{\mathbf{k}}\,\chi_k^{*}(\eta)
+ e^{-i\mathbf{k}\ \cdot\ \mathbf{x}}\;\hat{a}_{\mathbf{k}}^{\dagger}\,\chi_k(\eta)
\right)
\label{opex}
\end{equation}
where the expansion is written in terms of the conformal time, $\eta$, ${{\chi}_k}(\eta)$ is called the mode function, and each mode evolves independently through Eq. \ref{eomx}. ${{\hat{a}}^{\textsuperscript{\textdagger}}}_k$ and ${\hat{a}}_k$ are the well-known creation and annihilation operators satisfying the normalization condition described by the usual commutation relations:
\begin{align}
\left[\hat{a}_{\mathbf{k}},\,\hat{a}^{\dagger}_{\mathbf{k}'}\right]
=\delta(\mathbf{k}-\mathbf{k}') \\
\left[\hat{a}_{\mathbf{k}},\,\hat{a}_{\mathbf{k}'}\right]
=\left[a^{\dagger}_{\mathbf{k}},\,a^{\dagger}_{\mathbf{k}'}\right]=0
\label{com}
\end{align}
During our numerical calculations, we learned that if we rewrite the EoM in terms of dimensionless parameters, instead of dimensionful ones, the computation time decreases considerably. Hence, we rewrite the EoM, Eq. \ref{eomx}, by introducing the new dimensionless parameters:

\begin{align}
\Delta &\equiv H_{\mathrm{inf}}^{1/2}\,\chi,\\
\tau &\equiv H_{\mathrm{inf}}\,\eta,\\
\kappa &\equiv \frac{k}{H_{\mathrm{inf}}},\\
and\
\mu &\equiv \frac{m}{H_{\mathrm{inf}}}.
\label{dimlesspara}
\end{align}

Here $\Delta$ is the dimensionless auxiliary field, $\tau$ is the dimensionless conformal time, $\kappa$ is the dimensionless momentum (wave number), and $\mu$ is the dimensionless mass. Since in this study we are going to investigate the case of superheavy particles, the value of $\mu$ is of the order of $0.25$ to $10$. To write down the EoM in dimensionless form, it is also necessary to define the dimensionless curvature scalar as $C(\tau) \equiv \frac{R(\tau)}{{H_{inf}}^2}$. With these definitions in hand, we rewrite Eq. \ref{eomx} in the dimensionless form:
\begin{equation}
\frac{d^{2}\Delta_\kappa(\tau)}{d\tau^{2}}
+\Big[\,\kappa^{2}
+ \big(\xi-\tfrac{1}{6}\big)\,a^{2}C(\tau)
+ a^{2}\mu^{2}\,\Big]\Delta_\kappa(\tau)=0.
\label{main}
\end{equation}
This is the main EoM, which we are going to solve for two different eras of inflation and kination. We refer to the expression in the bracket as the dimensionless effective frequency of the field. This equation is a second-order linear differential equation in time, which has a two-dimensional space of solutions. These solutions are linearly independent, and if one finds a solution of the equation, for instance ${\chi}_k$, it is convenient to choose its complex conjugate, ${{\chi}^*}_k$, as the second general solution of the equation, and hence, the basis of the space of complex solutions becomes complete. So, the issue of finding the two independent solutions of the EoM reduces to finding only one solution (mode function), and picking that solution and its complex conjugate as the two bases or mode functions needed for the operator expansion, Eq.\ref{opex}. Using this mode function, the creation and annihilation operators are defined \cite{mukhanov2007introduction}, and hence, the vacuum state can be determined as a state that is annihilated by ${\hat{a}}_k$ for all modes. As a result, finding the solution of the EoM is equivalent to choosing a vacuum state. It is because of this fact that sometimes we use the terminology of the physical mode function and the physical vacuum instead of each other throughout our paper. It is essential to note that the mode function in the expansion must be found not as the general solution of the EoM, Eq.\ref{main}, but as the specific one. Hence, to reach the specific solution (mode function), one has to invoke the relevant initial conditions. The equation is second order, and hence two initial conditions are needed. Using these initial conditions and a general solution, a specific solution is found, and hence, the bases needed for the second quantization procedure are defined. As the mode function is selected, the vacuum state is defined, and hence, the physical particle states are obtained using the creation operator. We will refer to a specific mode function and its complex conjugate as "a set of mode functions" throughout this paper.

In the FLRW background with a time-dependent scale factor in its metric, a physical mode function considered at one time, ${\tau}_0$, may not necessarily be the physical mode function at another time, $\tau$. Due to the time-dependency of the background, the physical vacuum state at one time may contain particles from the viewpoint of an observer at a later time.

The relation between two different sets of physical mode functions defined at two different times is described using the Bogoliubov transformation. Ultimately, the square of the modulus of the Bogoliubov coefficient $\beta$ determines the particle production rate due to the time evolution of the background between the initial time ${\tau}_0$ and a later time $\tau$.  

\subsection{Bogoliubov Transformation}\label{bog}
Suppose there are two distinct slices of time, and different mode functions describe the physical particle states at each time. We denote these distinct mode functions as ${v_{\kappa}}(\tau)$ and ${u_{\kappa}}(\tau)$. First of all, these mode functions form a complete basis at two different times. Hence, each of them must satisfy the normalization condition as:
\begin{equation}
W\!\left[u_\kappa,u_\kappa^{*}\right]=W\!\left[v_\kappa,v_\kappa^{*}\right]=i
\label{norm}
\end{equation}
where the mathematical symbol ``$W[f, g]$" is the well-known Wronskian and is defined as $W[f, g] = fg'-f'g$, and as before,``prime" denotes the derivative with respect to the dimensionless conformal time, $\tau$. Since both of the sets $u_{\kappa}, {u^*}_{\kappa}$ and $v_{\kappa}, {v^*}_{\kappa}$ form a complete basis for the complex solution space of the EoM, one can expand $v_{\kappa}$ in terms of $u_{\kappa}$ and ${u^*}_{\kappa}$ as :
\begin{equation}
v_\kappa^{*}(\tau)=\alpha_\kappa\,u_\kappa^{*}+\beta_\kappa\,u_\kappa.
\label{vu}
\end{equation}

In this relation, ${\alpha}_{\kappa}$ and ${\beta}_{\kappa}$ are time-independent dimensionless coefficients, known as ``Bogoliubov coefficients". By substituting this expansion in the normalization condition, Eq. \ref{norm}, we obtain a constraint equation between the ${\alpha}_{\kappa}$ and ${\beta}_{\kappa}$:
\begin{equation}
|\alpha_\kappa|^{2}-|\beta_\kappa|^{2}=1,
\qquad
|\alpha_\kappa|\ge 1.
\label{alphabeta}
\end{equation}
Now we have two sets of mode functions in hand, and consequently, there would be two sets of creation and annihilation operators corresponding to each set of mode functions. We denote these sets as ${\hat{a}}_{\kappa}$ , ${{\hat{a}}^{\textsuperscript{\textdagger}}}_{\kappa}$ and ${\hat{b}}_{\kappa}$ , ${{\hat{b}}^{\textsuperscript{\textdagger}}}_{\kappa}$ for the $v_{\kappa}$ and $u_{\kappa}$, respectively. Hence, the operator expansion of the field, Eq. \ref{opex}, could be written using either $v_{\kappa}$ or $u_{\kappa}$, and their corresponding annihilation and creation operators. Using these two expansions for two different sets of modes, the Bogoliubov transformations are obtained as:
\begin{equation}
\hat{b}_{\mathbf{\kappa}}
=\alpha_\kappa\,\hat{a}_{\mathbf{\kappa}}
+\beta_\kappa^{*}\,\hat{a}^{\dagger}_{-\mathbf{\kappa}},\\
\qquad
\hat{b}^{\dagger}_{\mathbf{\kappa}}
=\alpha_\kappa^{*}\,\hat{a}^{\dagger}_{\mathbf{\kappa}}
+\beta_\kappa\,\hat{a}_{-\mathbf{\kappa}}.
\label{bogtrans}
\end{equation}
The Bogoliubov coefficients are time-independent numbers. Using the mode function  $u_{\kappa}$ or, correspondingly, the $b$ operators, one can build up the particle excited states using the creation operator. It is also the case for the $v_{\kappa}$ mode or the $a$ operators. We refer to these two sets of particle states as b-particles and a-particles. As a result, the Bogoliubov transformation, Eq. \ref{bogtrans}, demonstrates that the mean density of b-particles of mode $\kappa$ that the other observer $a$ detects in its own vacuum state is $n_{\kappa} = {\lvert {\beta}_{\kappa}\lvert}^2$. Suppose one seeks to calculate the GPP rate due to a transition in a time-dependent background (from one cosmological era to another). In that case, the two sets of mode functions or expansion operators must be chosen such that they describe the physical vacuum in each era. If the physical mode functions are chosen properly for each era, then the GPP is calculated through the ${\beta}_{\kappa}$ coefficient as:
\begin{equation}
|\beta_\kappa|
=
\left|
\frac{u_\kappa'\,v_\kappa-u_\kappa\,v_\kappa'}
     {\,v_\kappa\,v_\kappa^{*'}-v_\kappa'\,v_\kappa^{*}\,}
\right|_{\tau_0}
=
\left|
\frac{ W\!\left[v_\kappa,u_\kappa\right] }
     { W\!\left[v_\kappa,v_\kappa^{*}\right] }
\right|_{\tau_0}
=
\left|
\frac{ W\!\left[v_\kappa,u_\kappa\right] }{i}
\right|_{\tau_0}
=
\left|
W\!\left[v_\kappa,u_\kappa\right]
\right|_{\tau_0}
\label{beta1}
\end{equation}
The third equality is true if and only if the normalization condition Eq. \ref{norm} holds for the mode function $v_{\kappa}$. In the above equation, the time ${\tau}_0$ is supposed to be the time at which the gravitational particle production due to a transition between two cosmological era is calculated.

Hence, the main task to compute the GPP rate is to determine the physical mode functions corresponding to the physical vacuum, or equivalently, the physical particle states for the two different cosmological eras, before and after the transition, and then, using these mode functions, calculating the coefficient ${\beta}_{\kappa}$ through Eq. \ref{beta1}. The ${\beta}_{\kappa}$ or the Wronskian is calculated at a specific time. There are two kinds of calculations that we are going to follow in this paper: the numerical calculation and the analytical calculation for the inflation-kination transition. In the analytical calculation, we suppose that the transition from inflation to kination is almost sudden, and using a continuous piecewise function for the scale factor corresponding to two cosmological eras (inflation and kination), we derive the physical mode functions for each era, and finally, calculate the ${\beta}_{\kappa}$ and the GPP rate using the Wronskian in Eq. \ref{beta1} at the moment of the transition, ${\tau}_f$. The index $f$ demonstrates that this time corresponds to the end of inflation and the beginning of kination. It is obvious that such a model is only a toy model that will provide us with a quick and rough estimation of $T_{reh}$. \textbf{This is what we are going to do in Sec. \ref{anaresults}.} In contrast, in a more realistic manner, we know that the transition between two cosmological eras does not happen suddenly, but takes some time, which is known as the transition period or the transition time. If one considers the transition time in the calculation, the EoM can not be solved analytically from the beginning in the inflationary era to the final time in the kination era, as the physical mode function of inflation, and hence, it is necessary to invoke numerical calculations in this case. 

In order to perform numerical calculations, we first choose a function for the scale factor that continuously evolves from the inflation to the kination era. Secondly, using this scale factor function and also imposing the relevant initial conditions (known as the Bunch-Davis initial conditions which is, in fact, based on the notion of the adiabatic initial conditions), we solve the EoM from an initial time deep inside the inflationary era to a final time after the transition time in the kination era, numerically. This numerical solution with the appropriate initial conditions provides us with the physical mode function of the inflationary era. It is worth mentioning that with such a solution, the effect of the transition period is also considered and coded in the first physical mode function\footnote{In Sec. \ref{numresults}, during our surveying of the parameter space, we will also consider the effect of the transition time on the GPP rate and consequently, the $T_{reh}$. It turns out that the effect of the transition time is very important, and it is essential to consider its effect in the gravitational reheating studies.}. After that, the physical mode function of the kination era must also be derived. To this end, many previous studies in the literature\cite {pierre2023scalar, cembranos2020gravitational, capanelli2024gravitational, racco2024gravitational, verner2025nonminimal, cembranos2024vector, haro2019gravitational, haro2020gravitational, de2025note, chung2003classical} ] used the notion of WKB approximation for the adiabatic vacuum, where the K-vacuum, regardless of its phase, is considered the mode whose amplitude and time
derivative amplitude are equal to those of the WKB solution. Using these values, the GPP rate is calculated in the studies of gravitational reheating. As discussed in detail in the introduction, in the non-perturbative approach, the K-vacuum is not defined using the WKB approximation, which not only provides us with an approximate solution of the K-vacuum, but also its accuracy is fundamentally limited\cite{winitzki2005cosmological, mukhanov2007introduction}), but rather using the exact solution of the EoM in the kination era. Fortunately, in the case of the kination era, the EoM has an exact general solution that provides us with the opportunity to follow the approach proposed by \cite{bernard1977regularization, birrell1984quantum, ford2021cosmological, parker2009quantum, parker1976thermal, damour1996string}. This is the first pillar on which the non-perturbative approach is built, as mentioned in the introduction.

On the other hand, in the non-perturbative approach, the choice of the initial conditions is based on the notion of the adiabatic vacuum. As mentioned above, solving the EoM for the kination era only provides us with the general solution and hence the general form of the physical vacuum. To reach the exact form of the physical vacuum of the kination era (the K-vacuum), we have to impose valid initial conditions on the general solution. These adiabatic initial conditions for this era are referred to as the ``vacuum at short distances" in Ref. \cite{mukhanov2007introduction}. By imposing these initial conditions, the exact form of the K-vacuum is obtained analytically, and hence, using this vacuum and the vacuum of the inflationary era (obtained numerically), the GPP rate can be calculated using ${\beta}_{\kappa}$ through Eq. \ref{beta1}.

\subsection{The Physical Notion of Adiabatic Vacuum and Adiabaticity}\label{physadi}
In the discussions above, we argued that the notion of adiabatic vacuum is invoked in our procedure to impose physically meaningful initial conditions on the general solutions of the EoM for each era. On one hand, the notion of adiabatic vacuum is only valid in situations where a well-known inequality, known as the adiabaticity condition, holds. On the other hand, we mentioned that our procedure, the non-perturbative approach, is totally different from the other well-known procedure, where the physical mode function of each era is defined to be the adiabatic mode function itself, defined through the WKB approximation. So it seems important to make a clarification regarding the notion of the adiabatic vacuum and how this notion is invoked in our approach. Fundamentally, there is a profound difference between the well-known WKB adiabatic solution and the procedure that is used in our study. In addition, the notion of adiabaticity becomes very important when we are describing our numerical results, especially in the case of the effect of the transition time. Hence, to highlight such a difference in a clearer manner, and how the transition between two eras leads to GPP, let us discuss the physical notion of adiabaticity and the adiabatic vacuum in more detail.  

Firstly, we start with the historical notion of this terminology: If the effective frequency of the mode function in Eq. \ref{main}, ${\omega}_{\kappa}^2=\kappa^{2}
+ \big(\xi-\tfrac{1}{6}\big)\,a^{2}C(\tau)+ a^{2}\mu^{2}$ is a slowly-varying function of time for a specific era; such an era is called the adiabatic regime. The main constraint regarding the adiabatic regime is that ${\omega}_{\kappa}^2$ should be positive. During this regime, the evolution of the background is very close to the Minkowski case, with only a tiny difference, and hence, the adiabatic mode function corresponding to the adiabatic vacuum, which is obtained using the WKB mode function solution, can be used to define the physical vacuum and corresponding particle states. For further discussion, see App.\ \ref{wkbadi}. The slowly varying feature of a spacetime can be described quantitatively using the adiabaticity condition as $\lvert \frac{{\omega}_{\kappa}'}{{\omega}_{\kappa}^2}\lvert\ll1$. This condition demonstrates that the relative change of the effective frequency in the period of one oscillation, $\Delta\eta=\frac{2\pi}{{\omega}_{\kappa}^2}$, should be negligibly small.

If the $in$ and $out$ regimes are adiabatic ones, the in- and out-vacuum states are described through the adiabatic vacuum, defined through the WKB mode function. If one seeks to compute the GPP rate because of a transition between these two regimes, one should solve the EoM from an initial time in the $in$ regime just before the transition to a specific final time in the $out$ regime just after the transition period. Then, using the Bogoliubov transformations between the $in$ vacuum mode function computed at the final time and the adiabatic vacuum mode function of the $out$ era at that time, the GPP rate can be obtained. If the adiabaticity condition is not also violated noticeably during the transition time itself, the WKB approach can still be invoked to solve the EoM approximately from the initial time to the final time. In such a case, the GPP does not occur significantly.

So, the first note about the adiabatic vacuum and adiabaticity is that when one refers to the notion of an adiabatic mode function, it traditionally means that the physical mode function in each adiabatic era, where the adiabaticity condition holds, is defined using the WKB approximation. Of course, this notion of the adiabatic vacuum or mode function is not what we mean in this paper when we refer to the mode functions of each era.

Secondly, from the viewpoint of this study, the violation of the adiabaticity condition during the transition period is of particular interest. In this case, it is not important for us whether the $in$ and $out$ regimes are adiabatic or not. This is because, in our study, our center of attention is not the GPP during each era because of the time-varying background; instead, we seek to study the GPP obtained from the transition era.  In a situation where there is a transition time between two $in$ and $out$ eras, the evolution of the chosen physically meaningful positive-frequency mode function becomes important. To compute the GPP rate through $\lvert{\beta}_{\kappa}^2\lvert$, this evolution should be followed to the final time in the $out$ regime\footnote{This positive-frequency mode function is the coefficient of the annihilation operator in the mode expansion of the field and defines the physical vacuum state during the $in$ era.}. This mode function remains positive-frequency if the adiabaticity condition is not also violated noticeably even during the transition era. In this case, the transition time does not happen harshly, and hence, the positive-frequency mode has enough time to adapt gradually to the background at each time step and, in the end, adopt the physical mode function at the final time. In this case, there is no perceptible GPP at all. 

But if the background experiences a harsh transition during which the adiabaticity condition is violated, then a positive-frequency mode function at an earlier time is not necessarily a positive-frequency mode at a later time. Instead, it is a combination of positive-frequency and negative-frequency modes, with respect to the new background. In other words, the $in$ positive frequency can not adapt itself to the background evolution and tends to the physical positive-frequency mode function of the new background. So, it becomes an expansion of positive-frequency as well as the negative ones with respect to the physical mode function at later times. Then the physical mode functions of two eras can be expanded in terms of each other according to Eq. \ref{vu}, where the ${\beta}_{\kappa}$ coefficient determines the GPP rate. In fact, for a physical observer who is in the vacuum state at the final time, the $in$-vacuum state becomes an excited state with a noticeable occupation number. As a consequence, it is the violation of the adiabaticity condition, which is coded in the $in$ mode function during its passage through the transition period, that results in the noticeable GPP rate. This violation is the main reason for the difference between the vacuum mode functions of the $in$ and $out$ vacua, and such a difference yields a non-zero ${\beta}_{\kappa}$.

So, a second note about the notion of adiabaticity is that the violation of the adiabaticity condition during the transition era is a determinant factor in having a perceptible GPP rate. Now, it is time to clarify the fundamental difference between the notion of adiabatic mode function or vacuum invoked in our study and that in the WKB solution, and also highlight how we will return to the notion of adiabaticity when we are going to explain our numerical results:
\begin{itemize}
    \item  The notion of the adiabatic vacuum in our work is not used, as in the case in the WKB analysis mentioned above. The WKB mode function is used as an approximate solution of the EoM for an adiabatic regime, where the adiabaticity condition holds, and the background evolves slowly with time. The FLRW background is time-varying both during inflation and kination. For both of these eras, the adiabaticity condition may be violated for some mode frequencies. Hence, the WKB analysis becomes irrelevant for these eras. So in our case, we do not choose an approximate solution of the EoM as the physical mode function describing the vacuum of each era. Instead, we seek an exact non-perturbative solution to describe the vacuum state. But to find such a mode function, we invoke the adiabatic vacuum notion for the initial conditions of the EoM. For the inflationary era, in the limit of $\tau\rightarrow-\infty$, all modes lie deep inside the horizon and hence, in those circumstances, the adiabaticity condition holds for each mode. So we conclude that in that limit, the behavior of all modes should mimic the behavior of the mode function in Minkowski spacetime, or mode functions of the adiabatic vacuum. In the kination era, this situation happens in the limit of $\kappa\rightarrow\infty$. So, we do not claim that in these cosmological eras the physical vacuum mode function can be described using the WKB solution itself, but we found exact non-perturbative solutions that behave like adiabatic solutions in specific limits. As a result, the notion of the adiabatic vacuum provides us with meaningful physical initial conditions and nothing more.
    \item We want to emphasize that in this study our goal is computing the GPP rate and the resultant $T_{reh}$ solely because of the cosmological transition between inflation and kination eras. The GPP during the $in$ and $out$ eras themselves is not the subject of this study. Such a GPP, for instance, during the kination phase itself, is studied in some studies in the literature, such as Ref. \cite{lankinen2017gravitational}. In this study, we will not consider the GPP during inflation and kination, due to the violation of the adiabaticity condition and the time-dependency of the gravitational background during these eras; instead, we will concentrate only on the GPP resulting from the transition between these eras. 
    \item The GPP resulting from an inflation-kination transition is affected by the mass, gravitational coupling, and the duration of the transition time itself. We will examine these effects, and then we will invoke the adiabaticity condition as $\lvert \frac{{\omega}_{\kappa}'}{{\omega}_{\kappa}^2}\lvert\ll1$ to explain and justify our results. In fact, it is the violation of the adiabaticity condition that results in a noticeable GPP. We will explain in what is coming how each effect can violate this condition in its own way.
    \item It is important to note that the effective frequency of the field, ${\omega}^2_{\kappa}(\eta)$, should be a positive-valued number during the adiabatic regime. We also consider this constraint, but not because the inflation and kination eras are adiabatic regimes. We respect this constraint for two reasons: First, we want to impose the adiabatic vacuum prescription as our initial conditions at the beginning of our evolution in the inflationary era. Hence, the effective frequency should be positive at that time. Second, we are not going to study the GPP due to the tachyonic instability. Hence, to avoid such instability, we consider ${\omega}^2_{\kappa}(\eta)$ to be positive-valued throughout our evolution.
\end{itemize}

With these arguments in hand, we have to introduce a general model for the scale factor to derive our physical mode functions. Hence, our goal in the next subsections is to introduce a continuous function of the scale factor for our numerical analysis. The initial conditions that have to be imposed on the numerical solution of the EoM (for the inflationary era) are also presented. At the end, we derive the exact form of the physical adiabatic K-vacuum for the kination era, using the chosen continuous scale factor.

\subsection{The Scale Factor, Initial Conditions, and the K-vacuum Needed for the Numerical Analysis}\label{iniak}

There are many NO models of inflation in the literature\cite{felder1999inflation, spokoiny1993deflationary, peebles1999quintessential, ellis2021non, Haro:2019gsv, de2021review, dimopoulos2001models} with different transition times. Since our purpose is to survey the whole parameter space of the gravitational reheating, it turns out that the transition time, $\Delta\tau$, is one of the influential parameters in our study (it is obvious from the EoM, Eq. \ref{main}, that the other parameters whose effects must be considered are the mass of the scalar field, $\mu$, and the gravitational coupling, $\xi$.). In many studies of the GPP, the authors have chosen specific successful NO models for the inflation-to-kination transition, and then studied the GPP on such a background. Hence, it is obvious that the presented results are restricted to a specific value of $\Delta\tau$. To make it less arbitrary, and also to study the effect of transition time on the GPP rate, we will use a continuous function for the scale factor with a transition time $\Delta\tau$ as:
\begin{equation}
a^{2}(\tau)
=\frac{1}{2}\left[
\big(1-\tanh(\tfrac{\tau}{\Delta\tau})\big)\,\frac{1}{1+\tau^{2}}
+\big(1+\tanh\tfrac{\tau}{\Delta\tau}\big)\,(1+\tau)
\right].
\label{scalenum}
\end{equation}
This function provides us with the appropriate behavior of the evolution of the universe from the inflationary to the kination era, with an arbitrary transition time $\Delta\tau$. Any NO model of inflation can be included in this model if one sets its special transition time. It is important to note that this function is normalized so that the value of the scale factor at the end of inflation becomes equal to unity. Such a normalization is in harmony with the piecewise function that we will introduce for our analytical calculations\footnote{It is also important to mention that the normalization of the scale factor is quite arbitrary. In other words, different normalizations should not affect the physical results. Due to this fact, we have normalized our scale factors so that they become equal to unity. Such a normalization makes our calculations easier.}. In the limit of asymptotic past, far before the transition period, $\tau \rightarrow -\infty$, the $tanh$ tends to $-1$, and hence the asymptotic scale factor becomes ${a^2}(\tau) \rightarrow \frac{1}{1+{\tau}^2}\approx \frac{1}{{\tau}^2}$, which is the behavior of the scale factor corresponding to the inflationary era. In contrast, in the limit of the asymptotic future, far after the transition time,  $\tau \rightarrow \infty$, the $tanh$ tends to $+1$ and hence the asymptotic scale factor becomes ${a^2}(\tau) \rightarrow 1 + \tau \approx \tau$, corresponding to the behavior of the scale factor during the kination phase. This scale factor was introduced in \cite{hashiba2019gravitational, fujikura2023generation} to study the GPP for superheavy particles, the generation of DM, and also the baryon asymmetry of the universe in the context of quintessential models.

As discussed in the previous subsections, to compute the GPP rate and finally $T_{reh}$, we have to solve the EoM, Eq. \ref{main}, from an initial time deep inside the inflationary era to a final time at the beginning of the kination after the transition period using appropriate initial conditions. To this end, we note that if we go back in negative conformal time in the inflationary era, $\tau \rightarrow -\infty$, the size of the comoving cosmological horizon, $\frac{1}{aH}$, becomes greater than any comoving wavelength, or in other words, the physical wavelength becomes small in a way that every mode of the quantum fields lies deep inside the horizon. In such circumstances, the curvature of space-time becomes irrelevant for the modes, and all modes with different momenta do not feel the curvature of the universe. Hence, the functional form of their mode function must tend to the functional form of the well-known mode function of Minkowski space-time, ${{\Delta}_{\kappa}}(\tau) \rightarrow \sqrt{\frac{1}{2\kappa}}e^{-i\kappa\tau}$. Hence, the relevant initial conditions for solving the EoM are:
\begin{align}
\mathrm{IV}={\sqrt\frac{1}{2\kappa}}\,
e^{-\,i\kappa\tau}\\
\qquad
\mathrm{IV}'=-\,i\sqrt{\frac{\kappa}{2}}\,
e^{-\,i\kappa\tau}
\label{bd}
\end{align}
This notion of initial conditions is known as the Bunch-Davis initial conditions, relevant for the inflationary era. In fact, here we have used the notion of the adiabatic vacuum. This is because the basic idea of the adiabatic vacuum tells us that if the space-time background is a slowly changing function of time, or in other words, the momentum of the mode function is large enough (its wavelength is very small compared to the cosmological horizon), such a mode does not feel any changes in the background geometry with time. Hence, its behavior follows the functional form of the vacuum in the flat space-time. These conditions are known as adiabaticity conditions, and such a vacuum is known as the adiabatic vacuum. With these initial conditions in hand, we solve the EoM, Eq. \ref{main}, to a final time in the kination era. This solution provides us with the physical mode function (vacuum) of the inflationary era, in terms of which the operator expansion of the field is written. To calculate the Bogoliubov coefficient, ${\beta}_{\kappa}$, and hence the particle production rate, we have to choose a physical mode function (or equivalently, vacuum) for the kination era. To this end, we must also solve the EoM for the kination era. Fortunately, this solution can be found analytically due to the special form of the EoM during this cosmological era. 

First of all, we have to note that the cosmological scale factor in each era scales as $t^{\frac{2}{3(1+W)}}$ or ${\tau}^{\frac{2}{3W +1}}$ in terms of the cosmic and conformal time, respectively. Here $W$ is the equation of state parameter (EoS) for each cosmological era. Hence, for the kination with $\omega = 1$, the scale factor behaves as $a(\tau) \propto {\tau}^{\frac{1}{2}}$. On the other hand, the dimensionless Ricci scalar in Eq. \ref{main}, $C(\tau)$, can be written in terms of the scale factor as $C(\tau) = 6\frac{a^{\prime\prime}}{a^3}$, which for the kination era scales as $\frac{1}{{\tau}^3}$. Hence, in the effective frequency term ${\kappa}^2 + (\xi - \frac{1}{6})a^2C(\tau) + a^2{\mu}^2$, the second term becomes negligible compared to the bare mass term very soon after the transition time and can be ignored during the kination era. Although this procedure can be viewed as an approximation, it is sufficiently accurate. On the other hand, according to the function chosen for the scale factor in Eq. \ref{scalenum}, the factor $a^2(\tau)$ takes the form of $1 + \tau$ during kination. Taking into account all of these considerations, the EoM for the kination becomes:
\begin{equation}
\frac{d^{2}\Delta_{\kappa}(\tau)}{d\tau^{2}}
+\Big[\kappa^{2}+(1+\tau)\,\mu^{2}\Big]\,
\Delta_{\kappa}(\tau)=0.
\label{keom}
\end{equation}
Which yields the general solution of the form:
\begin{equation}
\Delta_{\kappa}(\tau)=
Q_{1}\,\sqrt{y}\,
H^{(1)}_{1/3}\!\left(\tfrac{2}{3}\,y^{3/2}\right)
\;+\;
Q_{2}\,\sqrt{y}\,
H^{(2)}_{1/3}\!\left(\tfrac{2}{3}\,y^{3/2}\right).
\label{kvg}
\end{equation}
In this equation, $Q_1$ and $Q_2$ are constant coefficients that have to be determined by imposing relevant final conditions. Also, $H^{(1)}$ and $H^{2}$ are the well-known Hankel functions of the first and the second kind, respectively. In addition, $y$ is a function which is defined as :
\begin{equation}
y=\frac{\kappa^{2}+\mu^{2}+\mu^{2}\tau}{\left(\mu^{2}\right)^{2/3}}.
\label{y}
\end{equation}
Despite the case of the inflationary era, where the initial conditions are imposed by going back in conformal time, $\tau \rightarrow -\infty$, it is not possible here to use such a limit for defining the relevant initial conditions. \textbf{In fact, we have to use another limit based on the notion of the adiabatic vacuum. Albeit, deep down, the notion of this limit is precisely the same as the limit used for the inflationary era, which is the fact that for modes deep inside the horizon, the evolution of space-time is irrelevant. But for the kination era, we refer to the fact that for high-frequency modes that lie inside the horizon with ${\kappa}^2 \gg a^2{\mu}^2$, the general behavior must follow the one for the mode in Minkowski space-time. This notion leads us to the adiabatic vacuum, and sometimes it is referred to as the ``vacuum at short distances" in the literature \cite{mukhanov2007introduction}.}

Imposing the aforementioned initial conditions in such a way that, in the limit of high-frequency modes, ${\kappa}^2 \gg a^2{\mu}^2$, the mode function and its first derivative follow the behavior of Minkowskian modes, the constant coefficients are determined, and the specific form of the mode function is obtained as :
\begin{align}
\Delta_{\kappa}^{\mathrm{kin}(n)}(\tau)
=\sqrt{\frac{\pi}{6}}\;\mu^{-1/3}\;
e^{
\,i\!\left\{\frac{2}{3}\frac{\kappa^{3}}{\mu^{2}}+\kappa-\frac{5\pi}{12}\right\}
}
\sqrt{y}\;
H^{(2)}_{1/3}\!\left(\tfrac{2}{3}\,y^{3/2}\right),\\
\qquad
y(\tau)=\frac{\kappa^{2}+\mu^{2}+\mu^{2}\tau}{\left(\mu^{2}\right)^{2/3}}.
\label{kv}
\end{align}
For the detailed steps in the path of derivation of the above mode function, see the App. \ref{kvder}. This mode function is obtained analytically by imposing the adiabatic conditions from the EoM for the kination era. Hence, it can be considered as the physical mode function that defines the physical vacuum of the kination era after the transition time. We refer to this mode as the K-vacuum (KV for short). Using the KV and the numerical solution, the particle production rate is calculated using the parameter ${\beta}_{\kappa}$. We present our results in Sec. \ref{numresults}. Our ultimate goal is to calculate $T_{reh}$ obtained from the GPP scenario, needed for the beginning of the hot Big Bang (HBB) era. To this end, in the next subsection, we are going to introduce the theoretical supplement essential to compute $T_{reh}$, as well as two possible reheating scenarios and their corresponding reheating temperature. 

\subsection{The Stability Relation}\label{treh}

The adiabatic notion of the vacuum can only be used if we have ${{\omega}_{\kappa}}^2>0$. This is explained in detail in \cite{mukhanov2007introduction, winitzki2005cosmological, kolb2024cosmological}. We use this assumption since we aim to use the notion of adiabatic vacuum as our initial conditions at relevent times, especially at the beginning of our evolution in the inflationary era. On the other hand, in this paper we do not consider production of particles due to the tachyonic instability. For these two reasons, we consider the square of the effective frequency, ${{\omega}_{\kappa}}^2>0$ in the EoM, Eq. \ref{main}, to be positive-valued, otherwise, tachyonic instability occurs. Returning to Eq. \ref{main}, the effective mass becomes:
\begin{equation}
M_{\mathrm{eff}}^{2}
=\left(\xi-\frac{1}{6}\right)a^2C(\tau)+a^2\mu^{2}.
\label{p1}
\end{equation}
The stability condition constrains us to $M_{eff}^{2}\geq0$. Using this constraint, we are going to derive a stability relation for each value of the bare mass. This relation implies:
\begin{equation}
\left(\xi-\frac{1}{6}\right)C+\mu^{2}\;\ge\;0.
\label{p2}
\end{equation}
This condition must be satisfied for both the inflationary and the kination eras. During the inflationary era, the Hubble parameter is considered to be approximately constant. As a result, due to the relation between the Hubble parameter and the Ricci scalar;
\begin{equation}
R=3(1-3W)H^2,
\label{p6}
\end{equation}
we see that the Ricci scalar is also a constant parameter. If one sets the EoS parameter of the inflationary model equal to $W=-1$, then the dimensionless curvature scalar becomes $C(\tau)=12$. Substituting this parameter in Eq. \ref{p2}, we obtain:
\begin{equation}
\xi\;\ge\;\frac{1}{6}-\frac{\mu^{2}}{12}.
\label{p3}
\end{equation}
This is the stability relation for the inflationary era. As we can see, this relation imposes a lower bound on the value of $\xi$ for a certain value of the bare mass of the $\Delta$-field. On the other hand, we know that according to Eq. \ref{p6}, the Ricci scalar is negative for $W>\frac{1}{3}$. Hence, if we consider (as we have done in this paper) that the EoS parameter during the kination era is equal to unity, the Ricci scalar just after the transition becomes $R_{in}=-6{H_{inf}}^2$, or in the language of the dimensionless parameters, we have $C_{inf}=-6$. Here, the index $in$ denotes that this value is just for the beginning of the kination era, just after the transition period. After that, as mentioned in the derivation of the KV, the curvature scalar damps very fast. Hence, if we consider the marginal case at the beginning of the kination era, and substitute $C_{in}=-6$ in Eq. \ref{p2}, we conclude that for the kination era:
\begin{equation}
\xi\;\le\;\frac{\mu^{2}+1}{6}
\label{p4}
\end{equation}
This relation imposes an upper bound on the value of $\xi$. Taking into account both Eqs. \ref{p3} and \ref{p4}, the stability relation is obtained as:
\begin{equation}
\frac{1}{6}-\frac{\mu^{2}}{12}\;\le\;\xi\;\le\;\frac{\mu^{2}+1}{6}
\label{p5}
\end{equation}
This relation demonstrates that for every certain value of the bare mass, the gravitational coupling must satisfy this inequality, and the possible range of $\xi$ gets wider as $\mu$ increases. It is an essential constraint if one wants to avoid the tachyonic instability, as well as to invoke the notion of the adiabatic vacuum as initial conditions to compute the resultant particle production rate. We will return to this stability relation several times during the next section.

\subsection{The Reheating Temperature}\label{treh}
The quantity ${\lvert {\beta}_{\kappa} \lvert}^2$ is the mean particle density for the mode $\kappa$, i.e. the mean particle number per momentum volume $d^3\bold{k}$ per spatial volume $d^3\bold{x}$. Hence, the number density of created particles is defined as:    
\begin{align}
n_{\phi}
=\int_{0}^{\infty} \frac{4\pi\,k^{2}\,dk}{(2\pi)^{3}}\;|\beta_{k}|^{2}\\
=\int_{0}^{\infty} \frac{4\pi\,H_{\mathrm{inf}}^{3}\,}{(2\pi)^{3}}\;
\kappa^{2}\,d\kappa|\beta_{\kappa}|^{2}
\label{n}
\end{align}
where we have written the above equation in terms of both dimensionful and dimensionless parameters\footnote{It is worth noting that the Bogoliubov coefficient $\beta$ is a Wronskian and hence is naturally dimensionless. As a result, it can be written in terms of both the dimensionful and dimensionless parameters without any coefficients.}. Note that ${\lvert {\beta}_{\kappa} \lvert}^2$ is isotropic in the $\kappa$-space. Since ${\lvert {\beta}_{\kappa} \lvert}^2$ is the particle density, the path to the definition of the energy density is straightforward. To this end, we notice that multiplying ${\lvert {\beta}_{\kappa} \lvert}^2$ to the factor $\sqrt{m^2 + k^2}$ provides us with the energy density of the mode $k$ per spatial and momentum volume. Hence, integrating over the momentum space yields the total energy density of created particles as:
\begin{equation}
\rho_\phi
=\int_{0}^{\infty}\!\frac{4\pi\,k^{2}\,dk}{(2\pi)^{3}}
\,\sqrt{m^{2}+k^{2}}\;|\beta_{k}|^{2}
= H_{\rm inf}^{4}\!
\int_{0}^{\infty}\!\frac{4\pi\,\kappa^{2}\,d\kappa}{(2\pi)^{3}}
\,\sqrt{\mu^{2}+\kappa^{2}}\;|\beta_{\kappa}|^{2}
\label{rho}
\end{equation}
which again is written in terms of both dimensionless and dimensionful parameters\footnote{Although at the end of the calculations all of the equations have to be written in terms of the dimensionful quantities, we also provide the reader with the dimensionless equations that have been used in the codes of our numerical analysis.}.

With these relations in hand, it is now time to write down the reheating temperature equations. There are, in fact, two scenarios of reheating in this case, based on the following important point. \textbf{As we will see in our numerical results, Fig. \ref{fig3}, the energy density of the created particles has a peak around the mass $\kappa \simeq \mu$, and is suppressed at high frequencies. This is completely in agreement with our physical intuition that the creation of particles with high momenta is less likely than lower momenta. On the other hand, the wavelength of these modes is very small compared to the cosmological horizon, due to their high momentum or wavenumber, and hence, the curvature of the space-time becomes irrelevant for them. On the other hand, in this study, we are going to consider the gravitationally produced superheavy particles whose momentum becomes negligible compared to their rest mass soon due to the expansion of the universe and cosmological redshift. Based on this reasoning, the created particles become non-relativistic soon after their creation in the kination era.} As a result, we can treat them as non-relativistic pressureless matter with EoS parameter $W = 0$. 

These particles are superheavy, and hence they will decay to the SM particles soon enough after their creation if one considers an appropriate coupling (decay channel) between them and the SM fields. So there are two possible reheating scenarios in this approach:

\begin{itemize}
    \item \textbf{Immediate decay and thermalization:} In this scenario, the scalar particles decay and thermalize immediately after their creation. This scenario of decay was considered in \cite{ford1987gravitational, kunimitsu2012higgs, nishi2016reheating}. Since the decay happens almost instantly at the beginning of kination, the energy density of SM radiation is equal to the energy density of the spectator field, $\phi$. At this time, the dominant energy component of the universe is still in the hands of the inflaton field. After decay, the energy density of the produced thermal bath decays like radiation with $W = \frac{1}{3}$ as $a^{-4}$, while the energy density of the inflaton during the kination scales as $a^{-6}$. As a result, at one specific time, these energy densities become equal. We refer to this time as reheating, and the temperature of the thermal bath at this moment is called the first reheating temperature, ${T^{(1)}}_{reh}$. This temperature can be obtained from the energy density of created particles resulting from our numerical analysis as:
\end{itemize}
\begin{equation}
T_{\rm reh}^{(1)}
=\left(\frac{10}{3\pi^{2}g_*}\right)^{\!1/4}
\frac{1}{M_{\rm Pl}^{\star} H_{\rm inf}}
\left(\rho_\phi^{\,I}\right)^{3/4}
=\left(\frac{10}{3\pi^{2}g_*}\right)^{\!1/4}
\frac{1}{M_{\rm Pl}^{\star} H_{\rm inf}}
\left(\rho_\phi^{\,I^{'}}\,H_{\rm inf}^{4}\right)^{3/4}
\label{t1}
\end{equation}

In the above equations, $g_{\star}$ is the total number of relativistic degrees of freedom contributing to the energy and entropy density of the SM thermal bath. ${M_{pl}}^{\star}$ is the reduced Planck mass, $H_{inf}$ is the constant value of the Hubble parameter during the inflationary era,, and ${{\rho}_{\phi}}^I$ is the energy density of gravitationally produced particles of the spectator field just after the transition time in the kination epoch. Similarly,${{{\rho}_{\phi}}^I}^{\prime}$ is the dimensionless energy density of created particles. To follow the derivation path of the $T_{reh}$ formulas, see App. \ref{trehder}.

\begin{itemize}
    \item \textbf{Decay at the time of equality:} In this scenario, the produced scalar particles decay at the time that their energy density dominates that of the inflaton. For the reasons discussed above, the created particles of the superheavy spectator field can be treated as a non-relativistic pressureless fluid and hence, their energy density scales as $a^{-3}$. At the equality time when the energy density of these particles becomes equal to that of the inflaton, these particles decay to the SM thermal bath and the universe reheats due to this decay. Since in this scenario, the decrease in the energy of the created particles is smaller than in the previous case, ${T^{(2)}}_{reh}$ is always greater than ${T^{(1)}}_{reh}$ and is obtained as:   
\end{itemize}
\begin{equation}
T_{\rm reh}^{(2)}
=\left(\frac{10}{\pi^{2}g_*}\right)^{\!1/4}
\frac{1}{\big(M_{\rm Pl}^{\star} H_{\rm inf}\big)^{1/2}}
\left(\rho_\phi^{\,I}\right)^{1/2}
=\left(\frac{10}{\pi^{2}g_*}\right)^{\!1/4}
\frac{1}{\big(M_{\rm Pl}^{\star} H_{\rm inf}\big)^{1/2}}
\left(\rho_\phi^{\,I^{'}}\,H_{\rm inf}^{4}\right)^{1/2}
\label{t2}
\end{equation}

Now, all of the essential theoretical supplements needed for our results are described. With these supplements in hand, we are going to present our results, over the whole parameter space, in the next section.

\section{Results of Numerical Calculation}\label{numresults}
\textbf{In this section, we present the main results of our numerical analysis by surveying the whole parameter space of the model during the inflation-kination transition.} It is essential to notice that although such a study can roughly be done using an analytical solution, as we will do in Sec. \ref{anaresults}, which provides us with a hasty estimation of $T_{reh}$ and the GPP rate, such solutions are obtained only if one considers the transition to be instantaneous. In realistic models of inflation, the transition to the post-inflationary era occurs through a transition period of time. Considering this fact about realistic models, the EoM can not be solved analytically, and hence, it is necessary to invoke a numerical approach. To this end, we solve Eq. \ref{main} numerically after substituting the adopted function for the scale factor with the transition time $\Delta\tau$ in Eq. \ref{scalenum}, together with the Bunch-Davis initial conditions. With the numerical solution in hand, we compute the particle production rate, ${\lvert {\beta}_{\kappa} \lvert}^2$, using the KV introduced in Eq. \ref{kv}. We present the results of the quantity ${\lvert {\beta}_{\kappa} \lvert}^2$ in relevant plots. As introduced in Eq. \ref{rho}, although the energy density stored in each mode of created particles is $\sqrt{{\mu}^2+{\kappa}^2}{\lvert {\beta}_{\kappa} \lvert}^2$, it is conventional in the literature to present the total integrand of the integral of ${\rho}_{\phi}$ as the energy density of each mode. For the sake of coordination with other studies of GPP, we will follow this definition and present the numerical results of the quantity $\frac{4\pi{\kappa}^2}{(2\pi)^3} \sqrt{{\mu}^2+{\kappa}^2}{\lvert {\beta}_{\kappa} \lvert}^2$ as the parameter ${\rho}_{\kappa}$. At the end, using the integral in Eq. \ref{rho}, Eq. \ref{t1}, and Eq. \ref{t2}, we present the resultant reheating temperature for both decay scenarios. 

The reduced Planck mass, ${M_{pl}}^{\star}$ is set as $2.44 \times 10^{18} GeV$\cite{nikandish2024signature}. The value of the Hubble parameter for the inflationary era, $H_{inf}$, is chosen to be $6.6 \times 10^{13} GeV$ in accordance with \cite{akrami2020planck, ade2016planck}. In addition, the number of relativistic degrees of freedom, $g_{\star}$, is chosen to be $106.75$\cite{saikawa2018primordial}. With these values for the constant parameters in hand and the procedure discussed above, we will solve our EoM numerically. The survey of the parameter space is done as follows: In the first subsection, we will study the effect of mass on the GPP and, as a result, on the reheating temperature. We will present our results for the first and second scenarios of $T_{reh}$ in separate plots. The second subsection is dedicated to the effect of the non-conformal coupling on the GPP rate, on the energy density spectrum of the produced particles, and finally, on the resultant reheating temperature. \textbf{This subsection is of particular importance, since in this study our main goal is to investigate the effect of non-conformal values of the coupling $\xi$ on GPP. It turns out that to study the GPP using the Bogoliubov procedure, there must be a constraint inequality on the value of $\xi$, depending on the bare mass of the scalar field, $\phi$. Hence, the coupling constant can take positive and negative non-conformal values, which amplify the gravitational production of scalar particles significantly, and subsequently increase the resultant $T_{reh}$, making this channel of reheating important.} Both of the two aforementioned parameters, $\xi$ and $\mu$, are model parameters of the theory of scalar fields, where the transition time is a characteristic feature of the inflationary model. Hence, in the last subsection of this section, we will study the effect of the different transition times (equivalently, different inflationary models) on the GPP rate and $T_{reh}$. It turns out that its effect is substantial, and these studies survey the full parameter space of the model. 

\subsection{The Effect of Mass on the GPP Rate and $T_{reh}$}\label{masseffect}
\begin{figure}
	\subfloat[]{\includegraphics[width=0.5\textwidth]{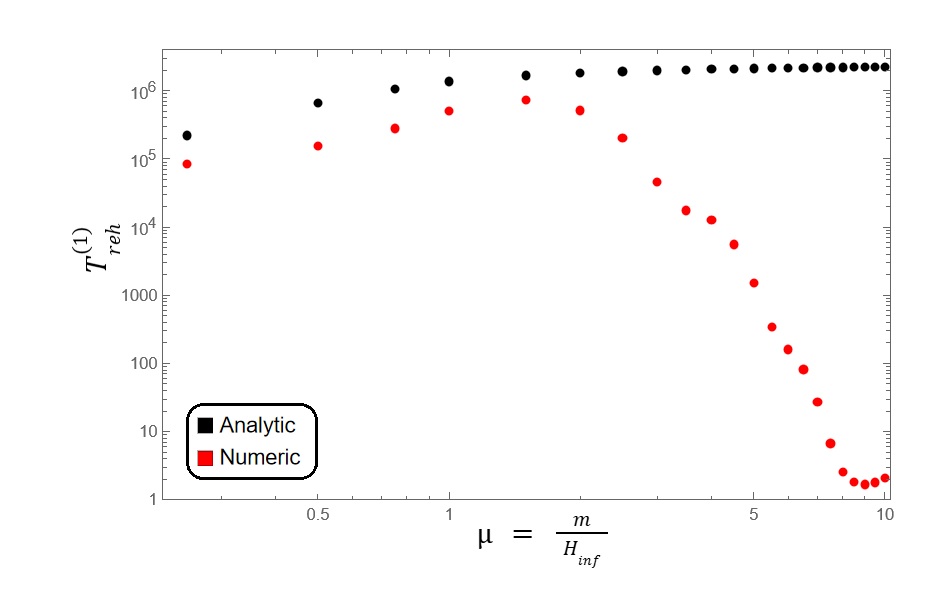}}
	\hfill
	\subfloat[]{\includegraphics[width=0.5\textwidth]{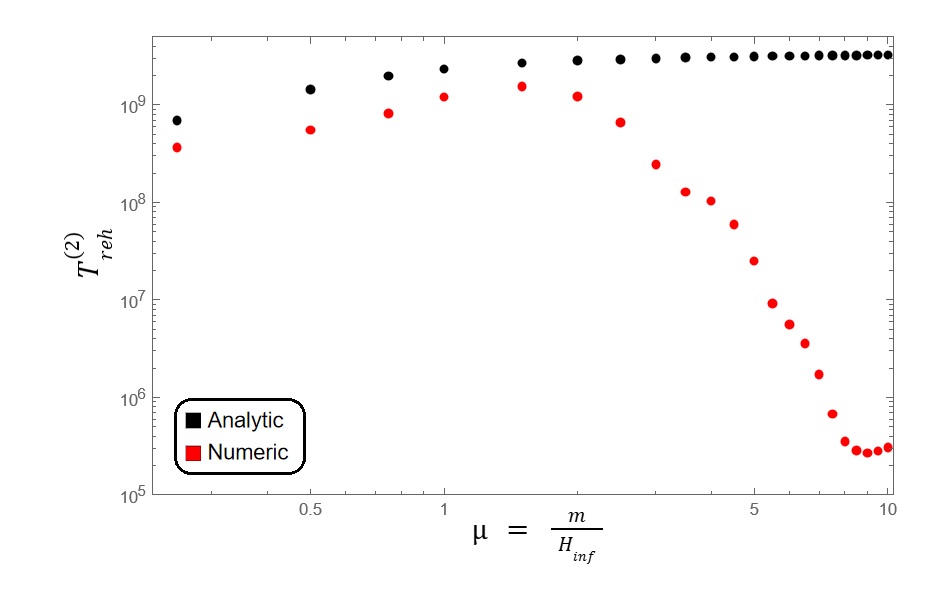}}
	\caption{The resultant reheating temperature of the (a) first scenario and (b) second scenario of decay, as a function of mass. In this plot, the effect of mass is studied in the case of $\xi=\frac{1}{6}$. The $red$ plot is for the numerical calculation with $\Delta\tau=1$, and the $black$ plot is for the analytical calculation.}\label{fig1}
\end{figure}

The first parameter that plays a role in the process of gravitational particle production of a superheavy scalar field is the bare mass of the scalar itself, $\mu$. To study the effect of mass alone, we will set the transition time to the value $\Delta\tau=1$. This specific value represents a general estimate for the transition time, which is consistent with many general well-known inflationary models. As a consequence, with this choice, we will stay on the side of generality. On the other hand, and more importantly, we will choose the case of conformal coupling for the value of $\xi$. With this choice, the term containing the gravitational coupling in the EoM disappears, and hence, there remains only the effect of mass that leaves its footprint on the GPP rate. In fact, we leave the consideration of the effect of the non-conformal values of $\xi$ for the next subsection. Hence, we will solve the EoM, setting $\xi=\frac{1}{6}$ and $\Delta\tau=1$, $\mu=0.25$ to $\mu=10$, from an initial time to a proper final time in the kination epoch after the transition time, and will find the numerical physical mode function, respecting the adiabatic initial conditions, for the inflationary era. Then, using this solution and also invoking the KV obtained in Eq. \ref{kv}, we will calculate the GPP rate, ${\lvert{\beta}_{\kappa}\lvert}^2$, and the resultant ${\rho}_{\kappa}$. Integrating over all values of the momentum space, the total energy density and hence the reheating temperatures of both scenarios are obtained. We have presented our numerical results in Fig. \ref{fig1} in $Red$ color. 

\textbf{The first thing that can be seen from Fig. \ref{fig1} is that the reheating temperature increases with increasing mass initially. After a specific value of mass, $\mu=1.5$, it starts to decrease, so that the lowest value of the temperature is obtained for the highest values of mass.} This behavior can be explained as follows: Returning to the definition of the energy density of each mode of $\kappa$, $\frac{4\pi{\kappa}^2}{(2\pi)^3} \sqrt{{\mu}^2+{\kappa}^2}{\lvert {\beta}_{\kappa} \lvert}^2$, we see that by increasing the bare mass, the energy density produced in each mode also increases, due to the term $\sqrt{{\mu}^2+{\kappa}^2}$. But, on the other hand, since increasing $\mu$ results in an increase in the effective frequency of the mode, the number density of produced particles will start to decrease. This, in turn, causes a total reduction of the GPP rate for greater values of mass. There is a reasonable mathematical interpretation regarding this issue:

Returning to the Eq. \ref{main}, the effective frequency of the field $\Delta$, is ${\omega}_{\kappa}^2=\kappa^{2}+ \big(\xi-\tfrac{1}{6}\big)\,a^{2}C(\tau)+ a^{2}\mu^{2}$. Using the fact that the adiabaticity condition is violated when $\lvert\frac{{\omega}'_{\kappa}}{{\omega}_{\kappa}^2}\lvert\gg 1$, we obtain the inequality:

\begin{equation}
\lvert\frac{{\omega}'_{\kappa}}{{\omega}_{\kappa}^2}\lvert=\lvert\frac{2(\xi-\frac{1}{6})aa'C(\tau)+(\xi-\frac{1}{6})a^2C'(\tau)+2aa'{\mu}^2}{({\kappa}^2+(\xi-\frac{1}{6})a^2C(\tau)+a^2{\mu}^2)^{\frac{3}{2}}}\lvert\gg1.
\label{adicon}
\end{equation}

This relation tells us why, for large values of mass, the GPP rate is suppressed. In the case of conformal coupling, the terms $a(\xi-\frac{1}{6})aa'C(\tau)$, $(\xi-\frac{1}{6})a^2C'(\tau)$ and $(\xi-\frac{1}{6})a^2C(\tau)$ are equal to zero. There is a term containing ${\mu}^2$ in the numerator, while there is also a term with ${\mu}^3$ in the denominator of the inequality. Hence, the parameter $\lvert\frac{{\omega}'_{\kappa}}{{\omega}_{\kappa}^2}\lvert$ is a function of bare mass with almost a functional form of $\frac{1}{\mu}$. So, for large values of mass, with large $\mu$, the adiabaticity condition is not violated significantly, and $\lvert\frac{{\omega}'_{\kappa}}{{\omega}_{\kappa}^2}\lvert$ does not grow enough. So the GPP is suppressed for these masses. This is the reason for such suppression due to the mass effect. So the spectrum of $T_{reh}$ increases initially and then starts to decrease for large values of mass. In this case, it has a maximum around $\mu=1.5$.

We are also able to provide our readers with a physically intuitive explanation: In this analysis, despite the case in the subsection on the effect of $\Delta\tau$, we consider $\Delta\tau$ to be of the order of unity for the sake of generality. As can be seen from Fig. \ref{scalebeh}, this choice provides us with a scale factor that behaves almost smoothly during the transition time. In the effective frequency, we have a term $a^2{\mu}^2$. For large values of mass, due to the smoothness of $a(\tau)$ in this case, the relative change of this term is negligible compared with the value of the effective frequency itself. So, the case with large masses behaves just like a high-frequency field mode. As we know, these modes are apathetic with respect to the changes in the background, and these changes in the curvature of spacetime are irrelevant for them. As a result, we do not expect much excitation for these values of mass. As a consequence, the GPP rate and the resultant $T_{reh}$ are suppressed, which in turn brings us back to this old proposition that in the study of reheating, the production of very heavy particles is suppressed because their production drains more energy compared with the lighter particles from the background. These reasons explain the behavior that is depicted in Fig. \ref{fig1} for the effect of mass on the produced reheating temperature in the conformal case and with $\Delta\tau=1$. In the next subsection, we will consider the non-conformal coupling, and we will see that the presence of the gravitational coupling changes this plot substantially.  

\subsection{The Effect of Non-conformal Gravitational Coupling on the GPP Rate and $T_{reh}$}\label{xieffect}
\begin{figure}
	\subfloat[]{\includegraphics[width=0.5\textwidth]{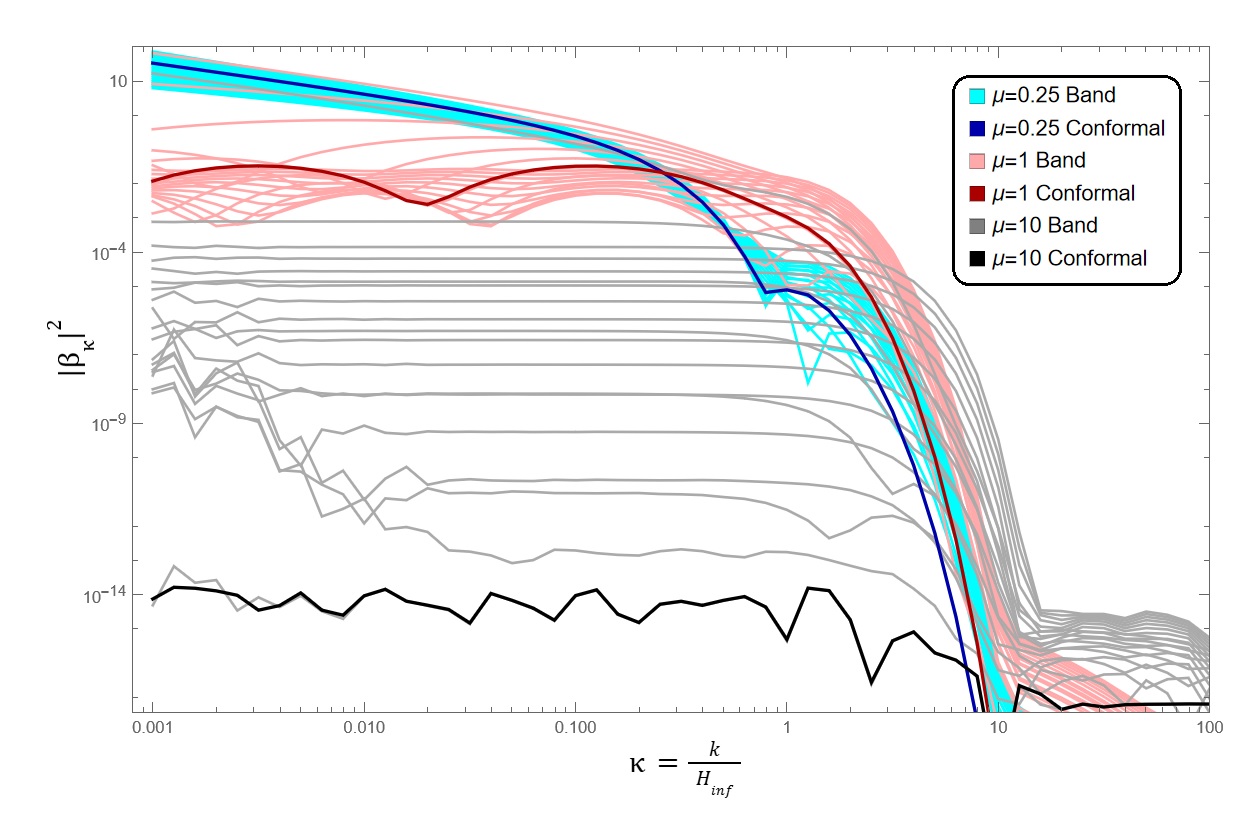}}
	\hfill
	\subfloat[]{\includegraphics[width=0.5\textwidth]{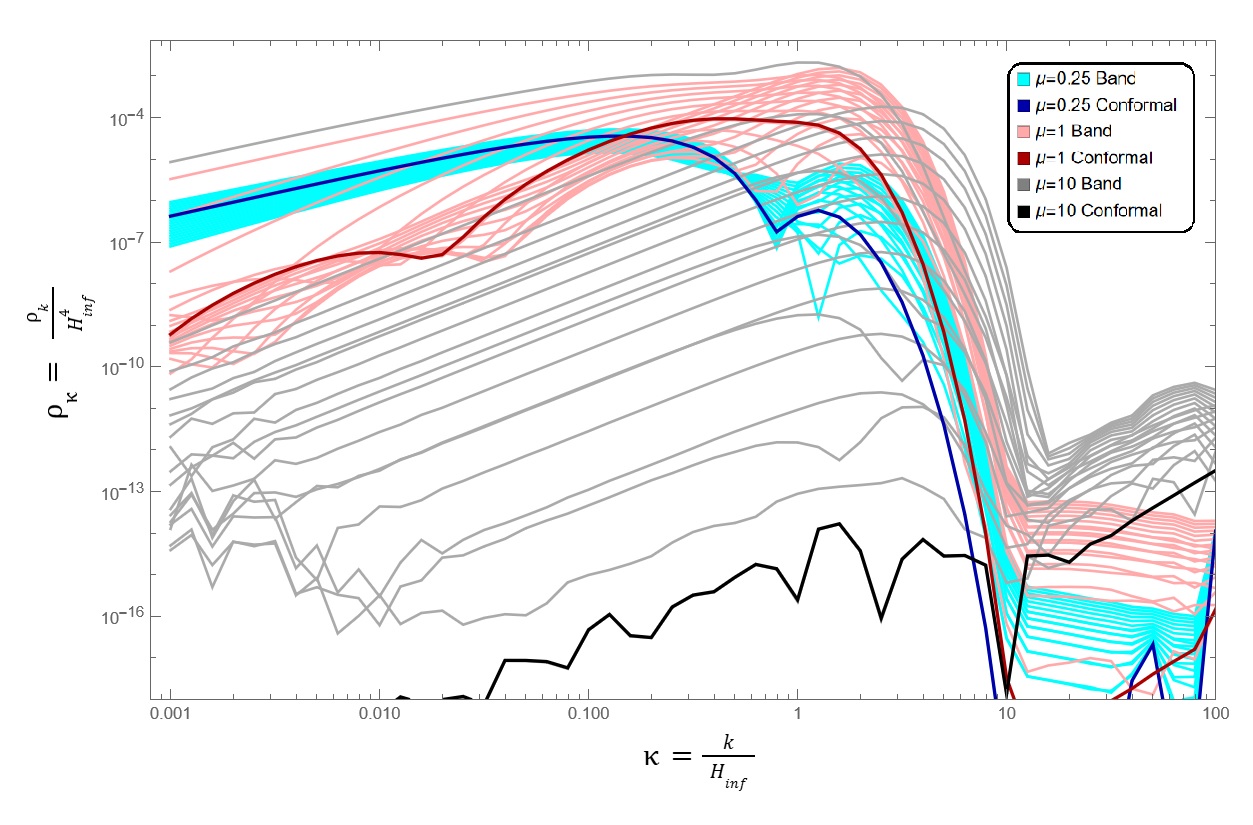}}
	\caption{The Band Plot: How the parameter space of the theory is broadened by considering the non-conformal values of $\xi$. (a) The GPP rate and (b) the produced energy density are plotted for $\mu= 0.25, 1, 10$ and $\Delta\tau=1$ in terms of the dimensionless momentum. In these plots, the GPP rate and the energy density spectra are depicted for the conformal case, as well as some non-conformal values of allowed $\xi$ from ${\xi}_{min}$ to ${\xi}_{max}$. As can be seen, considering the effect of non-conformal coupling, the parameter space is broadened and the spectrum can be very different from the conformal case. The difference is more perceptible for higher values of mass.}\label{fig2}
\end{figure}

In this subsection, we will consider non-conformal values for the gravitational coupling constant $\xi$. Using the stability relation, Eq. \ref{p5}, we will obtain the interval of the allowed values of $\xi$ for each value of bare mass $\mu$. For each mass, there are two values of $\xi$, ${\xi}_L$ and ${\xi}_R$, that correspond to the lowest and the highest values of $\xi$ allowed in the stability relation. In order to avoid the problem of tachyonic instability, we run our code from ${\xi}_{min}$ to ${\xi}_{max}$, which are very close to ${\xi}_L$ and ${\xi}_R$, respectively, but not precisely equal. We provide the reader with their values in Table. \ref{table} in App. \ref{tab}. Following the well-explained procedure of computing the GPP rate, the energy density of each mode, and the reheating temperature, we have studied the effect of non-conformal coupling for each value of the superheavy mass. The results are as follows:

As depicted in Fig. \ref{fig2}, which we will refer to as the \textbf{band plot}, we see that the space of solutions is expanded in a very interesting way. In the band plot, we have depicted the GPP rate and the energy density of each mass for the conformal case, also. We see that the GPP rate and the energy density can be very different for different allowed values of $\xi$. \textbf{Although the non-conformal values of $\xi$ do not considerably change the GPP rate and the produced energy density for low values of $\mu$, e.g. $\mu=0.25$, there is a substantial change for higher values of mass, like $\mu=10$.} It is due to the fact that by returning to the stability relation, Eq. \ref{p5}, for lower values of mass, ${\xi}_L$ and ${\xi}_R$ are very close to the conformal value. As a result, the parameter space for these masses does not expand enough and hence can not affect the GPP rate and the produced energy density to a considerable extent. The greater the value of the bare mass, the $\xi$ interval extends to wider values that are different from the conformal case. Hence, more different values for $\xi$ other than the conformal one are allowed, which in turn results in very different plots and changes the GPP rate significantly. 

For the lower values of the mass, $\xi$ can only acquire positive values, as shown in the Table. \ref{table} for masses greater than $\mu=1.5$, $\xi$ is allowed to have negative values. \textbf{The non-conformal coupling only changes the physical mode function of the inflationary era. For the kination era, the term $(\xi - \frac{1}{6})a^2C(\tau)$ damps very soon after the transition period. Since the curvature scalar is almost constant during inflation, $(\xi - \frac{1}{6})C(\tau)$ mimics the behavior of the constant bare mass term in this era. Hence, because of the gravitational coupling, the effective mass term, $(\xi - \frac{1}{6})a^2C(\tau) + a^2{\mu}^2$, acquires values that are significantly different from the effective mass term in the conformal case where $\xi=\frac{1}{6}$ and ${M^2_{eff}}=a^2{\mu}^2$. We refer to the modified mass term of the field during inflation, $(\xi - \frac{1}{6})C(\tau) +{\mu}^2$, as ICM (Inflationary Constant Mass). For positive values of $\xi$, the ICM increases, while for negative values it decreases compared with the bare mass of the field, $\mu$. } 

This change provides us with two mode functions corresponding to two different eras with two different effective frequencies. It is worth noting that in the conformal case, the difference between the time-dependent frequencies of two eras, ${\omega}^2_{\kappa}$, arises only from the difference in the behavior of the scale factor before and after the transition. By considering the non-conformal values of $\xi$, there appears another difference in the effective frequency, which is the difference between the ICM and bare mass. In the kination era, it is only the term ${\mu}^2$ that acts as the bare mass term of the field, while during inflation, the ICM plays this role. In fact, in this case, the difference in ${\omega}^2_{\kappa}$ is due to two effects: the difference between the bare mass and the ICM, which is due to the non-conformal coupling, and the difference in the behavior of the scale factor, which is due to the change of the spacetime because of the transition. In the conformal case, only the second effect was the reason for the difference between the physical mode functions of the two eras, and the bare masses were just the same. Hence, it is reasonable from our physical intuition that in the non-conformal case, the GPP, the produced energy density, and, at the end, the resultant reheating temperature experience substantial enhancement compared with the conformal case. From the mathematical point of view, the GPP rate is obtained through a mathematical Wronskian between the two mode functions. Both of these mode functions are solutions of the EoM. The Wronskian becomes larger if the difference between the functional forms of the two mode functions is noticeable. In the conformal case, the difference arises only because of the change in the background, which is coded in the scale factor. But in the case of non-conformal values of $\xi$, the difference becomes more significant, which in turn causes an amplified GPP rate.

There is another interesting result that can be obtained from Fig. \ref{fig2}. As can be seen, the GPP rate, or ${\lvert{\beta}_{\kappa}\lvert}^2$ equivalently, is suppressed for larger values of $\kappa$. This effect can be explained through the adiabaticity condition. Returning to Eq. \ref{adicon}, the quantity $\lvert\frac{{\omega}'_{\kappa}}{{\omega}_{\kappa}^2}\lvert$ becomes smaller for larger values of $\kappa$. This happens because of ${\kappa}^2$ in the denominator of the adiabaticity condition. Hence, for larger values of $\kappa$, $\lvert\frac{{\omega}'_{\kappa}}{{\omega}_{\kappa}^2}\lvert$ is not violated noticeably, and so the GPP is suppressed for these high-frequency modes. In fact, from our physical intuition, these high-frequency modes do not feel the curvature of spacetime in a perceptible manner, and hence, they mimic the behavior of modes in the Minkowski background. This behavior is referred to as``vacuum at short distances" in the literature.

\begin{figure}
	\subfloat[]{\includegraphics[width=0.48\textwidth]{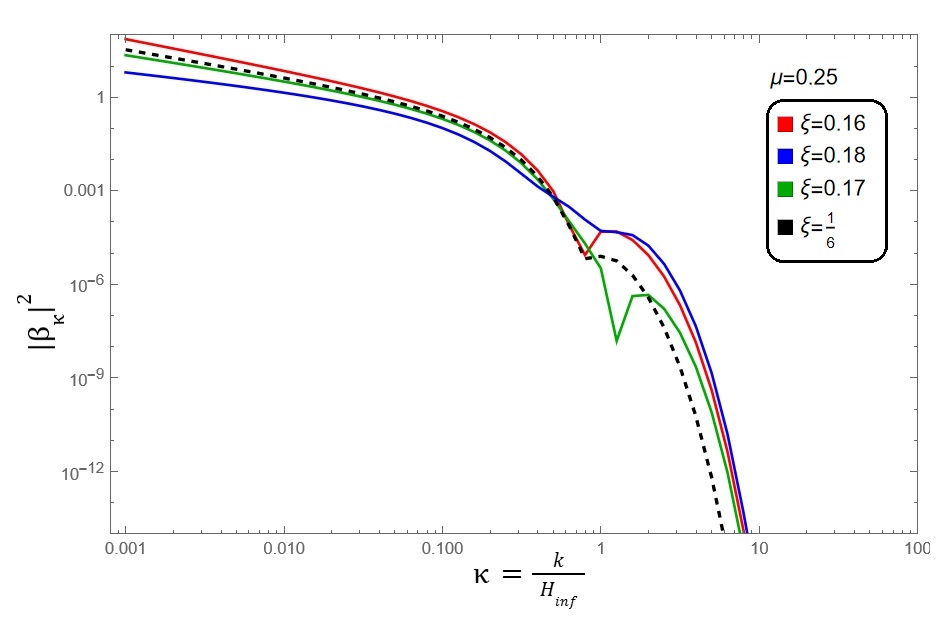}}
	\hfill
	\subfloat[]{\includegraphics[width=0.48\textwidth]{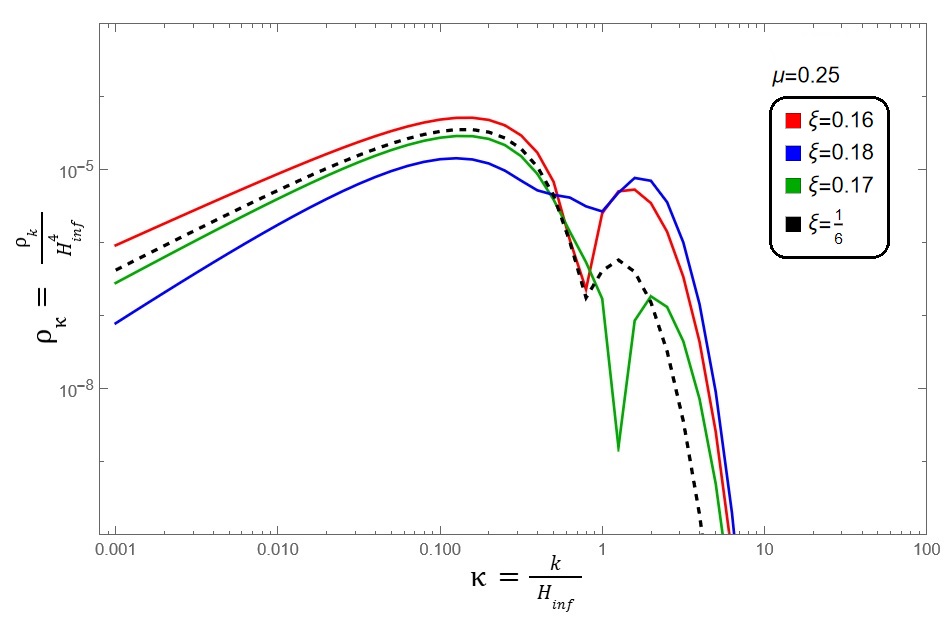}}\\
    \subfloat[]{\includegraphics[width=0.48\textwidth]{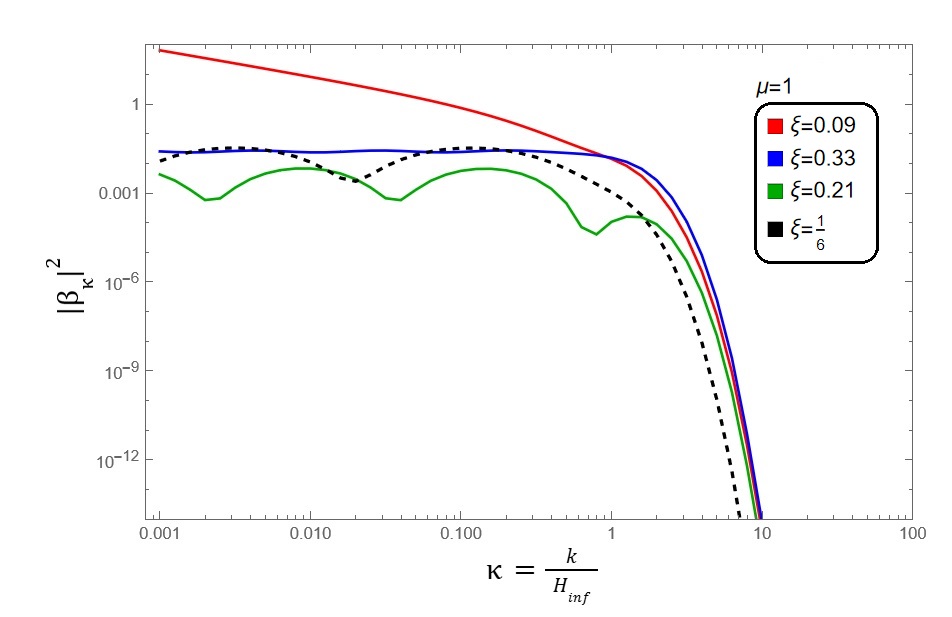}}
    \hfill
    \subfloat[]{\includegraphics[width=0.48\textwidth]{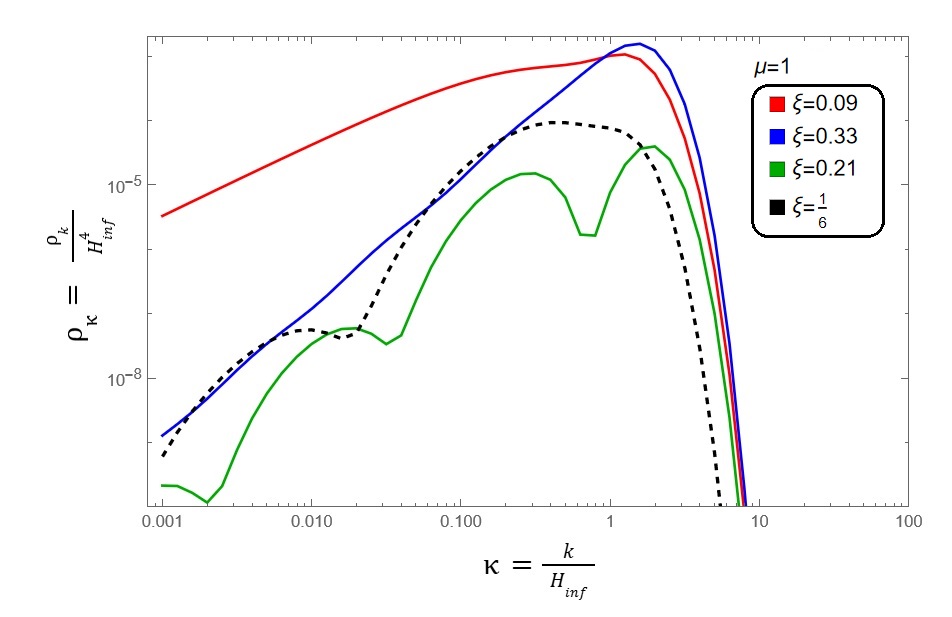}}\\
    \subfloat[]{\includegraphics[width=0.48\textwidth]{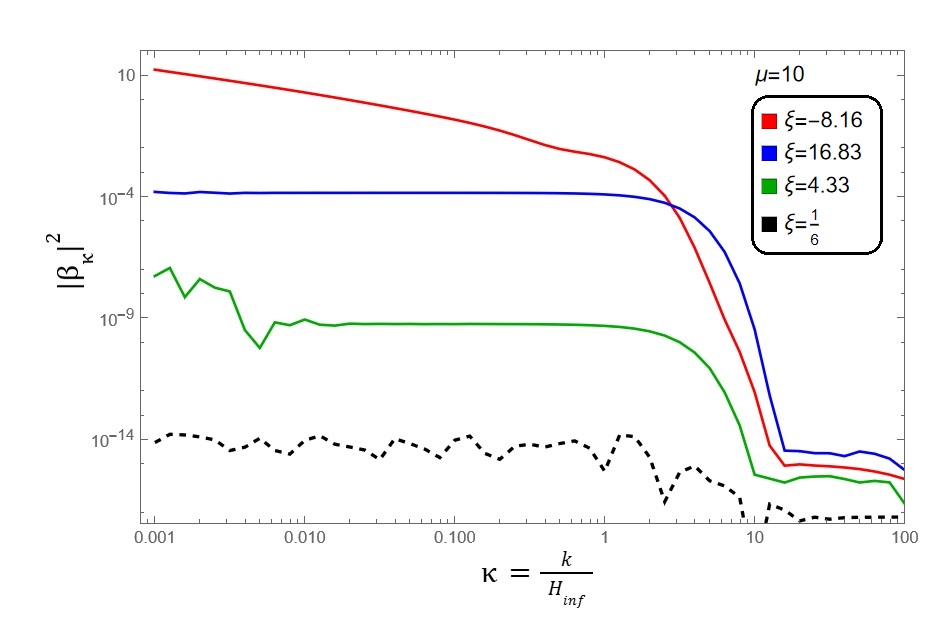}}
    \hfill
    \subfloat[]{\includegraphics[width=0.48\textwidth]{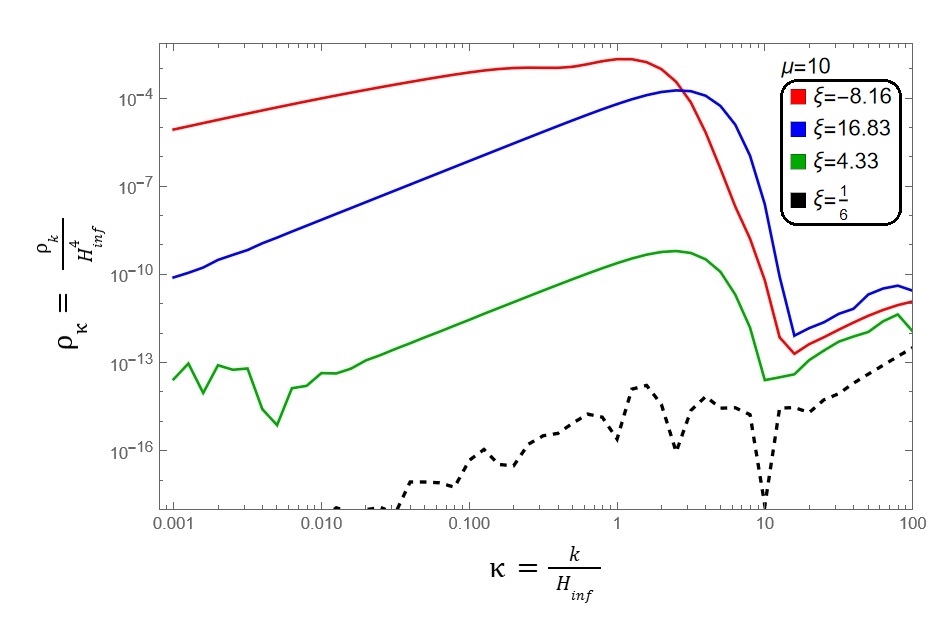}}
	\caption{The GPP rate and the produced energy density spectrum for $\mu= 0.25, 1, 10$ and $\Delta\tau=1$ for some interesting values of $\xi$ allowed for each mass. For each mass, three non-conformal couplings are depicted along with the conformal coupling. (a), (c) and (e) demonstrate the spectrum of the GPP rate, while (b), (d), and (f) are for the spectrum of the energy density.}\label{fig3}
\end{figure}

Although the band plot in Fig. \ref{fig2} demonstrates very clearly that, by changing the value of $\xi$ from the conformal case, the GPP rate and the energy density are affected substantially, we have depicted the GPP rate, ${\lvert{\beta}_{\kappa}\lvert}^2$, and the energy density of each mode for three sample values of $\xi$ and $\mu=0.25, 1, 10$, besides the plots of the conformal case in Fig. \ref{fig3}, for better understanding. These plots are less crowded and make the comparison between the conformal and non-conformal cases much easier. 

\begin{figure}
	\subfloat[]{\includegraphics[width=0.48\textwidth]{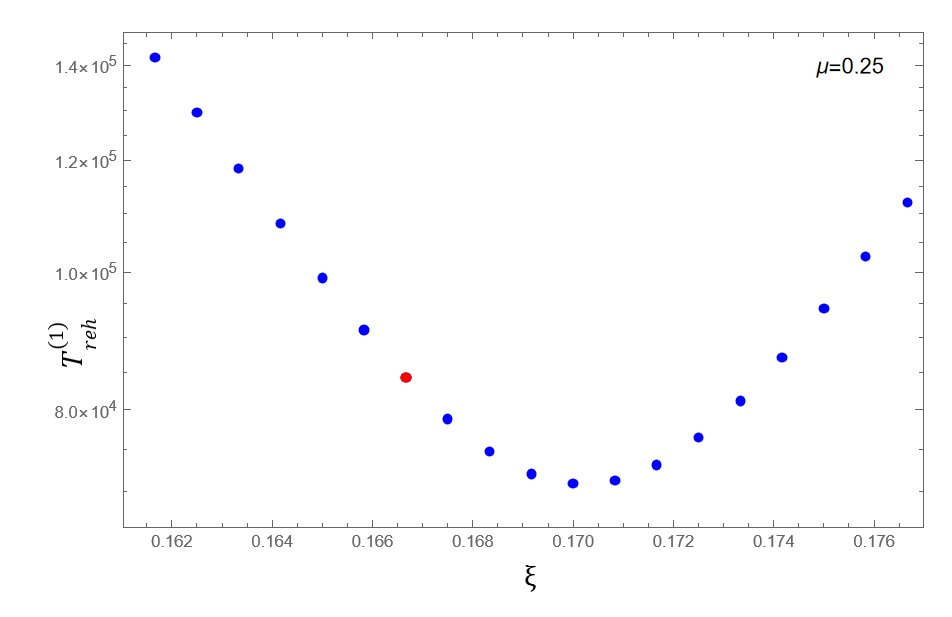}}
	\hfill
	\subfloat[]{\includegraphics[width=0.48\textwidth]{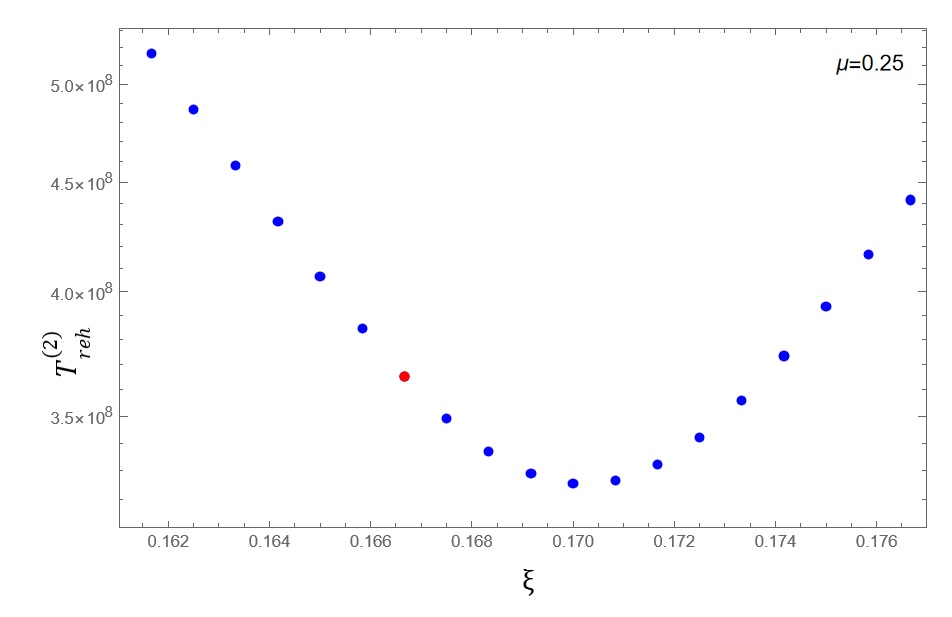}}\\
    \subfloat[]{\includegraphics[width=0.48\textwidth]{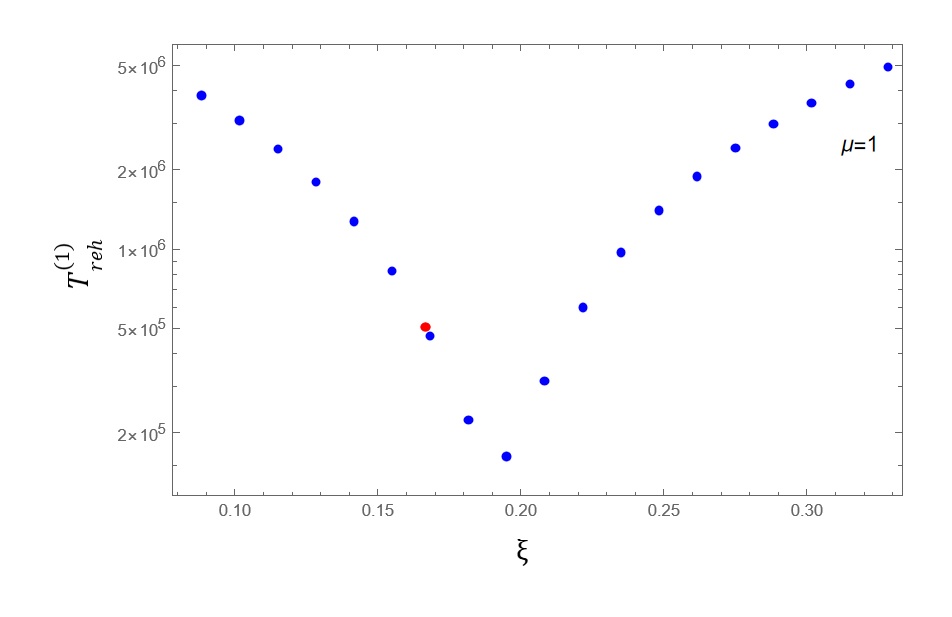}}
    \hfill
    \subfloat[]{\includegraphics[width=0.48\textwidth]{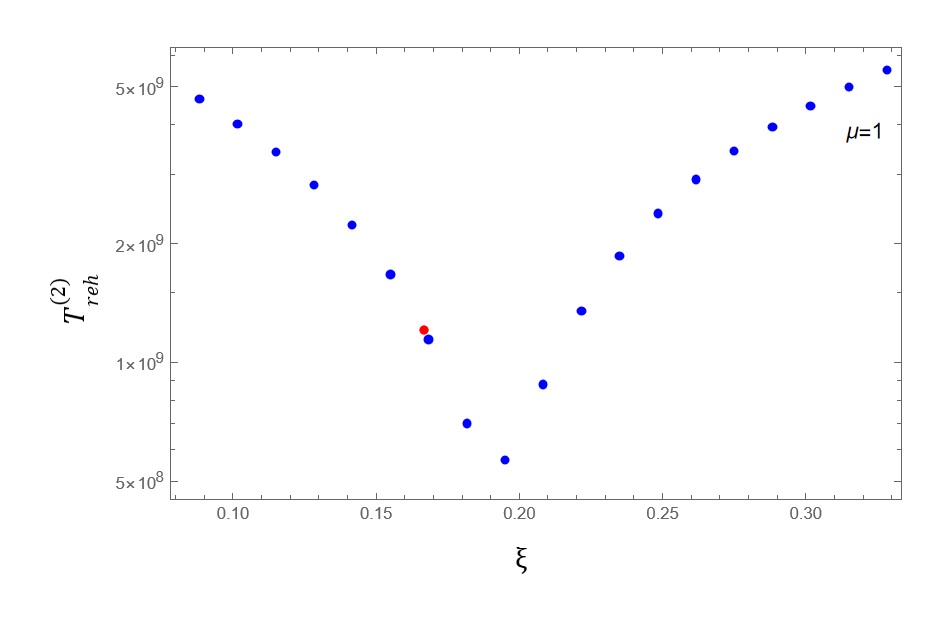}}\\
    \subfloat[]{\includegraphics[width=0.48\textwidth]{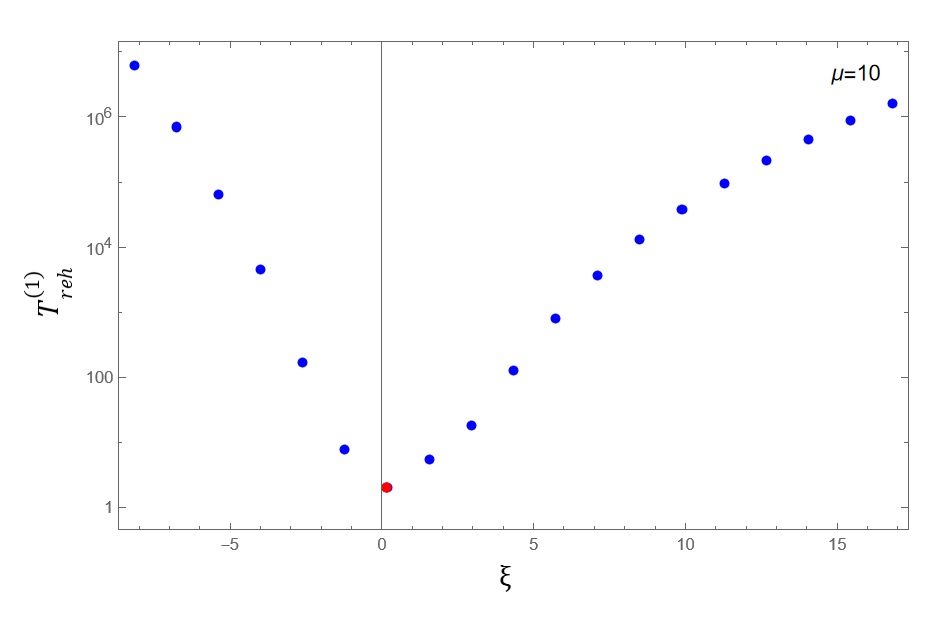}}
    \hfill
    \subfloat[]{\includegraphics[width=0.48\textwidth]{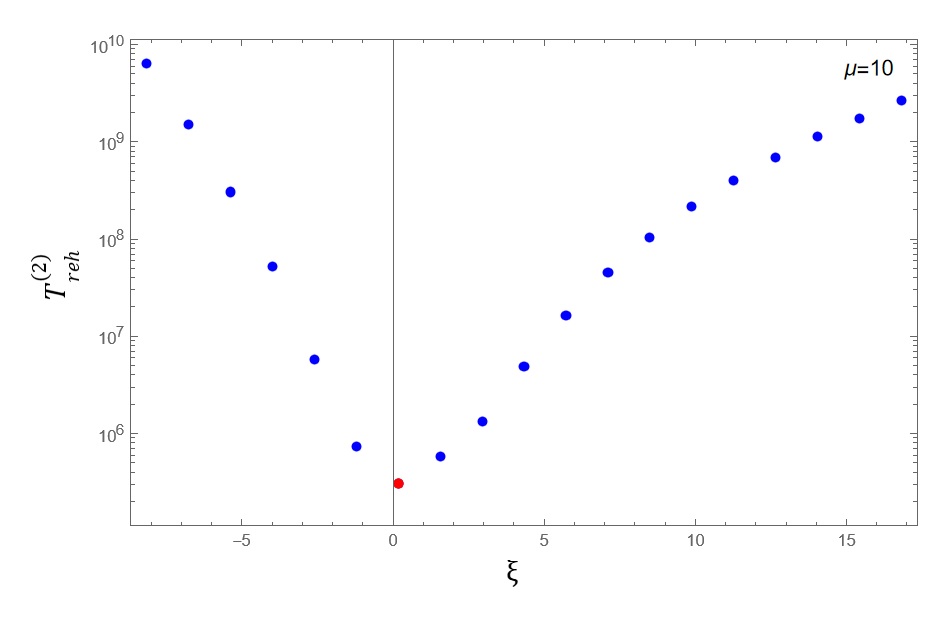}}
	\caption{The reheating temperature spectrum of the first and second scenarios of decay for (a), (b) $\mu=0.25$, (c), (d) $\mu=1$, and (e), (f) $\mu=10$, as a function of $\xi$. The temperatures of non-conformal values are shown in $blue$, while the temperature of the conformal case is depicted in $red$. As can be seen, any deviation from the conformal case causes a difference in the reheating temperature. Such a difference becomes greater as the $\xi$ tends to the ${\xi}_{L}$ and ${\xi}_{R}$. Hence, the highest reheating temperature in the parameter space of each mass corresponds to either ${\xi}_{min}$ or ${\xi}_{max}$. The pattern of the temperature spectrum could be different for different values of mass, as can be seen from this figure. All of these plots are depicted for $\Delta\tau=1$.}\label{fig4}
\end{figure}

In Fig. \ref{fig4}, the resultant reheating temperatures for both reheating scenarios are depicted for these three values of mass. As we can see, \textbf{the reheating temperature increases as $\xi$ acquires values near ${\xi}_L$ and ${\xi}_R$. The highest achievable reheating temperature is obtained either near ${\xi}_{min}$ or ${\xi}_{max}$.} \textbf{According to Fig. \ref{fig4}, it is also interesting to note that the conformal value of the coupling does not always correspond to the lowest possible reheating temperature. This fact clearly manifests itself for $\mu=0.25, 1$ in Fig. \ref{fig4}.}

\begin{figure}
	\subfloat[]{\includegraphics[width=0.5\textwidth]{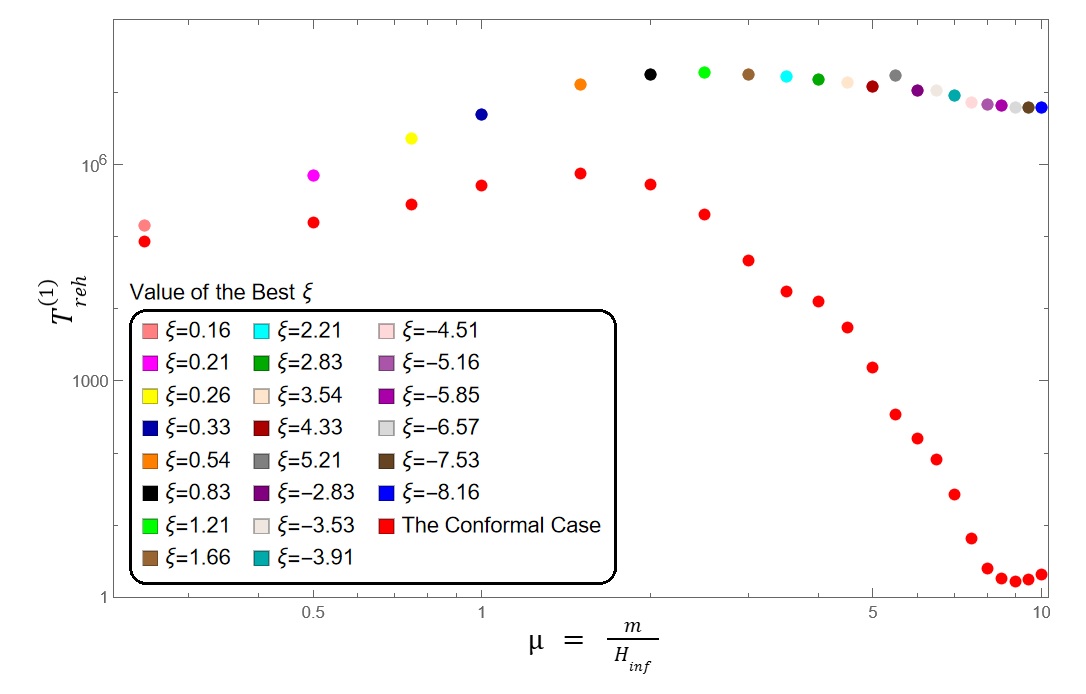}}
	\hfill
	\subfloat[]{\includegraphics[width=0.5\textwidth]{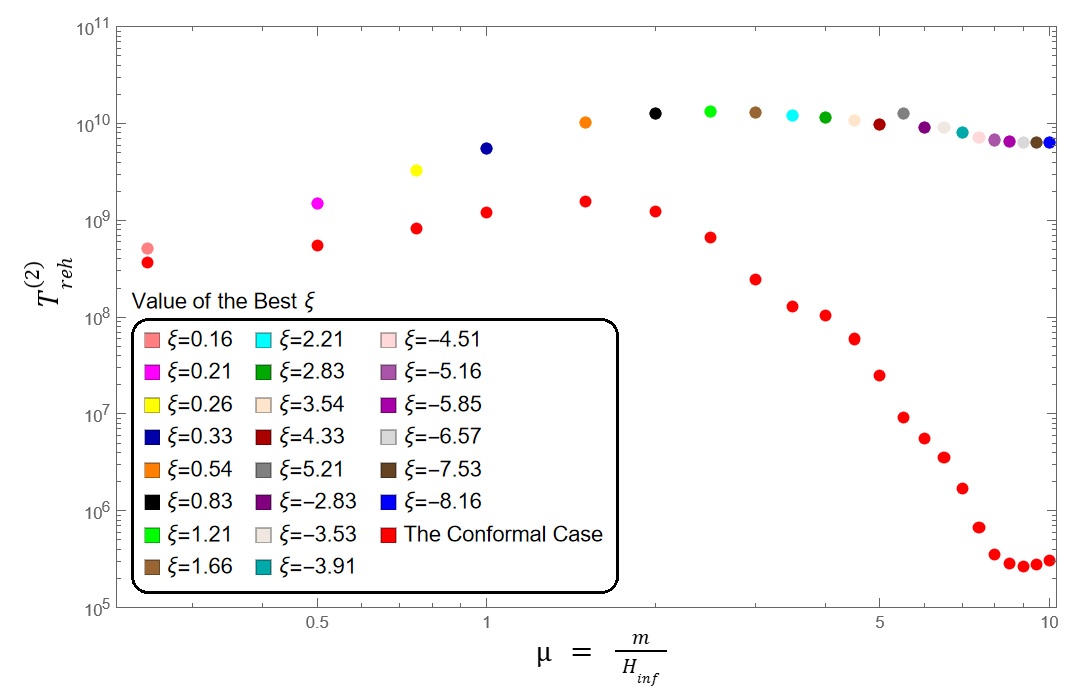}}
	\caption{The highest possible reheating temperature of each mass in the parameter space of $\xi-\mu$ for $\Delta\tau=1$. (a) is for the first and (b) for the second scenario of decay. For each value of mass, the highest reheating temperature corresponding to a specific value of $\xi$, which is either ${\xi}_{min}$ or ${\xi}_{max}$, is shown. To highlight the substantial effect of the non-conformal coupling in a clearer way, the temperature spectrum of the conformal case is depicted in $red$ as well. As can be seen, the non-conformal coupling compensates the suppression at higher mass values and brings the highest values of temperature of each mass to an equal order of magnitude.}\label{fig5}
\end{figure}

We have run our code for different values of mass and surveyed the whole parameter space of the gravitational coupling and mass. Using the results of these calculations, we infer which $\xi$ can result in the highest reheating temperature for each mass. We present our results in Fig. \ref{fig5}. A summary of relevant results is listed in Table. \ref{table} in App. \ref{tab}. In Fig. \ref{fig5}, we have also plotted the reheating temperature for the conformal case in $red$ (It is, in fact, just the $red$ plot in Fig. \ref{fig1}, where we considered the effect of mass). Having both of these plots together provides us with a deeper understanding of the effect of the non-conformal coupling on the reheating temperature. These results are very interesting, and we are going to explain them in what follows.

On the other hand, as discussed in the previous subsection, there must be a mass for which the reheating temperature can acquire the highest value. For the case of the effect of mass, Fig. \ref{fig1}, this highest temperature corresponded to $\mu=1.5$. If one considers the effect of non-conformal coupling, this value is shifted to $\mu=2.5$, as can be seen in Fig. \ref{fig5} or the numerical data in Table. \ref{table}. In fact, \textbf{considering the effect of non-conformal coupling, the highest achievable reheating temperature shifts from $\mu=1.5$ to $\mu=2.5$. This highest temperature is $1.84\times10^7 GeV$ and $1.32\times10^{10} GeV$ for the first and second scenarios, respectively, and corresponds to $\xi=1.2$.}

It is also obvious that \textbf{the effect of $\xi$ on the reheating temperature for lower masses\footnote{It is worth mentioning that in our case, we are considering the GPP scenario for the superheavy masses. As mentioned in Sec. \ref{theo}, the parameter $\mu$ is the dimensionless mass, and by rewriting in terms of dimensionful parameters, we have $m=\mu\times H_{inf}$. This means that, for instance, for $\mu=0.25$, the mass of the field is of the order of $1.65\times10^{13}$, which is a supermassive value. The terminology lower mass only refers to the low mass compared with the other masses considered in our study, such as $\mu=4, 6, 8, 10$.}, like $\mu=0.25, 0.5$ is not considerable.} This happens because, in the stability relation, Eq. \ref{p5}, the values of ${\xi}_L$ and ${\xi}_R$, or the ${\xi}_{min}$ and ${\xi}_{max}$, can not change that much. There is also another interesting phenomenon that happens when considering the non-conformal coupling. According to Table. \ref{table}, \textbf{there is a mass value $\mu=1.5$ for which the negative value of $\xi$ becomes possible for the first time.} For masses below $\mu=1.5$, the negative values of the gravitational coupling are not allowed because of tachyonic instability. For equal or masses greater than this mass, $\xi$ can acquire negative values as well as positive ones.

The mass $\mu=5.5$ is also of particular interest. \textbf{For masses below $\mu=5.5$, the highest possible reheating temperature happens for a positive ${\xi}_{max}$. But for masses above $\mu=5.5$, the largest possible reheating temperature corresponds to negative-valued $\xi$ with the greatest absolute value, which means ${\xi}_{min}$.} In particular, for $\mu=10$, $\xi_{min}=-8.16$ corresponds to the highest reheating temperature for this mass, with $6.10\times10^6GeV$ and $6.34\times10^9GeV$ for the first and second scenarios. As seen in Fig. \ref{fig5}, although the highest temperatures form a continuous shape, the highest reheating temperature of $\mu=5.5$ lies above the curve with an abnormal temperature of $1.72\times10^{7}$ and $1.27\times10^{10}$ compared with the other curves. However, this abnormal temperature is not large enough to overcome the reheating temperature of $\mu=2.5$ mentioned above.

As a result of this subsection, the effect of non-conformal values of gravitational coupling, allowed for the superheavy values of mass, on the temperature of the produced thermal bath is substantial and vital in a way that this effect should not be ignored in the studies of the GPP scenarios. Although this effect is more perceptible for higher values of mass, as can be seen from Fig. \ref{fig5}, the highest temperature obtained by considering the effect of non-conformal coupling corresponds to $\mu=2.5$ and its ${\xi}_{max}$. We want to draw your attention to the aforementioned result regarding the temperature of $\mu=2.5$, since this result is very important for our following purposes in the next subsection: \textbf{this will be changed if one considers the effect of the transition time.} We also have another result in hand which is important from the viewpoint of the purposes of the next subsection, that is, for masses greater than $\mu=5.5$, it is the $\xi_{min}$ with negative value that corresponds to the highest $T_{reh}$ possible for that mass. As we will see in the next subsection, \textbf{this result can also be changed if the effect of the transition time is allowed to enter the game.}   

It is also vital to note that although in this case, as well as the conformal case, the highest temperature was not produced in the case of the greater masses, the non-conformal coupling brings great masses back into the competition, as can be seen from Fig. \ref{fig5}. By considering the non-conformal gravitational couplings, there exists a coupling, mentioned as ${\xi}_{best}$, that produces the highest reheating temperature in the parameter space of each mass. Although in the end, the highest value of reheating temperature is still for a lower mass, $\mu=2.5$, Fig. \ref{fig5} demonstrates that because of the effect of non-conformal coupling, the highest temperatures of different masses are, roughly speaking, of the same order of magnitude and very close to each other. \textbf{Hence, considering the effect of non-conformal coupling has changed our results very fundamentally, and now it is feasible that the higher values of mass, abandoned in many studies due to their low reheating temperature, will return to the competition and may play the most important role in the gravitational reheating. Possibly, the highest reheating temperature in the parameter space is obtained in the case of the highest mass and a non-conformal coupling. If so, we have reached our main goal regarding the essential role of non-conformal coupling in studies of gravitational reheating.}  

In the two previous subsections, we considered the effect of mass and the gravitational coupling on the resultant reheating temperature. It is worth mentioning that \textbf{both of these parameters were related to the scalar spectator field} part of the whole theory with action written in Eq. \ref{action}. There remains another parameter, the transition time, which is a characteristic feature of the inflationary model. By studying the effect of that parameter on the GPP rate and the resultant reheating temperature, all of the parameter space of the whole model is studied, and, more importantly, this study is done in a completely general manner. On the one hand, we do not restrict ourselves to a particular inflation model, and on the other hand, the superheavy scalar field theory can easily be extended to any other field with any spin or potential.    

\subsection{The Effect of the Transition Time on the $T_{reh}$}\label{sec33}

This subsection studies the effect of the transition time on the GPP rate and the resulting reheating temperature in this particle production scenario. As we have mentioned above, this last remaining parameter is a characteristic feature of the inflationary model. During the last two subsections, we set the transition time to $\Delta\tau=1$, since most of the well-known inflationary models are consistent with this choice, and our goal was to advance our study path in a general manner as much as possible. Now, in this subsection, we are going to vary the value of the transition time and study its effect on the GPP rate. As we will see below, changing this remaining parameter yields the highest possible $T_{reh}$ in the GPP reheating channel.

To explain physically how the transition time affects the GPP rate, let us return to the subsection \ref{physadi} in Sec. \ref{theo}. We argued that if the transition period is accompanied by a violation of the adiabaticity condition, the $in$ mode function cannot follow changes in the gravitational background and mimic the behavior of the physical positive-frequency mode function of the new background at the final time. As a result, this positive-frequency $in$ mode function, which describes the $in$ physical vacuum state, is not a positive-frequency mode compared with the new background at a final time. Instead, it is a mixture of positive- and negative-frequency modes with respect to the final state of the background and its physical mode function. This situation occurs manifestly in the case where the cosmological background experiences a sharp, rapid, and harsh transition period between two eras. It is a famous quote that \textbf{the shorter the transition time, the greater the gravitational particle production rate.}

The physical explanation for this quote is that suppose there are two cosmological states, $in$ and $out$ states, and their corresponding physical positive-frequency mode functions that are connected to each other through a transition period. If the transition period is smooth and long enough, the adiabaticity condition is not violated significantly during this period, and hence the $in$ positive-frequency mode function evolves smoothly and can follow the evolution of the spacetime. Hence, this mode adapts itself to the evolution of the background and gradually mimics the behavior of the $out$ positive-frequency mode. So the GPP does not occur efficiently. On the other hand, if the transition period is short and sharp, the $in$ positive-frequency mode cannot adapt to these rapid changes in the background, and hence, after the transition it is a combination of positive- and negative-frequency mode functions with respect to the $out$ physical mode functions. So the GPP occurs significantly. In this case, the adiabaticity condition is violated noticeably, and the GPP is obtained through the $\beta$ coefficient. 

There is a physically intuitive example that can provide us with a pedagogical picture of what happens if the adiabaticity condition is violated because of a rapid and harsh transition time: Suppose we have a plastic bowl in hand with a metal ball resting in the depths of its center. In this case, the metal bowl is our field, and the center of the bowl represents the vacuum state. If we start to change the shape of the bowl smoothly and slowly, the ball can follow the changes of the bowl and follow its evolution so that it gradually moves toward the new bowl's depth (or vacuum). In other words, the metal ball was in the initial vacuum state at the initial time and will rest in the new vacuum state at a final time, only with tiny fluctuations around this new vacuum. These small fluctuations tell us that although there is an excitation with respect to the new vacuum, this excitation is small. This can be interpreted as a small GPP rate and a small gained energy. Now, if one starts to change the shape of the plastic bowl rapidly, the metal ball does not have enough time to mimic the changes and adapt itself to them and finally be situated in the new vacuum states. Instead, since the changes occur so rapidly and sharply, the ball is excited with respect to the new vacuum state of the new bowl shape. This can be interpreted as a significant GPP rate.

\begin{figure}
    \centering
	\includegraphics[width=1\textwidth]{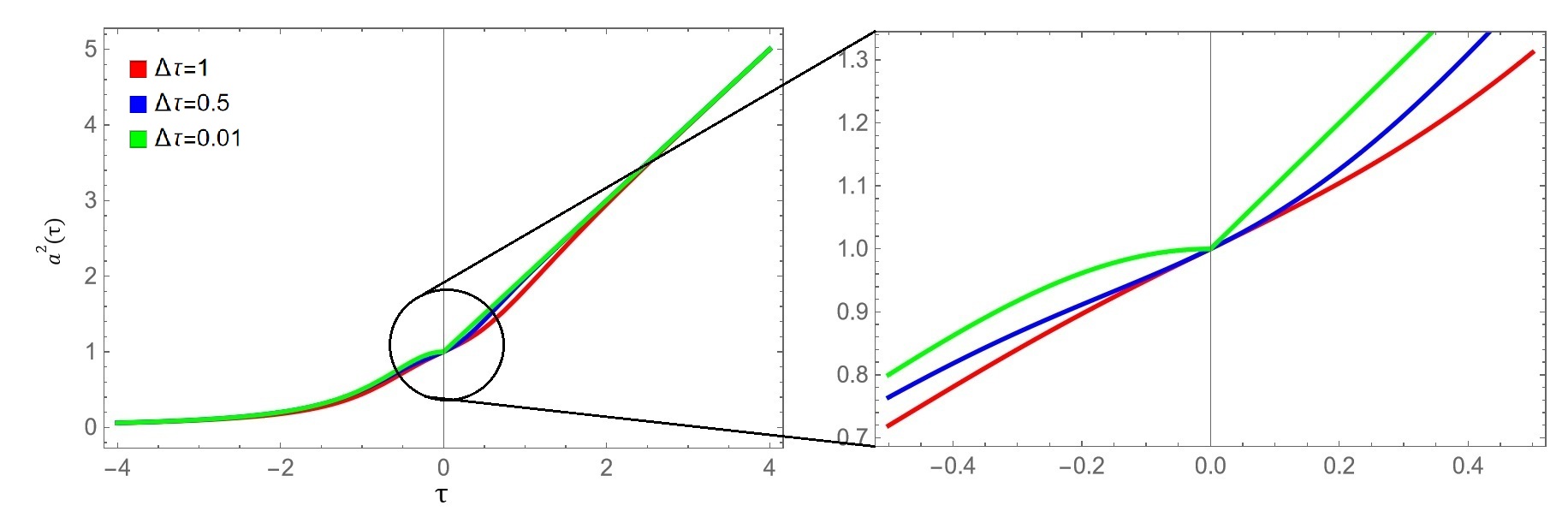}
	\caption{The behavior of the square of the scale factor proposed in Eq. \ref{scalenum}, $a^2(\tau)$, with respect to the dimensionless conformal time, $\tau$. This function is plotted for three different values of the transition time, $\Delta\tau$. By zooming in on the region of the transition time, it turns out that the reduction in $\Delta\tau$ results in larger values of the time derivative of $a(\tau)$, which causes a more severe violation of the adiabaticity condition.}\label{fig6}
    \label{scalebeh}
\end{figure}

This explains why a short and harsh transition time that violates the adiabaticity condition results in a significant, noticeable production of particles. Besides this intuitive explanation, we are going to present a mathematical justification of what happened due to the shortness of the transition time. Returning to the adiabaticity condition for this model of scalar field, Eq. \ref{adicon}, this inequality demonstrates that the adiabaticity condition is violated when the time derivative of the scale factor, $a(\tau)$, as well as the dimensionless curvature scalar, $C(\tau)$, become significant\footnote{There are two notes that we want to bring our readers' attention to: Firstly, the scalar curvature is a function of the Hubble parameter, and hence, a function of the scale factor. So, by reducing the transition time and increasing the time derivative of $a(\tau)$, the time derivative of the scalar curvature, $C'(\tau)$, grows significantly too. The second note is that although we have mentioned that the scalar curvature damps very soon and fast at the beginning of the kination era, just after the transition, it is not the case at the end of inflation and the beginning of the transition time. In fact, at that time, it has a noticeable value, and in the case of a harsh and rapid transition, its time derivative acquires large values too. Hence, these terms are in fact very important in the adiabaticity condition, especially for the case of non-conformal coupling.}. This happens when the transition occurs rapidly and hence the time derivative of $a(\tau)$ and $C(\tau)$ in the numerator gain large values. By reducing the transition, the shape of the scale factor near the transition time becomes different. If the transition time is of the order of $\Delta\tau=1$, the scale factor becomes a smooth function of time, while in the case of a rapid transition, the functional form of the scale factor changes so that its time derivative gains large values. This behavior is depicted in Fig.\ref{scalebeh} for different values of $\Delta\tau$ for the function of the square of the scale factor proposed in Eq. \ref{scalenum}. This figure, in conjunction with Eq. \ref{adicon}, explains why the reduction of the transition time results in the violation of the adiabaticity condition, which in turn causes an amplification in the GPP rate. So, for a shorter transition period, the evolution of the scale factor becomes more violent during the transition era, which results in the amplification of the GPP rate and its resultant reheating temperature. Now, with these intuitive and mathematical explanations in hand, we present and interpret our numerical results regarding the effect of the transition time.

Our first results are shown in Fig.\ \ref{fig6}. In these plots, we study the effect of the transition time for the conformal case. Hence, while we set $\xi=\frac{1}{6}$, the reheating temperatures of both scenarios are depicted for different values of the bare mass of the field. Again, the $Red$ line corresponds to $\Delta\tau=1$, and is the $red$ plot of "the effect of mass" in Fig. \ref{fig1}. The reheating temperature spectrum is also depicted for smaller values of the transition time. The following interesting results can be extracted from this figure.

\begin{itemize}
    \item \textbf{General Shape of the Spectrum is the same for different $\Delta\tau$:} In the case of $\Delta\tau=1$ and conformal coupling, as depicted in Fig.\ \ref{fig1}, there was a maximum in $T_{reh}$ spectrum around $\mu=1.5$. This maximum is a characteristic feature in this plot. The GPP rate and resultant $T_{reh}$ increase for smaller values of mass and then decrease for larger masses because of an almost behavior of $\lvert \frac{{\omega}_{\kappa}'}{{\omega}_{\kappa}^2}\lvert$ as $\frac{1}{\mu}$. This behavior can also be seen in Fig.\ \ref{fig6} for shorter values of $\Delta\tau$. But the point is that the value of the mass corresponding to the maximum $T_{reh}$ is shifted to larger masses. This maximum point can be seen in Fig.\ \ref{fig6} for $\Delta\tau=1,\frac{1}{2}$ and even $\frac{1}{100}$. But for $\Delta\tau=\frac{1}{1000}$, it is shifted to larger values of mass so that it lies outside the range of our plot. It is because of this fact that in our study, we only consider the masses of our spectator field to be $0.25\le\mu\le10$. 
    \item \textbf{Effect of $\Delta\tau$ is perceptible for large masses but negligible for the small ones:} According to the adiabaticity condition, Eq. \ref{adicon}, this behavior seen in Fig.\ \ref{fig6} can be explained by understanding the role of different terms in $\lvert \frac{{\omega}_{\kappa}'}{{\omega}_{\kappa}^2}\lvert$: In the case of conformal coupling, the first two terms in the numerator, $2(\xi-\frac{1}{6})aa'C(\tau)$ and $(\xi-\frac{1}{6})a^2C'(\tau)$, vanish and hence the transition time leaves its footprint on $2aa'{\mu}^2$. According to Fig.\ \ref{scalebeh}, by reducing the transition time, the derivative of the scale factor, $a'(\tau)$, grows substantially while its value, $a(\tau)$, does not change noticeably. Hence, by reducing the transition time, $2aa'{\mu}^2$ in the numerator grows, while, the denominator of $\lvert \frac{{\omega}_{\kappa}'}{{\omega}_{\kappa}^2}\lvert$ remains almost the same. Now, in the case of small values of mass, the effect of the large value of $a'(\tau)$ is weakened since it is multiplied by ${\mu}^2$. Hence, the adiabaticity condition is not violated in a noticeable manner, and as a result, one shall not expect noticeable changes in $T_{reh}$. However, the situation is totally different for the larger masses. In fact, a large value of mass acts like a supporting army that makes the effect of the derivative of the scale factor more perceptible, so the adiabaticity condition is violated more severely in cases with rapid and short transition time. By this justification, it becomes easy to understand why the effect of the reduction of the transition time is more perceptible in the case of greater values of mass.
    
\end{itemize}

\begin{figure}
	\subfloat[]{\includegraphics[width=0.5\textwidth]{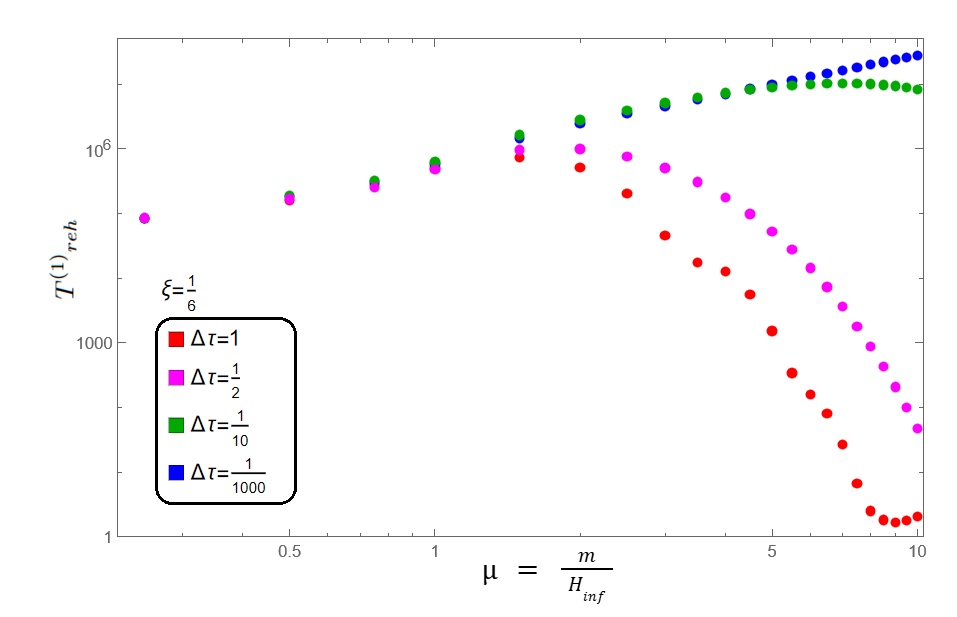}}
	\hfill
	\subfloat[]{\includegraphics[width=0.5\textwidth]{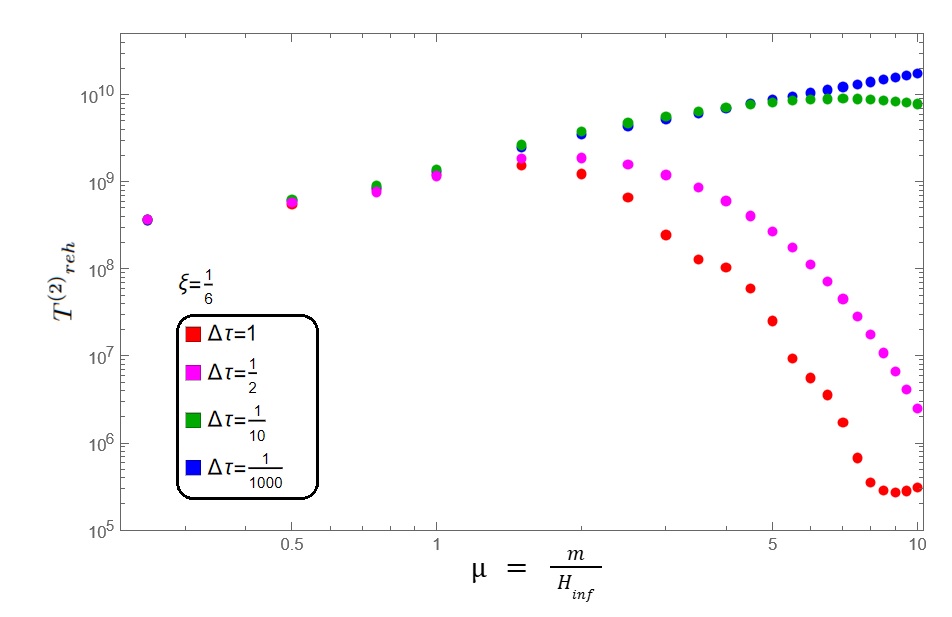}}
	\caption{The effect of the reduction in the transition time: (a) for the first and (b) for the second scenario of decay. These plots show the spectrum of the reheating temperature as a function of $\mu$ for $\xi =\frac{1}{6}$ and $\Delta\tau= 1, 0.5, 0.1, 0.001$. As shown, the effect of reducing the transition time is more pronounced for higher mass values, and it is almost negligible for small masses.}\label{fig6}
\end{figure}

\begin{figure}
	\subfloat[]{\includegraphics[width=0.5\textwidth]{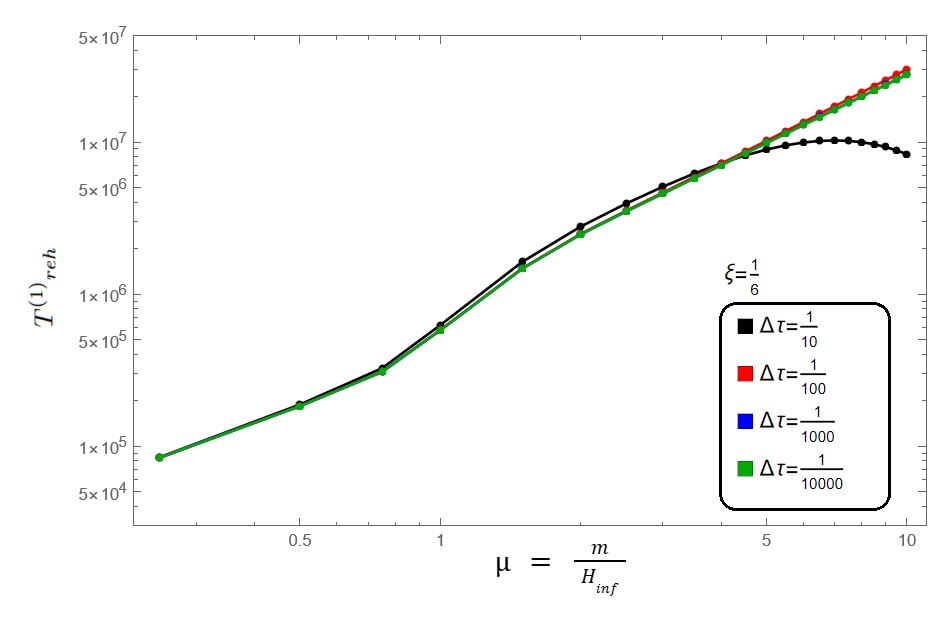}}
	\hfill
	\subfloat[]{\includegraphics[width=0.5\textwidth]{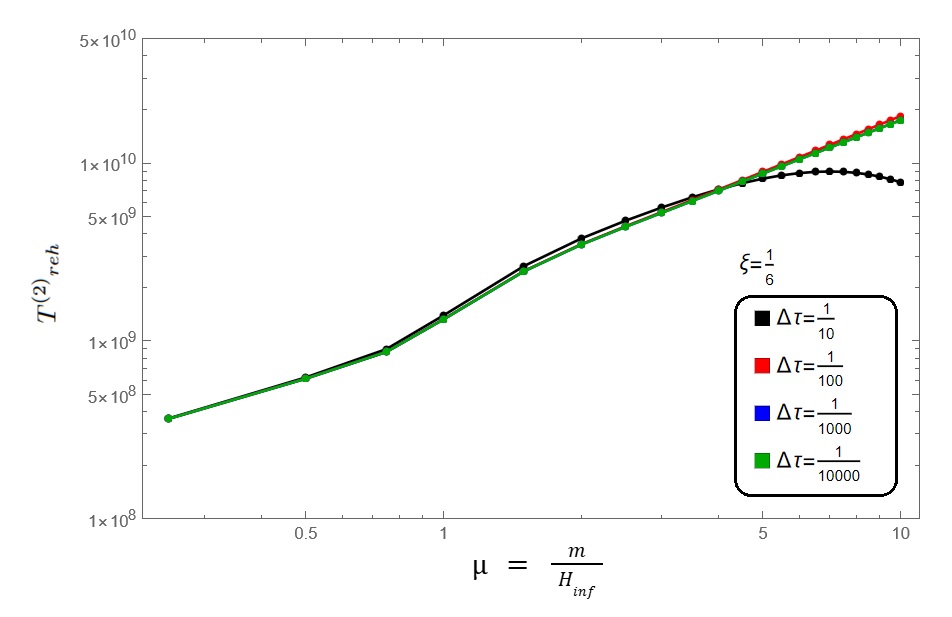}}
	\caption{The behavior of the spectrum of $T_{reh}$ due to the reduction of the transition time. In these plots, the reheating temperature is plotted for different values of mass, in the case of the conformal coupling. As can be seen, for masses up to $\mu=10$, the spectrum of the reheating temperature does not change anymore with the reduction of the transition time to values smaller than $\Delta\tau=0.01$. In these plots, (a) is for the first and (b) for the second scenario of decay.}\label{fig7}
\end{figure}

With the above results in hand, $\mu=10$ seems to deserve more attention in this subsection. Before going further, we produced another figure to answer our personal curiosity. During our code runs, an important question sparked in our minds regarding whether the reheating temperature will increase to unlimited numerical values if one reduces the transition time to an arbitrary infinitesimal value. To answer this question, we ran our code in the conformal case for several small values of $\Delta\tau$ and masses up to $\mu=10$, which retain the field as a spectator. We depicted our results in Fig. \ref{fig7}. As can be seen, although there is a tangible change in the reheating temperature plots when the transition time is reduced from $0.1$ to $0.01$, there is almost no change when we reduce the transition time from $0.01$ to $0.001$ or even less, to $0.0001$. Hence, we conclude that \textbf{there is an interesting result regarding the reduction in the transition time. Although it is a well-known physical fact that the reduction in the transition time results in an increase in the GPP rate and the reheating temperature, our study reveals that for any definite mass, such an increase becomes inconspicuous by reducing $\Delta\tau$ below a specific value. The larger the mass, the smaller this specific value of $\Delta\tau$.} Our careful considerations assured us that the aforementioned results are valid for the non-conformal cases, as well. Therefore, the effect of the reduction in the transition time causes a greater increase in the GPP rate and the reheating temperature for $\mu=10$. Also, there is no considerable change in $T_{reh}$ when one decreases $\Delta\tau$ to values smaller than $0.001$ for this interval of masses that we have considered, which is $0.25\le\mu\le10$.      

Besides, we will continue our survey in the parameter space. We have conducted two additional considerations that yielded very fascinating results. First of all, as discussed above, and as can be seen from Figs. \ref{fig6} and \ref{fig7}, generally speaking, the reduction in the transition time results in an increase in the reheating temperature, or equivalently, the produced total energy density. But we have to keep in mind that these results are obtained for the conformal coupling. Our considerations reveal that although this increase is a general behavior in this case, the detailed behavior can be quite different for the positive and negative values of the non-conformal coupling constant.

\begin{figure}
	\subfloat[]{\includegraphics[width=0.5\textwidth]{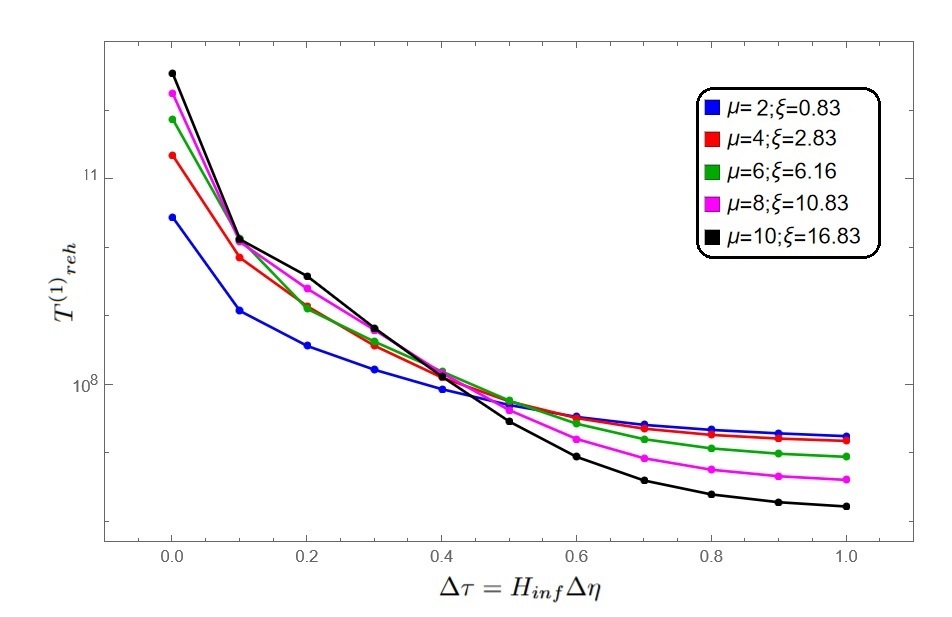}}
	\hfill
	\subfloat[]{\includegraphics[width=0.5\textwidth]{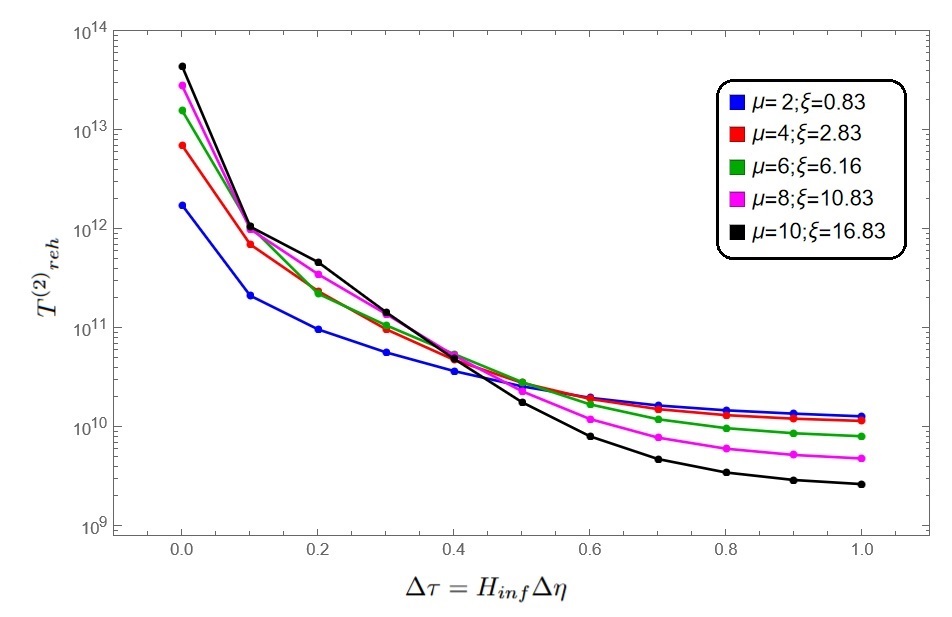}}
	\caption{The pattern of the increase in the reheating temperature due to the reduction of the transition time for positive values of $\xi$ for $\mu=2, 4, 6, 8, 10$. As can be seen in the case of positive-valued $\xi$, the reheating temperature increases permanently with the reduction of the transition time. (a) is for the first and (b) for the second scenario of reheating.}\label{fig8}
\end{figure}

It is shown in Refs. \cite{ford1987gravitational, ford2021cosmological} that if the universe experiences a transition from inflation to the RD era with the transition time equal to $\Delta t$, then the energy density of created particles becomes of the order of $\rho=\frac{(\xi - \frac{1}{6})^2H^4}{8{\pi}^2a^4}ln[\frac{1}{H\Delta t}]$. This relation tells us that any reduction in the transition time causes an increase in the energy density. But it is worth mentioning that these studies, which were done by Ford, were considered in the case that the gravitational coupling is very close to the conformal values, and not the non-conformal case with values of $\xi$ very different from $\frac{1}{6}$\footnote{It is because in those studies the GPP rate was calculated through the perturbative approach, which forced them to stay very close to the conformal value of $\xi$ in the parameter space.}. To check whether or not this increase is always happening for the non-conformal case, we run our code for five different masses and their ${\xi}_{max}$, which is a positive-valued number, and examine the effect of the reduction of $\Delta\tau$ on the resultant reheating temperature for each one. As can be seen from Fig. \ref{fig8}, \textbf{the reduction in the transition time absolutely yields an increase in the resultant reheating temperature when the positive values of non-conformal coupling are considered.} Hence, the general behavior of the reheating temperature spectrum is in accordance with the well-known physical fact discussed above. Besides, there is also another very important result that can be extracted from Fig. \ref{fig8}. 

\textbf{We mentioned earlier that in the case of conformal coupling, Fig. \ref{fig6}, it seems that the effect of the transition time is more sizable on the reheating temperature of $\mu=10$. We can see manifestly in Fig. \ref{fig8} that it is also the case for the non-conformal values of coupling $\xi$.} In fact, although the reheating temperature of $\mu=10$ with $\xi={\xi}_{max}$ was not the highest one in considerations with $\Delta\tau=1$, as can be seen in Table. \ref{table}, after reducing the transition time, it increases to higher values compared with those of the other masses and their corresponding non-conformal couplings, $\xi={\xi}_{max}$. More importantly, it lies on a higher step of reheating temperature value compared with that of the mass of $\mu=2.5$, as this mass and its ${\xi}_{max}$ correspond to the highest reheating temperature when we have set $\Delta\tau=1$, as mentioned earlier. Hence, it seems that we have to bring our attention to the highest value of mass with its $\xi={\xi}_{max}$, since it is one of the main candidate points in the parameter space that may yield the highest reheating temperature in the GPP scenario. 

There is a mathematical justification for this phenomenon. According to the adiabaticity condition, Eq. \ref{adicon}, it is obvious that, just like in the case where the effect of the reduction of $\Delta\tau$ became important for large values of mass, due to the term $2aa'{\mu}^2$ in the numerator, the non-conformal values of $\xi$ act just the same as the mass term in $\lvert \frac{{\omega}_{\kappa}'}{{\omega}_{\kappa}^2}\lvert$. By considering non-conformal values of the coupling, and more importantly, $\xi={\xi}_{max}$ or $\xi={\xi}_{min}$, the other terms in the numerator of the adiabaticity condition containing the time derivative of the scale factor and the curvature scalar, namely $2(\xi-\frac{1}{6})aa'C(\tau)$ and $(\xi-\frac{1}{6})a^2C'(\tau)$, become important. Just like in the case of the large masses, the values of the non-conformal coupling and the large value of the time derivative of the scale factor will have a constructive collaboration. For values of coupling different from $\xi=\frac{1}{6}$, the coefficient $(\xi-\frac{1}{6})$ grows, which together with $a'(\tau)$, makes these terms important. The reduction in $\Delta\tau$ already makes these terms large, but in the case of conformal coupling, these terms were equal to zero. By varying the gravitational coupling from the conformal case, these terms start to grow, and in the case of $\xi={\xi}_{max}$ or $\xi={\xi}_{min}$ they reach their maximum absolute values. In the case of a short and sharp transition time and a non-conformal value of coupling, just like the case of large values of mass, these terms become noticeable for two reasons: First, the significant values of time derivatives of $a(\tau)$ and $C(\tau)$\footnote{Which is a function of the Hubble parameter, $H$, and as a consequence, a function of $a(\tau)$ and its derivative.}, and second, the large value of $(\xi-\frac{1}{6})$ that is multiplied by these derivative terms.   

Again, it is not the case for smaller values of mass. In fact, for smaller values of mass and their non-conformal coupling values, the effect of the transition time is again less perceptible compared with the case of large mass and their non-conformal coupling. It is due to the fact that for these masses, the stability inequality, Eq. \ref{p5}, forced the values of $\xi$ to stay close to the conformal value. This, in turn, makes the coefficient $(\xi-\frac{1}{6})$ small and hence, the terms $2(\xi-\frac{1}{6})aa'C(\tau)$ and $(\xi-\frac{1}{6})a^2C'(\tau)$ ineffective. So, one does not expect a noticeable increase in $T_{reh}$ for these values of mass due to the reduction of $\Delta\tau$, even in the non-conformal case.  

\begin{figure}
	\subfloat[]{\includegraphics[width=0.5\textwidth]{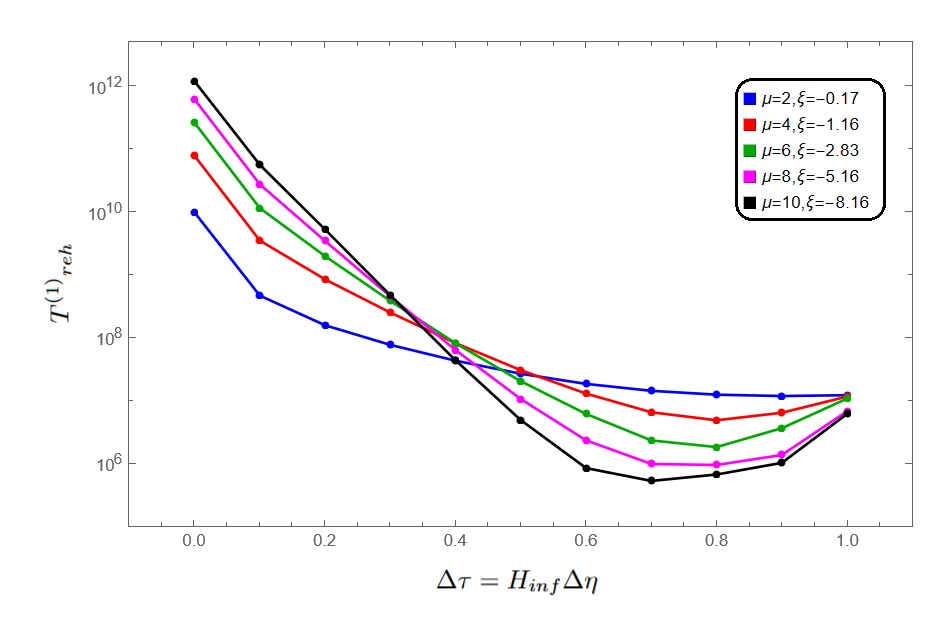}}
	\hfill
	\subfloat[]{\includegraphics[width=0.5\textwidth]{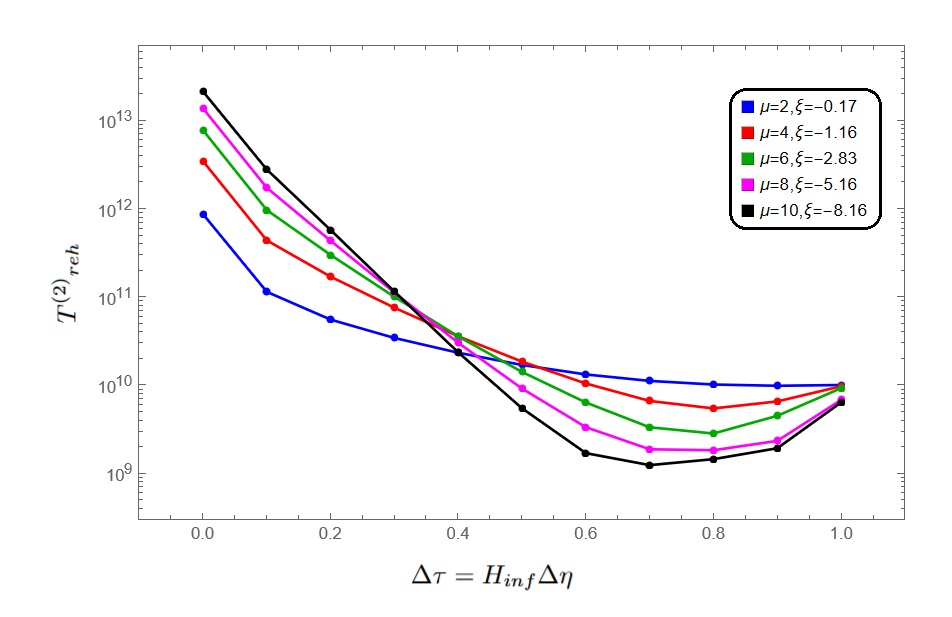}}
	\caption{The pattern of the increase in the reheating temperature due to the reduction of the transition time for negative values of $\xi$ for $\mu=2, 4, 6, 8, 10$. As can be seen in the case of a negative-valued $\xi$, the reheating temperature, at first, experiences a reduction due to the reduction in the transition time. But, after a specific value of $\Delta\tau$, it starts to increase with the reduction of the transition time. Such a pattern is totally different from the case of positive values of $\xi$. (a) is for the first and (b) for the second scenario of reheating.}\label{fig9}
\end{figure}

There also remains another point that should be considered in this case: The effect of the transition time on the reheating temperature of different masses with their corresponding couplings, $\xi={\xi}_{min}$, \textbf{where ${\xi}_{min}$ is a negative-valued coupling}. We have considered this, just like the cases with a positive-valued gravitational coupling constant, $\xi={\xi}_{max}$. The results are depicted in Fig. \ref{fig9}. \textbf{We have done these calculations for the same masses as in Fig. \ref{fig8}} to provide our readers with the opportunity to compare the results of positive and negative values of $\xi$\footnote{We want to mention that in these runs of our code, we have considered some of the masses that are equal to or greater than $\mu=2$. This is due to the fact that, as mentioned above and can be seen from Table. \ref{table}, the mass $\mu=1.5$ is the first value of mass that is allowed by Eq. \ref{p5} to acquire negative coupling constants. Hence, we choose $\mu= 2, 4, 6, 8, 10$ for the values of mass in this investigation to study the effect of $\Delta\tau$ and non-conformal coupling together and compare the effect of $\Delta\tau$ on positive and negative values of $\xi$. We have done our calculations for ${\xi}_{min}$ and ${\xi}_{max}$, since these values correspond to the highest values of reheating temperature for each mass, as can be seen from Fig. \ref{fig4}.}.  

These results are fascinating, since they challenge our preconceived belief that any reduction in transition time surely increases the reheating temperature. As shown in these plots, \textbf{for negative values of $\xi$, reducing the transition time also reduces the reheating temperature at first. After a specific value of $\Delta\tau$, the reheating temperature starts to grow again. Hence, although the reduction in the transition time increases the reheating temperature for positive values of coupling, it is not the case for negative values of $\xi$.} This result is very important and deserves more attention in the studies of GPP scenarios. These results are obtained by invoking the functional form of the scale factor introduced in Eq. \ref{scalenum}; hence, this reduction in $T_{reh}$ may be an effect of this specific choice or not. So it seems very important and interesting to examine whether this phenomenon is a generic behavior of negative gravitational couplings or it happens only in our case and due to the choice of this functional form of the scale factor. As a consequence, it deserves more attention in an independent study, especially for models with transition times not small enough, to consider whether the negative values of $\xi$ cause a reduction in the reheating temperature and whether this result is a generic consequence of negative values of $\xi$ or not. 

Although in the case of negative couplings, there is a reduction in the reheating temperature by reducing the transition time to a specific value, Fig. \ref{fig9} tells us that, as well as in the case with positive $\xi$, \textbf{the highest mass, $\mu=10$ in our case, still generally experiences a greater increase in the reheating temperature by reducing the transition time to smaller and smaller values.} In addition, in the case of negative coupling, again, $\mu=10$ with $\xi={\xi}_{min}$ experiences the greatest increase in its resultant reheating temperature and hence, its reheating temperature lies at a higher value in comparison with the other masses and their negative couplings. As in the case of positive $\xi$, this result can be explained and justified mathematically, using the adiabaticity inequality in Eq. \ref{adicon}. By substituting negative values of $\xi$ rather than the conformal one, the first two terms in the numerator become negative numbers with large absolute values. Hence, these terms cause the absolute value of $\lvert\frac{{\omega}'_{\kappa}}{{\omega}^2_{\kappa}}\lvert$ to take values greater than unity and violate the adiabaticity condition in a noticeable manner. This, in turn, causes a significant GPP rate and a perceptible increase in $T_{reh}$.   

So, \textbf{although in the case of $\Delta\tau=1$, the highest reheating temperature corresponded to $\mu=2.5$ and its ${\xi}_{max}$, by reducing the transition time, the reheating temperature of $\mu=10$ obtains higher values for both positive and negative values of ${\xi}_{max}$ and ${\xi}_{min}$, in comparison with that of other masses and their ${\xi}_{max}$ and ${\xi}_{min}$.} Hence, there emerges an important question: which of the positive or negative $\xi$ corresponding to $\mu=10$ yields the higher reheating temperature? This question is important for our study, since one of our main goals is to determine in which part of the parameter space we can reach the highest possible reheating temperature. As discussed in detail in Sec. \ref{intro}, one of the main challenges that the gravitational reheating scenario has faced was the insufficiency of this mechanism for reheating. Hence, it is a matter of importance for our study to determine the highest reheating temperature that can happen in the parameter space of gravitational reheating and show that this temperature is high enough to reheat the universe and start the RD era. In fact, by calculating the highest possible temperature and the part of the parameter space corresponding to it, we will show that this reheating mechanism is a very efficient mechanism for reheating the universe, despite many of the studies in the literature.   

To this end, we want to compare the final reheating temperature corresponding to ${\xi}_{max}$ and ${\xi}_{min}$ of $\mu=10$ after the reduction of the transition time to $\Delta\tau=0.001$. We remember from the previous subsection, and also from Table. \ref{table}, that for $\mu=10$, it was the ${\xi}_{min}$ with a negative value that yielded the highest temperature when we set $\Delta\tau=1$. As we have seen, it was also the case for the masses greater than $\mu=5.5$. On the other hand, we have seen throughout this subsection that at the end, both of the temperatures corresponding to ${\xi}_{max}$ and ${\xi}_{min}$ of $\mu=10$ are increased by reducing the transition time to $\Delta\tau=0.001$. Now the question is \textbf{which of them gains a higher value from such a reduction of the transition period.}

\begin{figure}
	\subfloat[]{\includegraphics[width=0.3\textwidth]{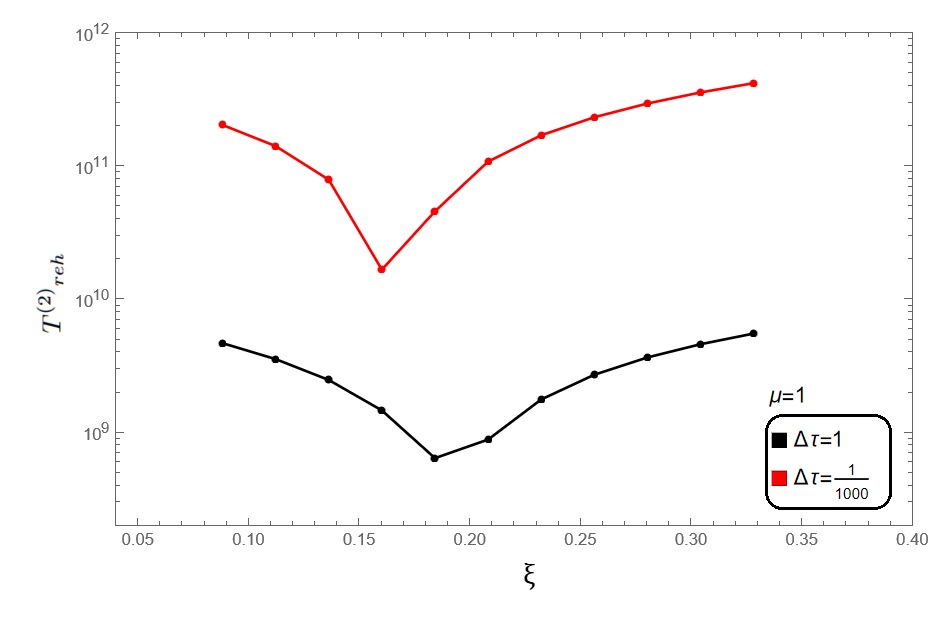}}
	\hfill
	\subfloat[]{\includegraphics[width=0.3\textwidth]{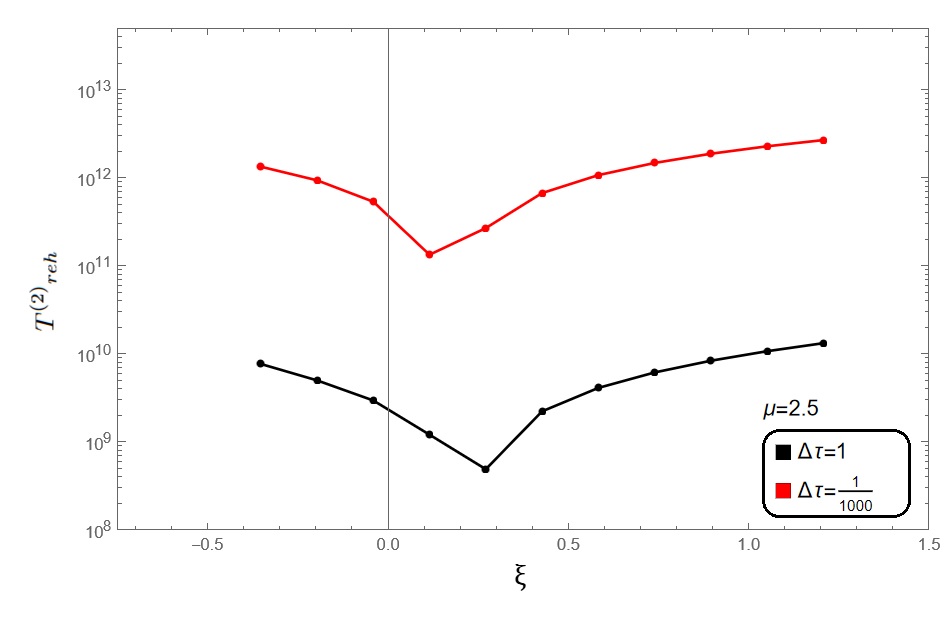}}
    \hfill
	\subfloat[]{\includegraphics[width=0.3\textwidth]{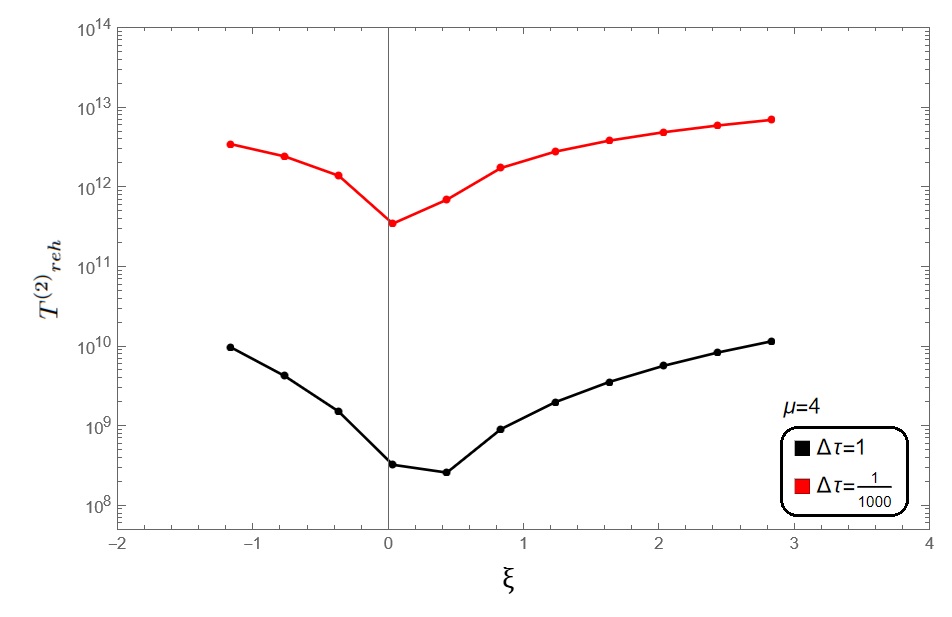}}\\
    \subfloat[]{\includegraphics[width=0.3\textwidth]{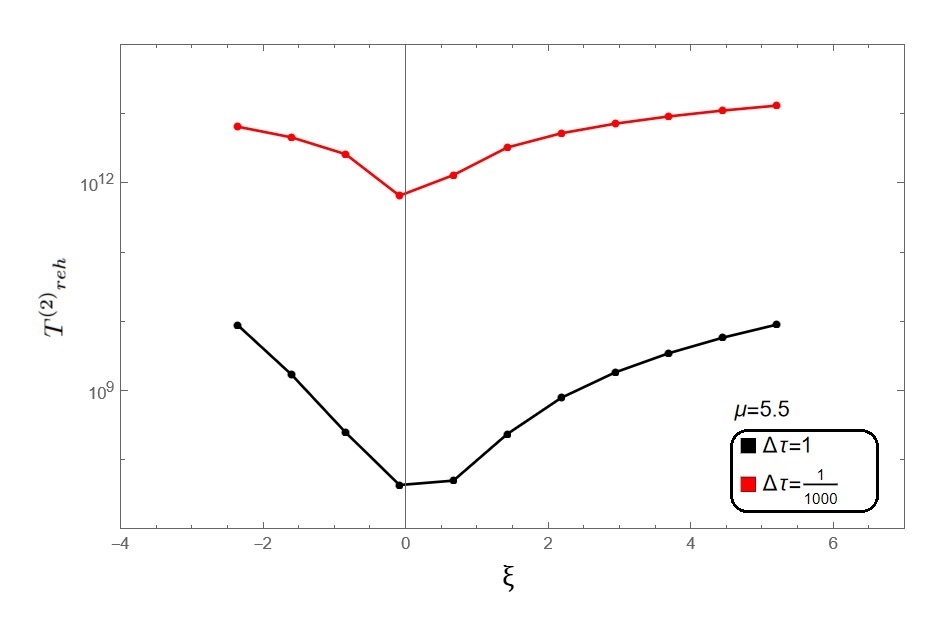}}
    \hfill
    \subfloat[]{\includegraphics[width=0.3\textwidth]{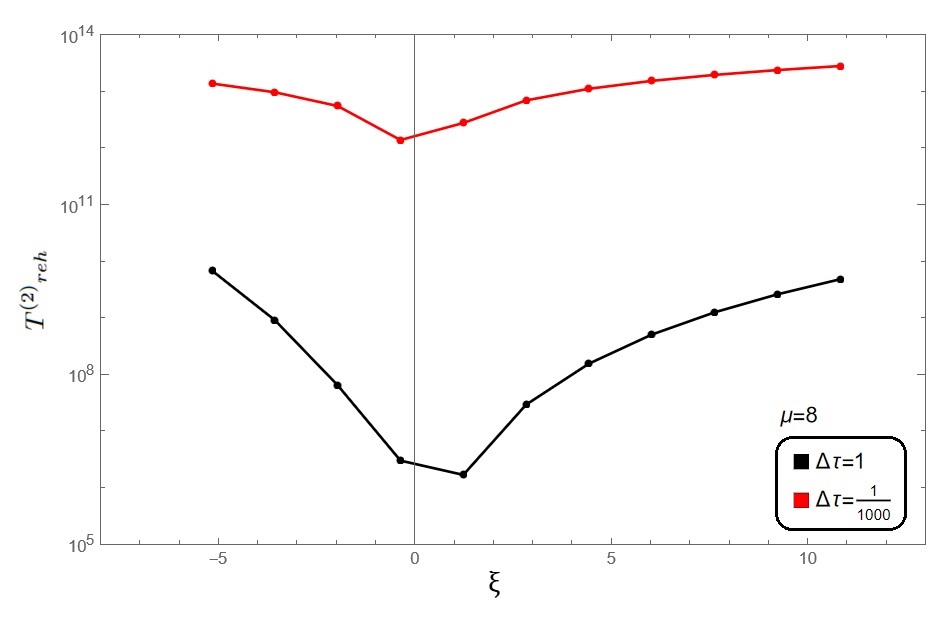}}
     \hfill
    \subfloat[]{\includegraphics[width=0.3\textwidth]{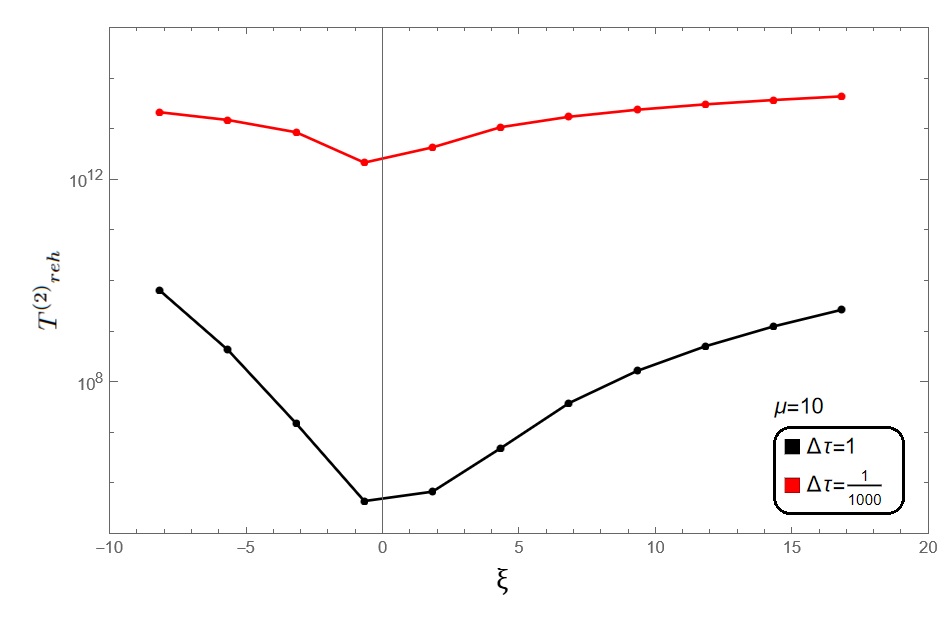}}\\
	\caption{The reheating temperature of the second scenario as a function of $\xi$ for six different values of masses, $\mu= 1, 2.5, 4, 5.5, 8, 10$. In each plot, the spectrum is plotted for $\Delta\tau=1$ and $\Delta\tau=0.001$ in $black$ and $red$, respectively. As can be seen, either the highest possible reheating scenario in the case of $\Delta\tau=1$ corresponds to ${\xi}_{min}$ or ${\xi}_{max}$; with reducing the transition time, the highest temperature corresponds to ${\xi}_{max}$. It is important to note that this result is also valid in the case that the highest temperature corresponds to ${\xi}_{min}$ with a negative value. Also, in cases like $\mu=1, 2.5, 4, 5.5$, where the highest temperature already corresponds to ${\xi}_{max}$ in the case of $\Delta\tau=1$, this behavior is obtained too. But, in these cases, the difference between the temperature of ${\xi}_{max}$ and ${\xi}_{min}$ becomes greater with the reduction of the transition time. Since the general behavior of the first and second reheating temperatures is just the same, in this plot, we have only presented the results of the second temperature.}\label{fig10}
\end{figure}

The behavior of the reheating temperature as a function of allowed values of $\xi$ for three different values of mass and, for the case $\Delta\tau=1$, was plotted in Fig. \ref{fig4}. As we can see, the highest reheating temperature occurs for ${\xi}_{max}$ and ${\xi}_{min}$ at the beginning and the end of the allowed interval of $\xi$. It is obvious from this plot and also from Table. \ref{table} that for masses below $\mu=5.5$, it is ${\xi}_{max}$ that yields the highest temperature, and the value of ${\xi}_{max}$ is always positive. For masses greater than $\mu=5.5$, the ${\xi}_{min}$ corresponds to the highest reheating temperature in the parameter space. For these masses, ${\xi}_{min}$ is negative valued. To answer our important question regarding whether this behavior remains unchanged by reducing the transition time, or in other words, whether the positive-valued $\xi$ can dominate the negative value by reducing the transition time or not, it would be very interesting if we produce such plots as in Fig. \ref{fig4} for $\Delta\tau=0.001$ and compare the results with those of $\Delta\tau=1$. We are already sure from the aforementioned results that both the temperatures of ${\xi}_{min}$ and ${\xi}_{max}$ are increased. But our motivation to produce plots in Fig. \ref{fig10} is to examine, at last, which one can provide us with the highest reheating temperature.

As can be seen, in Fig. \ref{fig10}, we examine six values of mass, $\mu=1, 2.5, 4, 5.5, 8, 10$. These plots demonstrate that for all of the values of mass, by reduction of $\Delta\tau$, ${\xi}_{max}$ dominates the ${\xi}_{min}$ in the competition of producing the highest temperature. More importantly, the results reveal that the behavior in which the reheating temperature of ${\xi}_{max}$ dominates that of ${\xi}_{min}$ by reduction of transition time is quite general, since we also study this behavior for $\mu=1$, which is below $\mu=1.5$, for which the negative values of $\xi$ are not allowed. As can be seen from the results of Fig. \ref{fig10}, the $T_{reh}$ of ${\xi}_{max}$ dominates that of ${\xi}_{min}$ also for this mass. Although for the masses below $\mu=5.5$, ${\xi}_{max}$ plays the role of ${\xi}_{best}$, this role remains unchanged by reduction of the transition time. For higher values of mass, this result is repeated in an interesting manner. We have seen from Fig. \ref{fig4} that for masses greater than $\mu=5.5$ it is the ${\xi}_{min}$ with a negative value that produces the highest temperature in the case of $\Delta\tau=1$. But if one reduces the transition time to values such as $\Delta\tau= 0.01, 0.001$, the GPP rate, the energy density, and, as a result, the reheating temperature of ${\xi}_{max}$ grow in a way that it dominates that of ${\xi}_{min}$. Since Fig. \ref{fig8} tells us that the increase in the reheating temperature by reduction of transition time is more intense in the case of $\mu=10$, we can conclude that in the end, it is $\mu=10$ with its ${\xi}_{max}$ that can provide us with the highest possible reheating temperature in the parameter space of the gravitational reheating scenario.

In this subsection, by considering the effect of the transition time, we found the highest reheating temperature possible. We have seen that although in the case of $\Delta\tau=1$, the highest possible reheating temperature was obtained for $\mu=2.5$ with $\xi=1.21$, which was $1.84\times 10^7 GeV$ and $1.32\times 10^{10} GeV$, corresponding to the first and second reheating scenarios, the highest reheating temperature possible in the parameter space of our model corresponds to $\mu=10$, with its ${\xi}_{max}= 16.83$. This happens for the transition time of the order of $\Delta\tau=0.001$, and is $3.46\times 10^{12} GeV$ and $4.34\times 10^{13} GeV$ for the first and second scenarios of decay, respectively. These reheating temperatures are high enough to prepare the universe to start the HBB era, and demonstrate that the channel of gravitational coupling can, in fact, be considered a very efficient mechanism of reheating. In the last section, we will derive an analytical formula, just as a mathematical tool, that showcases in an exquisite manner the physics behind our numerical calculations discussed in this section.

\section{Results of Analytical Calculation}\label{anaresults}
In this section, we present another function for the scale factor, which is exactly solvable without the need for numerical calculations. This model should be regarded as a parallel model, independent of our primary model in the previous subsection, which also represents the transition from inflation to kination, but instantly. In fact, this model is far from a realistic transition and must be regarded only as a toy model, whose mission is to provide us with a rough and quick estimation of $T_{reh}$. Although the numerical calculation provides us with a complete picture of the effect of the three free parameters of the theory, $\mu$, $\xi$, and $\Delta\tau$, on the GPP rate and $T_{reh}$, it is always beneficial to try to find an analytical description for the model under study. Such an analytical description will be important from the viewpoint of our physical intuition, as it will provide us with a deeper physical understanding of what is happening in the system of equations, as well as the role of each parameter in the resultant gravitational particle production rate and $T_{reh}$. In fact, this analysis can shed light on how each parameter plays its role and affects the physical behavior obtained in more realistic numerical analysis. But it is important to note that usually such analytical solutions are built based on simplified assumptions, and hence they can only provide us with a rough mathematical solution. So, these analytical solutions should be considered neither as an exact expression of the results of a physical model nor as an effort to reproduce the numerical results using exact mathematical formulas. The reason is obvious: The EoM can not be exactly solved in situations where there is a transition period. 

Hence, in this section, our primary goal is to derive an analytical formula for the quantity ${\lvert{\beta}_{\kappa}\lvert}^2$, which is responsible for the GPP rate using a different function for the scale factor. Although \textbf{we tried to use a solvable scale factor whose general behavior is similar to the numerical one} in Eq. \ref{scalenum}, there will, of course, be differences. \textbf{Hence, the calculations of this section should be considered as an independent model from our primary one in this paper, which was presented in the previous section. In addition, although an instant transition is usually considered as the limiting behavior for models with a transition time in the limit of $\Delta\tau\rightarrow0$, in this case such a consideration is not relevant. It is due to the fact that the functional form of the scale factors have differences and hence, the limiting behavior can not be manifestly obtained.} This analysis is done for the sake of completeness and to present a rough description of the GPP channel of reheating in the non-perturbative procedure. \textbf{Such an analysis can be considered as the first step in studying gravitational reheating in a specific model, which provides us with a physical picture of what is happening, as well as an estimation for $T_{reh}$ to check whether the channel of reheating under study is efficient or not. But, as we will present in what follows, the resultant plots do not fit the numerical plots with acceptable accuracy and also can not be considered as their limiting behavior. This, in turn, demonstrates the necessity of numerical calculations and reveals the fact that the analytical calculation should not be regarded as a replacement to the numerical one.} This analysis must be considered only as a helpful tool for better understanding and rough estimation of what is going on.

Since the analytical formula can not be obtained and the EoM, Eq. \ref{main}, can not be solved if we consider the transition period in the evolution function of the scale factor, our first simplifying assumption is that the transition from the inflationary era to the kination must occur instantly. Hence, the function proposed for the evolution of the scale factor in Eq. \ref{scalenum} is not relevant for the analytical calculations of this section, and we have to propose another function. To this end, we will consider the following piecewise function for the scale factor:
\begin{equation}
a(\tau)=
\begin{cases}
-\dfrac{1}{\tau}, & \tau<-1,\\[6pt]
\left(-\dfrac{2}{\tau_f}\,\tau+\dfrac{1}{\tau_f^{2}}+2\right)^{1/2}, & \tau>-1,
\end{cases}
\label{scalean}
\end{equation}

\begin{equation}
a^{2}(\tau)=
\begin{cases}
\dfrac{1}{\tau^{2}}, & \tau<-1,\\[6pt]
-\dfrac{2}{\tau_f}\,\tau+\dfrac{1}{\tau_f^{2}}+2, & \tau>-1.
\end{cases}
\end{equation}

In this equation, the conformal time corresponding to the end of inflation (we will refer to this time as the final time in what follows) and the beginning of the kination is denoted as ${\tau}_f$. This piecewise function is built under these assumptions:
\begin{itemize}
    \item The EoS parameter during the kination epoch is considered to be $W =1$, and also $W = -1$ for the inflationary era. 
    \item \textbf{At the end of inflation, ${\tau}_f$, the scale factor and its first time derivative have to be continuous.}
    \item As well as the proposed function of the scale factor for the numerical analysis, \textbf{this scale factor is normalized so that at the end of inflation, $(\tau = {\tau}_f)$, its value becomes unity.}  
    \item The final time, corresponding to the end of inflation, is a negative-valued number. This is because, in general, the conformal time is negative during the inflationary epoch. According to this and previous assumptions, the final time must be set equal to ${\tau}_f = -1$ at the end of calculations.  
    \item The transition between inflation and kination occurs instantly. From the viewpoint of inflation, this means that the inflationary potential must be like a step function.    
\end{itemize}
Beginning with these assumptions, one can easily obtain the functional form of the scale factor as it is written in Eq. \ref{scalean}. For the details of the procedure for obtaining the piecewise functional form of the scale factor, and how the aforementioned assumptions are used in this procedure, see App. \ref{anscder}. It is important to keep in mind that this scale factor is derived so that it has an acceptable level of correspondence with the numerical scale factor. But of course, there are differences. Using this scale factor, we are going to solve the EoM for the inflationary and kination eras, separately. Although the result of this calculation provides us with the general solutions of the EoM, using the appropriate initial conditions, again based on the notion of the adiabatic vacuum, we can find the specific solution of the EoM for each era. These specific solutions are, in fact, the physical mode functions that define the physical vacuums corresponding to the inflation and kination epochs. Using these mode functions, we can calculate the quantity ${\lvert{\beta}_{\kappa}\lvert}^2$ responsible for the GPP rate at a specific time, which in this case is the final time, ${\tau}_f$. Hence, we are going to calculate the GPP rate analytically. We start by calculating the exact form of the physical mode function for the inflationary era. It is worth mentioning that here, we are only going to present the final results of analytical calculations. For detailed calculations containing the main mathematical steps and formulas used during the derivations of equations, see App. \ref{bes} and App. \ref{ander}.

Using the functional form of the scale factor for the inflationary era from Eq. \ref{scalean}, the EoM for this era becomes:
\begin{equation}
\frac{d^{2}\Delta_{\kappa}(\tau)}{d\tau^{2}}
+\left[\kappa^{2}+\frac{1}{\tau^{2}}M^{2}\right]\Delta_{\kappa}(\tau)=0
\label{eominf}
\end{equation}
where
\begin{equation}
M^{2}=\mu^{2}+12\,\xi-2
\label{M}
\end{equation}
is the ICM of the spectator field during inflation. To obtain the above equations, we have used the fact that the curvature scalar, $C(\tau)$, remains constant throughout the inflationary era with the value $C(\tau) = 12 = cte$. Solving Eq. \ref{eominf}, the general solution of the EoM is obtained as:
\begin{equation}
\Delta_{\kappa}(\tau)=
\frac{1}{2}\!\left(C_{1}+\frac{C_{2}}{i}\right)\!\sqrt{-\tau}\;
H^{(1)}_{\nu}\!\left(-\kappa\tau\right)
+\frac{1}{2}\!\left(C_{1}-\frac{C_{2}}{i}\right)\!\sqrt{-\tau}\;
H^{(2)}_{\nu}\!\left(-\kappa\tau\right)
\label{gensolinf}
\end{equation}
In this solution, as is expected, there are two constants, $C_1$ and $C_2$, that have to be determined using the initial conditions. This is due to the fact that the EoM, Eq. \ref{main}, is a second-order differential equation in conformal time\footnote{It is important to note that the conformal time is a negative-valued parameter during the inflationary era and hence, the $\sqrt{-\tau}$ part of the general solution is a real-valued function of time.}. As we have mentioned in Sec. \ref{theo}, the relevant initial conditions that have to be imposed on this general solution are the Bunch-Davis initial conditions for the inflationary era. Invoking these conditions from Eq. \ref{bd}, \textbf{which are based on the notion of the adiabatic vacuum}, the special solution of the EoM for the inflationary era is obtained as:
\begin{equation}
\Delta_{\kappa}^{\mathrm{inf}}(\tau)=
\frac{1}{2}\sqrt{-\tau}\;
e^{\,i\frac{2\nu+1}{4}\pi}\;
H^{(1)}_{\nu}\!\left(-\kappa\tau\right)
\label{bdvac}
\end{equation}
where the parameter $\nu$ is defined as:
\begin{equation}
\nu \equiv \frac{1}{2}\sqrt{\,1-4M^{2}\,}
\label{nu}
\end{equation}

This mode function, known as the Bunch-Davis mode function, is considered to be the physical mode function that defines the vacuum state for the inflationary era (We will refer to this mode function as the Bunch-Davis vacuum or the BDV for short.). There remains an important point that is worth turning our attention to: in the constant exponential that appears in the BDV, $e^{i\frac{2\nu+1}{4}\pi}$, the parameter $\nu$, defined through Eq. \ref{nu}, can be a real or complex number. It is obvious from Eq. \ref{nu} that for smaller values of the $M$, the parameter $\nu$ can take real values, while its values can be complex if one considers greater values for mass or $\xi$. Hence, the exponential $e^{i\frac{2\nu+1}{4}\pi}$ may be only a constant phase, or a real number that is essential to consider for the true normalization of the BDV. This normalization must be held so that $W[{\Delta}_{\kappa},{{\Delta}_{\kappa}}^{\star}]=i$. Now, the same procedure has to be done for the kination stage. Again, we have to start with the piecewise function of the scale factor in Eq. \ref{scalean}, and use the functional form corresponding to the kination stage. As a result, the curvature scalar, or the dimensionless Ricci scalar, becomes:
\begin{equation}
C(\tau) \equiv \frac{6}{a^{3}}\frac{d^{2}a}{d\tau^{2}}
= -\,\frac{6}{\tau_f^{2}}
\left(-\frac{2}{\tau_f}\,\tau + \frac{1}{\tau_f^{2}} + 2\right)^{-3}
\label{ckin}
\end{equation}
Substituting this function and the relevant form of the scale factor in the EoM, Eq. \ref{main}, we obtain:
\begin{equation}
\frac{d^{2}\Delta_{\kappa}(\tau)}{d\tau^{2}}
+\Bigg[
\kappa^{2}
-\frac{6\big(\xi-\tfrac{1}{6}\big)}{\tau_f^{2}
\left(\frac{-2}{\tau_f}\,\tau + \frac{1}{\tau_f^{2}} + 2\right)}
+\left(\frac{-2}{\tau_f}\,\tau + \frac{1}{\tau_f^{2}} + 2\right)\mu^{2}
\Bigg]\Delta_{\kappa}(\tau)=0
\label{eomkin1}
\end{equation}
This equation can not be solved analytically, and hence, we have to use a valid, reasonable approximation similar to that one in the case of numerical calculation. In that case, in the procedure of deriving the KV, we argued that since, during the kination stage, the curvature scalar damps very quickly, it is possible not to consider the term $(\xi-\frac{1}{6})a^2C$ in the EoM, compared to the term $a^2{\mu}^2$. A similar Argument can be valid for this case too. Since, after the final time, the kination stage starts instantly, the curvature scalar starts its damping very soon, and hence, again, it is justified to ignore the $(\xi-\frac{1}{6})a^2C(\tau)$ in the EoM. Hence, the final form of the EoM for the kination era becomes:
\begin{equation}
\frac{d^{2}\Delta_{\kappa}(\tau)}{d\tau^{2}}
+\Bigg[
\kappa^{2}
+\left(-\frac{2}{\tau_f}\,\tau + \frac{1}{\tau_f^{2}} + 2\right)\mu^{2}
\Bigg]\Delta_{\kappa}(\tau)=0
\label{eomkin}
\end{equation}
Now this equation can yield an analytical general solution as :
\begin{equation}
\Delta_{\kappa}(\tau)=
T_{1}\,\sqrt{x}\;H^{(1)}_{1/3}\!\left(\tfrac{2}{3}\,x^{3/2}\right)
+ T_{2}\,\sqrt{x}\;H^{(2)}_{1/3}\!\left(\tfrac{2}{3}\,x^{3/2}\right)
\label{kgensol}
\end{equation}
where the parameter $x$ is defined as:
\begin{equation}
x(\tau)\equiv
\frac{\kappa^{2}+2\mu^{2}+\dfrac{\mu^{2}}{\tau_{f}^{2}}
-\dfrac{2\mu^{2}}{\tau_{f}}\,\tau}
{2^{2/3}\!\left(\dfrac{\mu^{2}}{\tau_{f}}\right)^{2/3}}
\label{x}
\end{equation}
Again, we invoke the notion of the adiabatic vacuum for the kination era\footnote{As we have mentioned before, this notion of adiabatic vacuum is sometimes referred to in the literature as the ``vacuum at short distances."}, which demonstrates that for the high-frequency modes with short wavelength, ${\kappa}^2\gg a^2{\mu^2}$ or ${\frac{{\kappa}^2}{a^2{\mu}^2}}\rightarrow\infty$, the behavior of the mode must tend to the one for the modes in the Minkowski space-time. Imposing these conditions on the general solution in Eq. \ref{kgensol}, the specific solution of the EoM for the kination era is obtained as:
\begin{equation}
\Delta_{\kappa}^{\mathrm{kin}}(\tau)=
\sqrt{\frac{\pi}{6}}\left(\frac{2\mu^{2}}{-\tau_{f}}\right)^{-1/6}
e^{i\ 
\!\left((-\tau_{f})\frac{\kappa^{3}}{3\mu^{2}}
+(-\tau_{f})^{3/2}\kappa-\frac{5\pi}{12}\right)
}
\,\sqrt{x}\;
H^{(2)}_{1/3}\!\left(\tfrac{2}{3}\,x^{3/2}\right)
\label{kspsol}
\end{equation}
where again the function $x$ is defined through Eq. \ref{x}. This special solution is considered to be the physical mode function that defines the physical vacuum for the kination era, using which the physical particle number is defined. We will refer to this mode function as the analytical KV or AKV for short. Since both the BDV and AKV are solutions of the EoM, we have two sets of normalized bases in the space of the solutions of the EoM, which are ${{\Delta}_{\kappa}}^{inf}$, ${{{\Delta}_{\kappa}}^{inf}}^{\star}$ on one hand, and on the other hand, ${{\Delta}_{\kappa}}^{kin}$, ${{{\Delta}_{\kappa}}^{kin}}^{\star}$. Hence, each of these bases can be written as an expansion in terms of the other one. We write this expansion for the BDV in terms of AKV as:
\begin{equation}
\Delta_{\kappa}^{\mathrm{inf}}(\tau)
=\alpha_{\kappa}\,\Delta_{\kappa}^{\mathrm{kin}}(\tau)
+\beta_{\kappa}\,\Delta_{\kappa}^{\mathrm{kin^{*}}}(\tau)
\label{exp}
\end{equation}
and if the mode functions are normalized to $i$, the Bogoliubov coefficient ${\beta}_{\kappa}$ can be calculated through:
\begin{equation}
\beta_{\kappa}
= W\!\left[\Delta_{\kappa}^{\mathrm{kin}},\,
\Delta_{\kappa}^{\mathrm{inf}}\right]_{\tau}
,\qquad
W[f,g]\equiv f\,g' - f'\,g
\label{bet}
\end{equation}
Now, with BDV and AKV in hand, we must calculate their first-time derivative with respect to the dimensionless conformal time. Then, using the Bogoliubov transformation, we can compute the GPP rate. The resultant formula by setting the final time to ${\tau}_f=-1$ for ${\beta}_{\kappa}$ becomes:
\begin{align}
|\beta_{\kappa}|=\Bigl|e^{\tfrac{1}{2} i \nu \pi}\Bigr|\,
\Bigl|\,
\sqrt{\frac{\pi}{6}}\,(2\mu^{2})^{-1/6}\,x_f^{1/2}
H^{(2)}_{1/3}\!\left(\frac{2}{3}x_f^{3/2}\right)\times...\,\\
\qquad
...\Big\{\frac{-1}{4}\sqrt{\pi}H^{(1)}_{\nu}(\kappa)
-\frac{\kappa}{4}\sqrt{\pi}\!\left[H^{(1)}_{\nu-1}(\kappa)-H^{(1)}_{\nu+1}(\kappa)\right]\Big\}...\,\\
\qquad
...-\sqrt{\frac{\pi}{6}}\,(2\mu^{2})^{1/6}
\Big\{\frac{1}{4}x_f^{-1/2}H^{(2)}_{1/3}\!\left(\frac{2}{3}x_f^{3/2}\right)...\,\\
\qquad
...+\frac{1}{4}x_f\!\left[H^{(2)}_{-2/3}\!\left(\frac{2}{3}x_f^{3/2}\right)
-H^{(2)}_{4/3}\!\left(\frac{2}{3}x_f^{3/2}\right)\right]\Big\}
\sqrt{\pi}\,H^{(1)}_{\nu}(\kappa)\,\Bigr|
\label{bettafin}
\end{align}
where 
\begin{equation}
x_f \equiv \frac{\kappa^{2}+\mu^{2}}{\big(2\mu^{2}\big)^{2/3}}
\end{equation}

\begin{equation}
\nu \equiv \frac{1}{2}\sqrt{\,1-4M^{2}\,}
\end{equation}

\begin{equation}
M^{2} \equiv \mu^{2}+12\,\xi-2
\end{equation}
To follow the details and steps in the derivation procedure of these equations, see the App. \ref{ander}. The square of this quantity is exactly the GPP rate computed through the analytical calculations. 

\begin{figure}
	\subfloat[]{\includegraphics[width=0.5\textwidth]{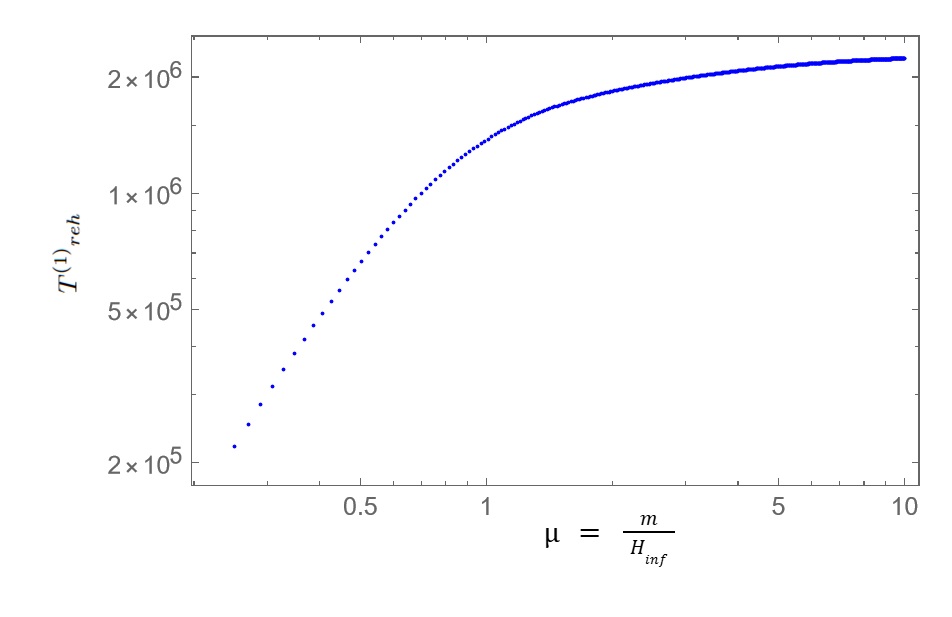}}
	\hfill
	\subfloat[]{\includegraphics[width=0.5\textwidth]{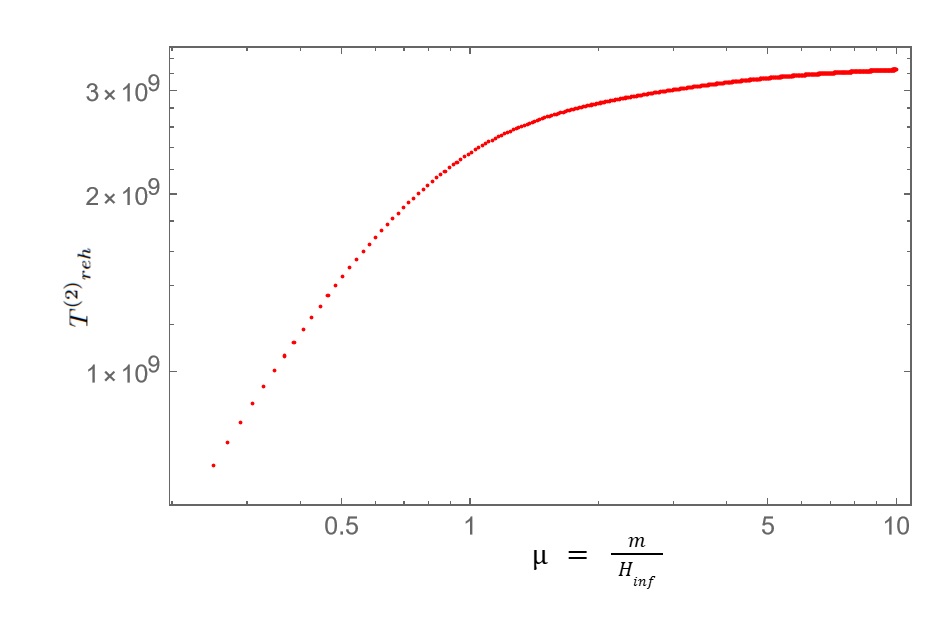}}
	\caption{The reheating temperature as a function of mass in the conformal case obtained by analytical calculation. In this figure, (a) is for the first scenario and (b) is for the second scenario of reheating. In both scenarios, the reheating temperature permanently increases by increasing the mass.}\label{fig11}
\end{figure}

Now, with this formula in hand, we are going to plot the GPP rate, the energy density of produced particles, as well as the reheating temperatures of both scenarios. We will compare the resultant plots with those computed by the numerical approach to emphasize the necessity of more realistic numerical examination on the one hand, and on the other hand, show that the analytical study is still very helpful to provide us with a better understanding of how each parameter affects the results mathematically. As our first plot of results in this section, we provide the reheating temperature of both scenarios for the conformal case in Fig. \ref{fig11}. Before going to interpret the plots, we have to note again that in the analytical calculations, there are no effects due to the transition time. Here, the transition between the inflation and kination eras is considered to happen instantly. The plots in Fig. \ref{fig11} are exactly the plots depicted in black in Fig. \ref{fig1}. Those plots were also produced using the analytical formulas obtained in this section. We presented them in Fig. \ref{fig1} to provide our readers with a plot that compares the results of both numerical and analytical studies. Of course, we should keep in mind that we did our best to find an analytical scale factor whose behavior is as similar as it can be to that of the numerical one; these numerical and analytical plots are depicted for two independent models of the scale factor. Nevertheless, this comparison is helpful from the viewpoint of our physical intuition. The difference between the two plots of Fig. \ref{fig1} and Fig. \ref{fig11} is that we produced Fig. \ref{fig11} with more points (steps of the code run). Our motivation to produce this new plot is that we want to extract important results from it. 

First, in the case of numerical calculation with $\Delta\tau=1$, we have seen that in the conformal case, the gravitational production of particles is suppressed for high values of mass. In fact, at first, for low values of mass, $T_{reh}$ increases by increasing the mass and then, it decrease for large values of the mass. As can be seen, this is not the case in the analytical calculation of the reheating temperatures in the conformal case. In this case, as Fig. \ref{fig11} demonstrates, the reheating temperature, resulting from the GPP scenario, is not suppressed for higher values of mass. Instead of such a suppression, the resultant $T_{reh}$ always increases with increasing mass. So, in the numerical calculations, there exists a specific value of mass, around $\mu=1.5$, which corresponds to the maximum $T_{reh}$ achievable in the case of $\xi=\frac{1}{6}$ and $\Delta\tau=1$. But there is no such maximum in analytical calculations; instead, $T_{reh}$ permanently increases with mass. The existence of this extremum point in the numerical plot and its absence in the analytical one is a vital difference between the numerical and analytical calculations. This difference is highlighted more in Fig. \ref{fig1}, where we have plotted the results of both calculations in one plot. \textbf{We must stress that the result "$T_{reh}$ increases by increasing the mass" is obtained in the mass interval under consideration in our study, i.e. $0.25\le\mu\le10$. We did not examine higher values of mass.}

Although there is vital difference between the results of analytical calculation with that one of numerical calculation with $\Delta\tau=1$, it seems the behavior of increasing of $T_{reh}$ with mass in this case is similar to the case of short transition time in numerical calculations. This plot demonstrates that the GPP occurs more intensely for higher values of mass. It is the opposite of the case of the numerical calculation for $\Delta\tau=1$, but it is in accordance with the result obtained in the numerical calculation for short and rapid $\Delta\tau$, for instance $\Delta\tau=0.01$ or $\Delta\tau=0.001$. From our physical intuition discussed in the previous section, there is a physical justification for this result. As we have seen in the previous section, the effect of a rapid, harsh transition time is more perceptible for higher values of mass. In our analytical calculations, we are using a piecewise scale factor, which demonstrates that the transition between inflation and kination eras happens very sharply and instantly. So this result is well in accordance with our physical intuition that such an instant transition leaves its effect on larger values of mass more significantly and hence, it is reasonable that $T_{reh}$ increases by increasing the mass in this analysis.     

\begin{figure}
	\subfloat[]{\includegraphics[width=0.5\textwidth]{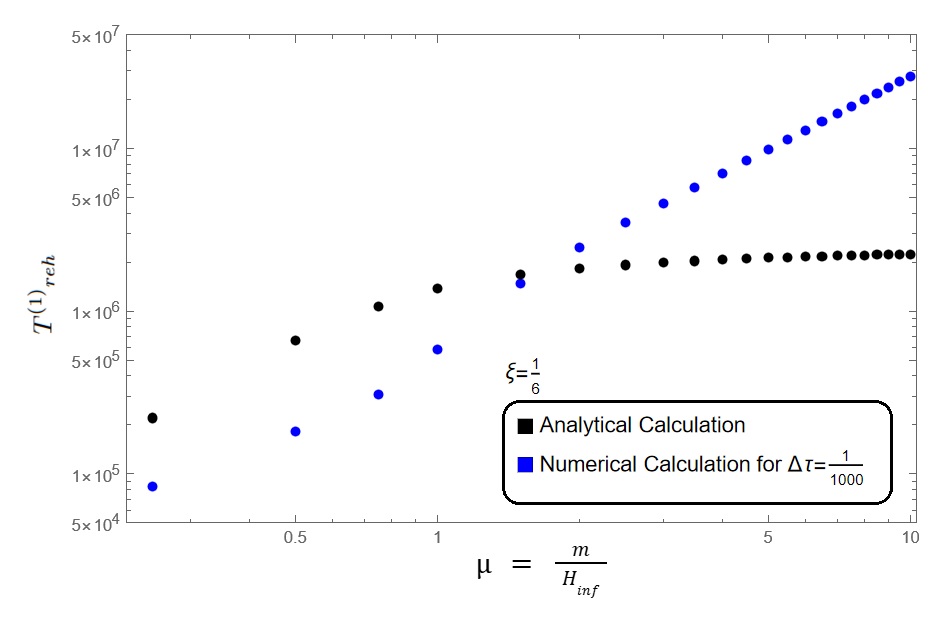}}
	\hfill
	\subfloat[]{\includegraphics[width=0.5\textwidth]{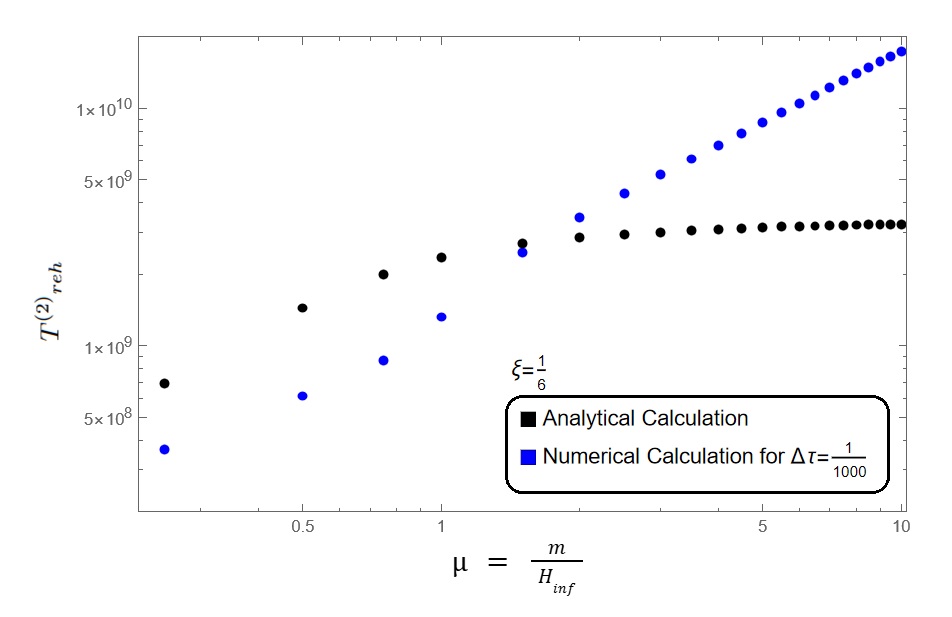}}
	\caption{The comparison between the instant transition and the numerical solution with $\Delta\tau=0.001$. In this figure, (a) is for the first and (b) is for the second scenario of decay. The color $black$ corresponds to the analytical solution, where $blue$ is for the numerical one. This plot contradicts our physical intuition regarding the fact that, in the case of analytical consideration, the transition happens instantly and hence we expect its plot to lie at higher values than the numerical one. But we have to keep in mind that these two analyses are done for two different models of scale factors for the inflation-kination transition.}\label{fig12}
\end{figure}

We argued above that the justification for the behavior obtained in Fig. \ref{fig11} is just like the explanation that we presented in the previous section for the effect of the reduction of the transition time, which, in fact, is the violation of the adiabaticity condition during the transition. Hence, it would be interesting to compare the resultant reheating temperatures of the conformal case obtained from the analytical calculations and also from the numerical calculations, but with $\Delta\tau=0.001$. We have done such a comparison, and the resultant plots are depicted in Fig. \ref{fig12}. This figure clearly demonstrates that, although the transition in the case of analytical calculation is considered to happen instantly, the particle creation and the resultant reheating temperature, in the case of numerical calculations with a very short transition time, lie at higher values. At first glance, maybe such a result contradicts our physical intuitions, since we expect that the reheating temperature of the analytical case should be greater than the numerical one, even for a short transition time. But there may be a justification for this plot that rests in the functional forms of our chosen scale factors. 

From our physical intuition, we expect that the sharper the transition time, the higher $T_{reh}$ for greater masses. In the case of analytical calculation, the transition happens instantly and hence, we expect that the resultant $T_{reh}$ takes values higher than in the case of the numerical one with a short but non-zero transition time. But, as can be seen from Fig. \ref{fig12}, the results contradict this expectation. In fact, the reason may be in the difference between the two different scale factors invoked for the numerical and analytical calculations. As we have mentioned above, these two models should be considered as two independent models of transition. In the previous section, we used the numerical scale factor introduced in Eq. \ref{scalenum} for our numerical calculations. That scale factor behaves as $a^2(\tau)=\frac{1}{1+{\tau}^2}$ during inflationary era and as $a^2(\tau)=1+\tau$ during kination. That scale factor also provides us with a transition time whose duration was tuned by the factor $\Delta\tau$. It is obvious that its limiting behaviors are, in fact, different from those of the analytical one in Eq. \ref{scalean}. Although the analytical scale factor is obtained in a way that its general behavior is almost the same as the numerical one, there are differences that make our results different, as is seen in Fig. \ref{fig12}. We have no way out of such a difference. For making the analytical scale factor, although we forced ourselves to make the behavior of the two scale factors become just the same, we have to consider many constraints discussed above, and these constraints cause such a difference. \textbf{This is exactly the main reason for our emphasis above that the analytical study can only provide us with a rough estimation of GPP rate and $T_{reh}$, and so, the necessity of numerical calculations is inevitable.}

It would be interesting in future studies if one finds functional forms for a numerical and an analytical scale factor that behave exactly the same and confirm the claim above that the resultant $T_{reh}$ from analytical calculation lies above the numerical one, even with shorter and shorter transition times. In fact, from our physical intuition, we must expect that the analytical calculation behaves as an upper limit for the numerical one. So by limiting $\Delta\tau$ to smaller and smaller values, $\Delta\tau\rightarrow0$, the plot of numerical $T_{reh}$ tends to that of the analytical one and mimics its behavior.

\begin{figure}
	\subfloat[]{\includegraphics[width=0.3\textwidth]{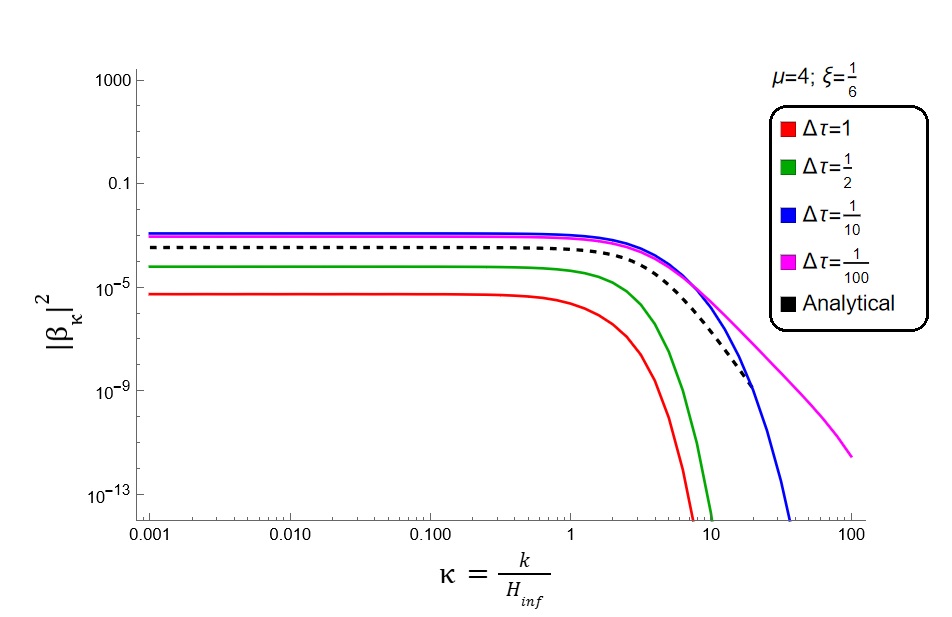}}
	\hfill
	\subfloat[]{\includegraphics[width=0.3\textwidth]{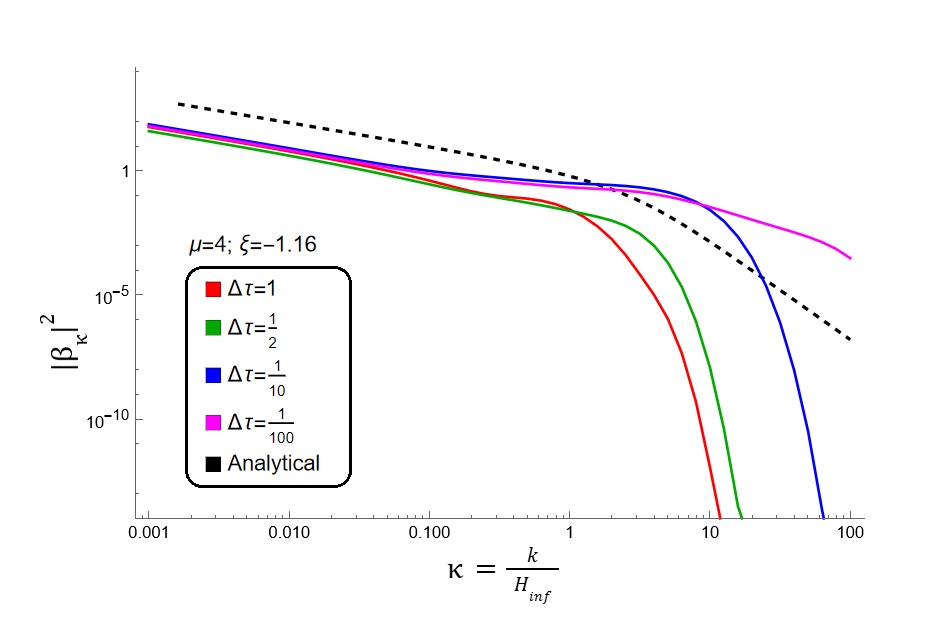}}
    \hfill
	\subfloat[]{\includegraphics[width=0.3\textwidth]{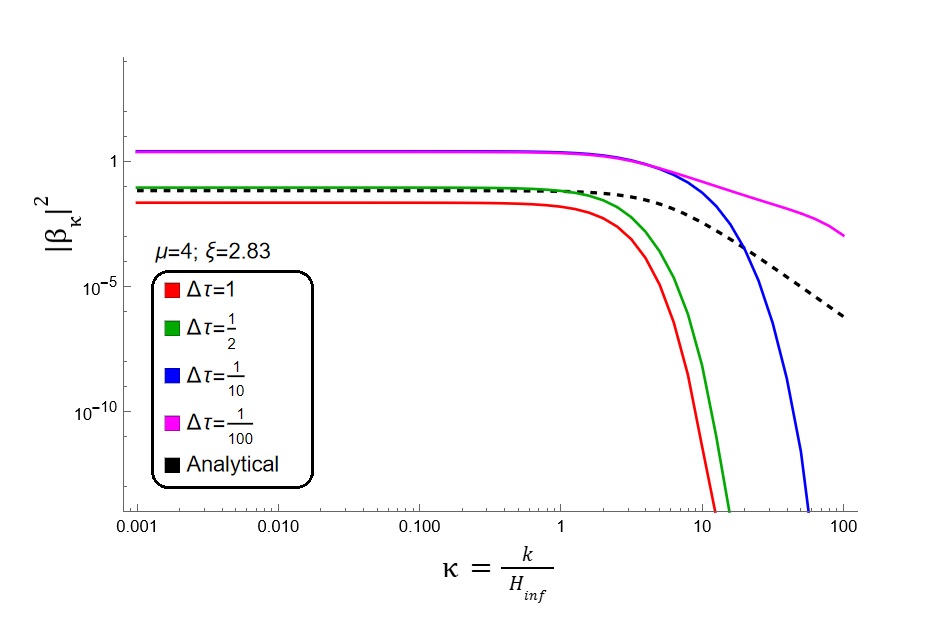}}\\
    \subfloat[]{\includegraphics[width=0.3\textwidth]{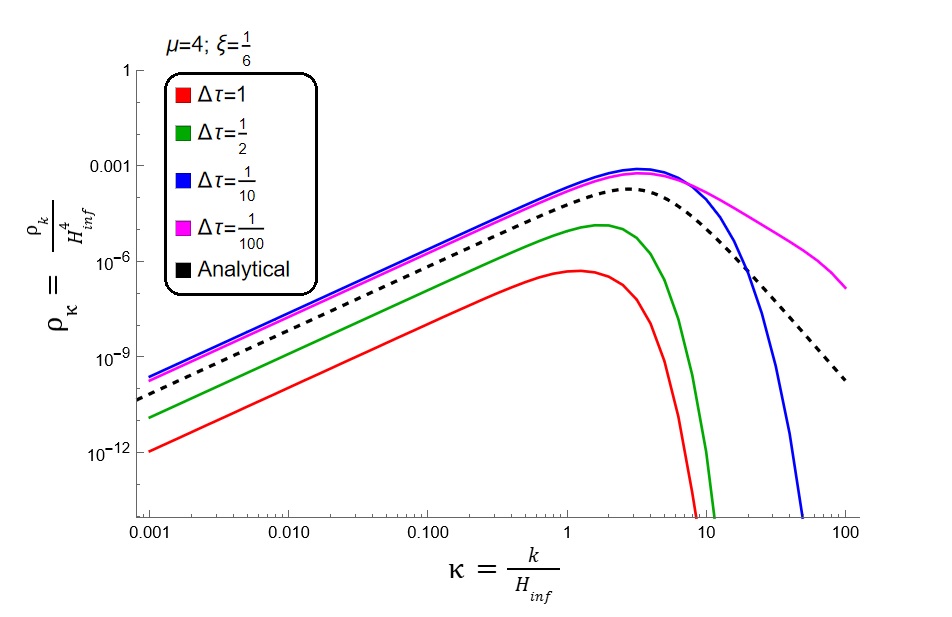}}
    \hfill
    \subfloat[]{\includegraphics[width=0.3\textwidth]{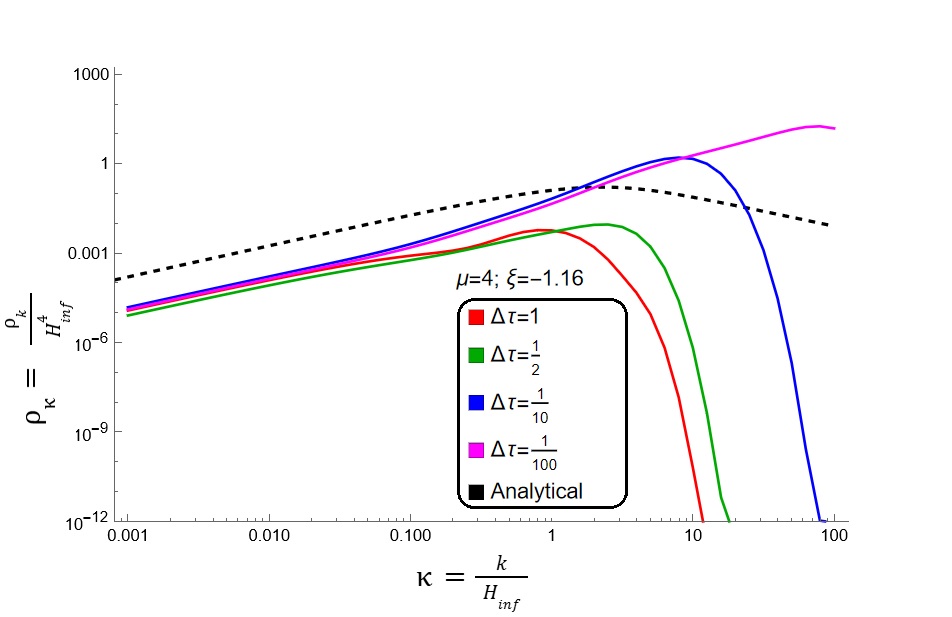}}
     \hfill
    \subfloat[]{\includegraphics[width=0.3\textwidth]{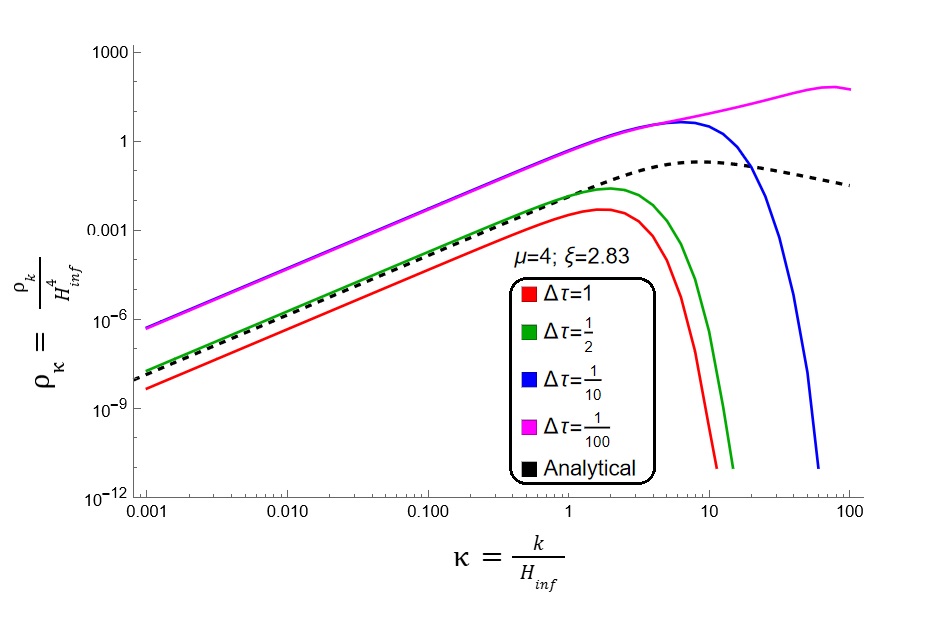}}\\
	\caption{Comparison between the numerical solutions with different transition times and the analytical solutions in the case of $\mu=4$. In these plots, in (a) and (d), the GPP rate and the energy density are depicted for $\xi=\frac{1}{6}$. (b) and (e) are for $\xi=-1.16$, and (c) and (f) are for $\xi=2.83$. These plots demonstrate that the analytical calculation must be regarded only as an independent parallel model for a rough estimation of the GPP rate. To obtain the exact GPP rate for a specific model of transition, numerical analysis is necessary.}\label{fig13}
\end{figure}

\begin{figure}
	\subfloat[]{\includegraphics[width=0.3\textwidth]{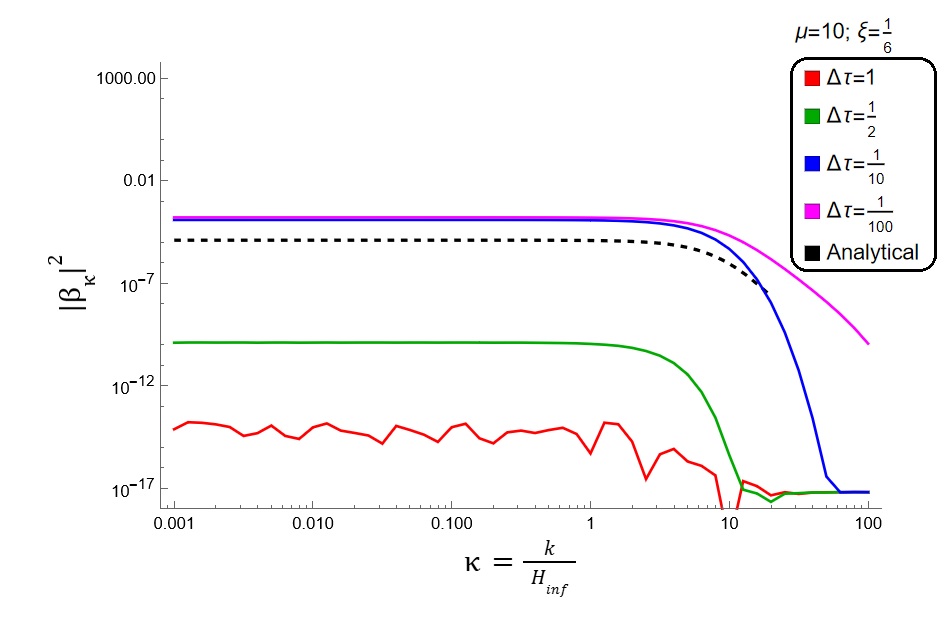}}
	\hfill
	\subfloat[]{\includegraphics[width=0.3\textwidth]{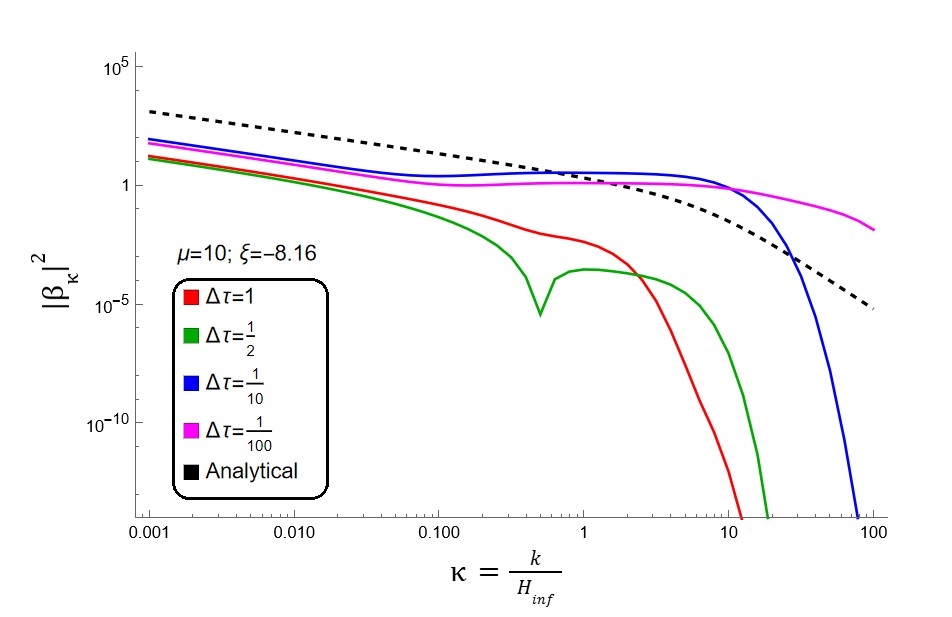}}
    \hfill
	\subfloat[]{\includegraphics[width=0.3\textwidth]{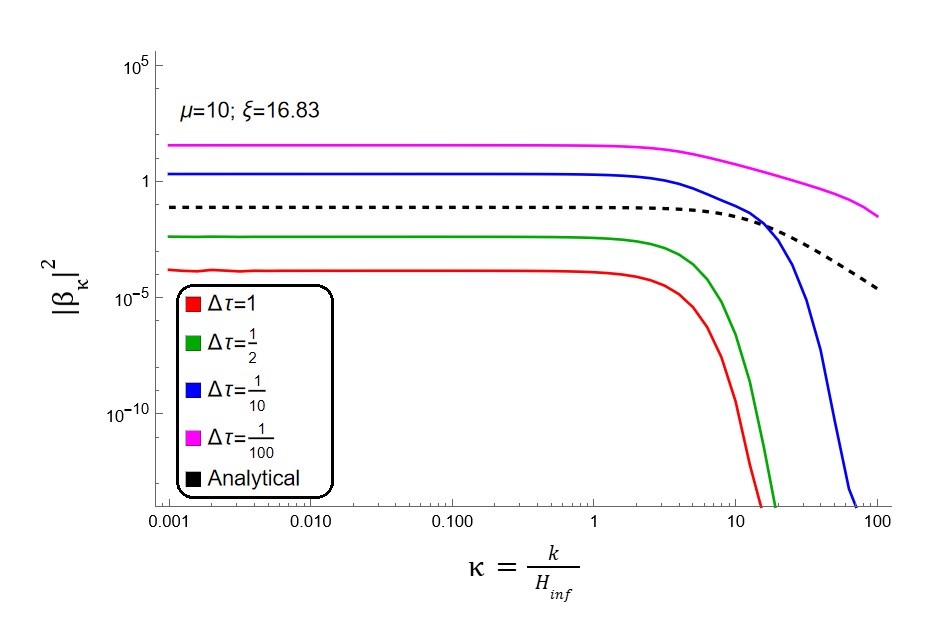}}\\
    \subfloat[]{\includegraphics[width=0.3\textwidth]{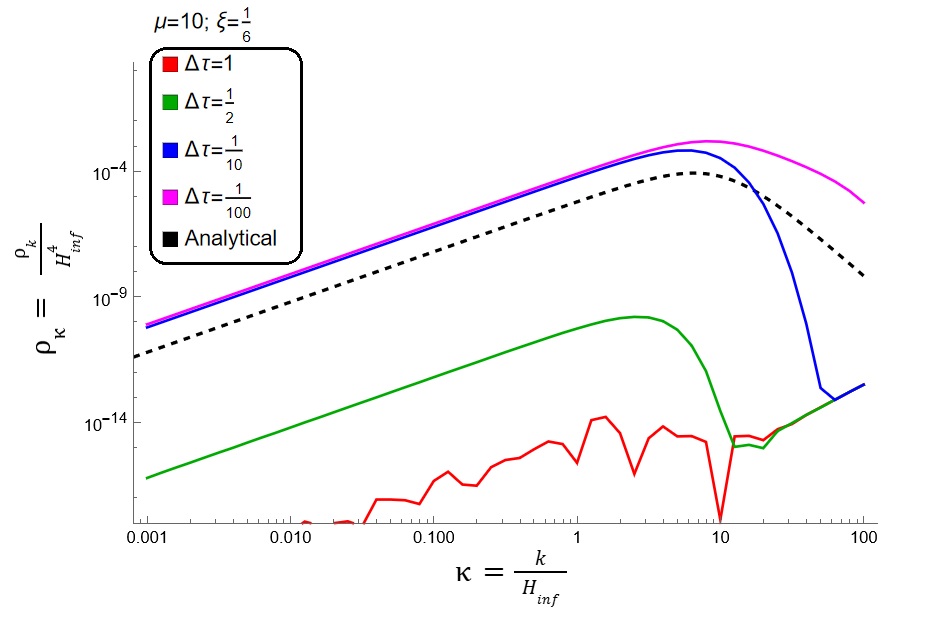}}
    \hfill
    \subfloat[]{\includegraphics[width=0.3\textwidth]{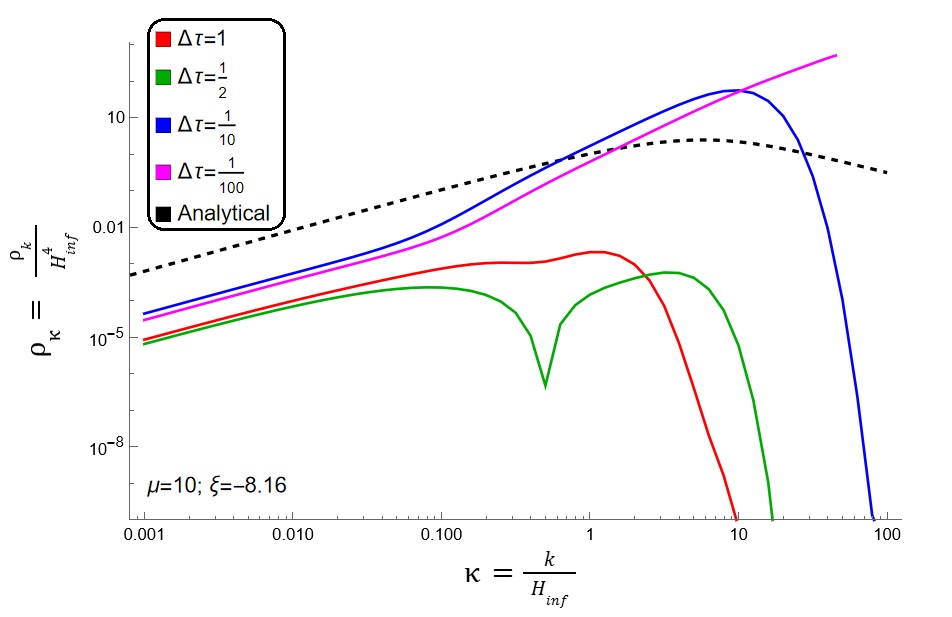}}
     \hfill
    \subfloat[]{\includegraphics[width=0.3\textwidth]{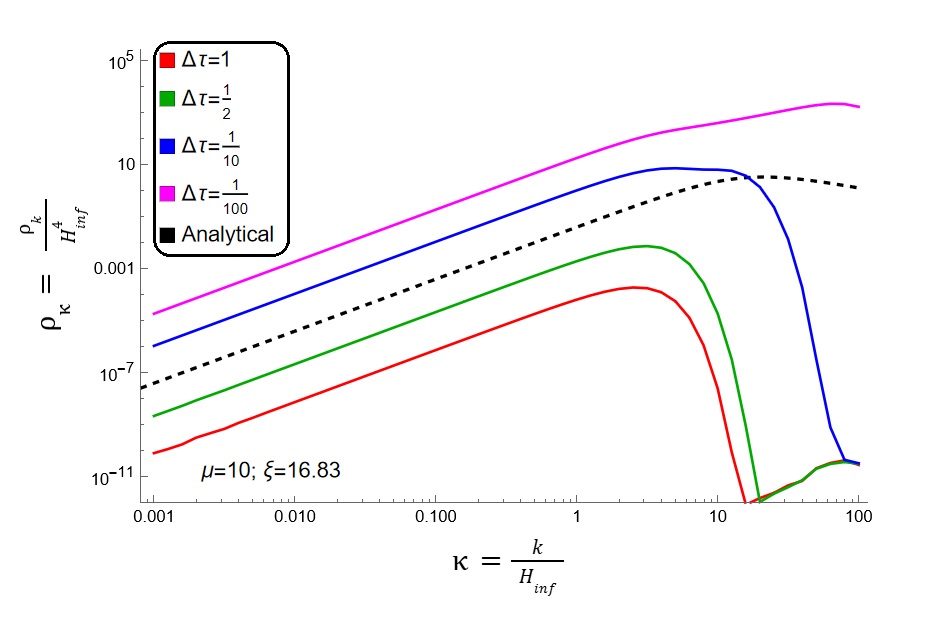}}\\
	\caption{Comparison between the numerical solutions with different transition times and the analytical solutions in the case of $\mu=10$. In these plots, in (a) and (d), the GPP rate and the energy density are depicted for $\xi=\frac{1}{6}$. (b) and (e) are for $\xi=-8.16$, and (c) and (f) are for $\xi=16.83$. These plots again contradict our physical intuition regarding the fact that the analytical calculations should provide us with the highest GPP rate and produced energy density. The reason for this contradiction lies in the difference between the functional form of the two scale factors chosen for numerical and analytical calculations.}\label{fig14}
\end{figure}

Such a phenomenon can also be seen from the spectrum of GPP rate, ${\lvert{\beta}_{\kappa}\lvert}^2$, and the energy density spectrum, ${\rho}_{\kappa}$. For a better comparison, we have plotted these quantities for $\mu=4$ and $\mu=10$ for the conformal case, their ${\xi}_{min}$ and ${\xi}_{max}$ in Fig. \ref{fig13} and Fig. \ref{fig14}. As can be seen, in these plots, for each mass and each value of the gravitational coupling, we have plotted the analytical spectrum, as well as the numerical one for some different transition times.

These figures elegantly demonstrate that the analytical calculation can be considered only as an independent model for a rough and quick estimation of the GPP rate and the resultant $T_{reh}$. Nevertheless, these figures tell us that in some cases, the general behavior of the analytical results may have similarities to the numerical one and so, the analytical calculation still can be considered as a not too bad approximation for the GPP calculations. Such a calculation may not tell us about the exact values of the GPP rate and $T_{reh}$, but it can help us to understand mathematically how each parameter plays its role in gravitational reheating and also provide us with a hasty estimation of $T_{reh}$ in a gravitational background that is caused by a specific model. This mathematical realization happens through the analytical formulas obtained in this section for $\lvert{\beta}_{\kappa}\lvert^2$.

By these considerations, we have covered the whole parameter space of the gravitational reheating scenario for the superheavy particles in an inflation-kination transition. We study this scenario of reheating using both numerical calculations and an analytical model. In this section, we explained the analytical way to study this issue. The details of our analytical calculation are presented in the appendices of our paper. Our analytical results show that this path of study of the GPP scenario is also helpful for a better and deeper understanding of this gravitational channel of reheating.   

\section{Conclusion}\label{con}

In this paper, we have studied the issue of gravitational reheating in the context of superheavy scalar particles within an inflation-kination background. This model consists of three free parameters that build up a 3D parameter space. Each point in this parameter space corresponds to a specific model of the scalar field that lives in the background of an inflationary model in the early universe. Two of these parameters are for the scalar field, which are the mass of the scalar field, $\mu$, and its coupling to the gravitational background, $\xi$. The other parameter, $\Delta\tau$, which is the transition time between inflation and kination, is the characteristic feature of the inflationary model. \textbf{We had three main goals in this study}: \textbf{First}, to examine the effect of non-conformal gravitational coupling on the GPP rate and its resultant reheating temperature. In fact, we aimed to investigate how the presence of non-conformal coupling affects the GPP rate and its resultant reheating temperature. \textbf{Second}, we want to survey a wide range of the parameter space of this model discussed above. Such a survey provides us with a complete picture of the issue of the gravitational production of superheavy particles, and more importantly, how different parameters act together, how they can change each other's results, and finally, what role each one plays in providing the universe with the highest possible reheating temperature. Our \textbf{third} goal was to determine the highest possible reheating temperature in this scenario. This goal is also of great interest in the literature, as it has been one of the main questions in early universe studies for two or three decades: whether gravitational reheating can serve as a sufficient scenario for reheating in the early universe. In other words, is this scenario capable of providing the universe with a reheating temperature necessary for the beginning of the hot Big Bang era? These questions seem more important if the scenario of the inflationary background is that of the NO models.

To reach these goals, we solved our EoM numerically. Our calculations were based on the notion of the adiabatic vacuum as our physically meaningful initial conditions. In addition, we have used the new non-perturbative approach to find the physical mode function corresponding to the physical vacuum of each era. This approach defines the vacuum state through the exact solutions of the EoM, and hence, the particle number can be defined without any ambiguity and approximations. \textbf{The results of our numerical calculations were classified into three different parts}. At \textbf{first}, for the case of the conformal coupling and the general case of $\Delta\tau=1$, we have examined the effect of the bare mass of the scalar field on the reheating temperature. As expected, the resultant reheating temperature was suppressed for higher values of mass, and there was a peak in the spectrum of reheating temperature around $\mu=1.5$. This is because, for these masses, the adiabaticity condition is not violated noticeably. In common words, this effect is sometimes interpreted as: the production of these values of mass requires extracting more energy from the background, and hence, the reheating temperature due to their production is suppressed.

In the \textbf{second} part of our calculations, we considered the effect of the presence of the gravitational coupling with non-conformal values of the coupling constant, $\xi$, with $\Delta\tau=1$ for the sake of generality. Our numerical calculations reveal the fact that by considering non-conformal values of the coupling, the resultant reheating temperature can gain high values for the greater masses, in a way that their reheating temperature becomes comparable with other masses. There is a stability condition that determines which values of non-conformal coupling are considered in our study for each value of mass. Using this stability condition, we derived the allowed parameter space in our study for which the adiabatic vacuum prescription can be used as our initial condition as well as provides us with the avoiding of tachyonic instability. 

The results of this section reveal that by considering the non-conformal values of $\xi$, generally speaking, the reheating temperature is increased as the coupling constant deviates from the conformal value. According to the stability relation, there are naturally ${\xi}_L$ and ${\xi}_R$ corresponding to the lower and higher allowed values of $\xi$. So, there must be ${\xi}_{min}$ and ${\xi}_{max}$ chosen very close to ${\xi}_L$ and ${\xi}_R$, so that the calculations avoid tachyonic instability and the adiabatic notion is relevant. In the case of $\Delta\tau=1$, the negative values of $\xi$ become allowed for $\mu>1.5$. For masses below $\mu=5.5$, it was the ${\xi}_{max}$ of each mass with positive value that resulted in the highest possible value of $T_{reh}$, while for masses greater than $\mu=5.5$, it was ${\xi}_{min}$ with negative value that produced the highest temperature. By considering the highest possible $T_{reh}$ in the parameter space of each mass, we see that the increase in the reheating temperature is more perceptible for the higher values of mass, in a way that, roughly speaking, for $\mu>1$, the resultant highest reheating temperature of all masses becomes of the order of $10^7 Gev$ and $10^{10}GeV$ for the first and second scenario of decay, respectively. So we conclude that although in the conformal case, the increase in the value of mass for large masses results in a decrease in the reheating temperature, the non-conformal coupling compensates this reduction in a way that, by considering its effect, the highest possible reheating temperature of each mass, regardless of the value of the mass, becomes of the same order. Of course, this compensation is not so that the highest reheating temperature in the parameter space becomes the temperature of $\mu=10$. In fact, in this step, the highest reheating temperature becomes $1.84\times10^7 GeV$ and $1.32\times 10^{10} GeV$ for the first and second scenarios, corresponding to $\mu=2.5$ with $\xi=1.21$. But as mentioned earlier, the highest temperature of $\mu=10$ was close to these values.

With this consideration in hand, in the last step of our numerical calculations, we considered the effect of a reduction in the transition time, $\Delta\tau$. As expected from physical intuition and mathematical justifications from the adiabaticity condition, any reduction in the transition time will lead to an increase in the GPP rate and its resultant reheating temperature. This is sometimes mentioned as a common and simple interpretation: if the transition occurs with a shorter transition period, or almost instantly, the background provides the field with a greater budget of energy, which in turn results in a more intense production of its particles. 

First of all, we did our calculations for the case of conformal coupling. The results reveal that, as expected, the reheating temperature increases with the reduction of $\Delta\tau$, and this increase is more perceptible for the higher masses. After that, we repeat our calculations for the case of the non-conformal couplings. Our results demonstrate that despite the above physical expectation, the pattern of the increase is different for the positive and negative values of $\xi$. For a positive value of $\xi$, the increase in the reheating temperature occurs permanently, while for negative values, the behavior of the spectrum of the reheating temperature is such that it decreases at first and then starts to increase after a specific value of the transition time. We also found that for each value of mass and its relevant $xi$ interval, the increase of the reheating temperature is always greater for the ${\xi}_{max}$ than the ${\xi}_{min}$ in a way that $T_{reh}$ corresponding to ${\xi}_{max}$ lies at a higher value than ${\xi}_{min}$. More importantly, it has also happened for $\mu>5.5$, where the ${\xi}_{min}$ corresponded to the highest temperature in the case of $\Delta\tau=1$. We also found that such an increase in the reheating temperature does not happen anymore if we continue to reduce $\Delta\tau$ to an amount smaller than a specific value, for instance, and roughly speaking, smaller than $\Delta\tau=0.001$. In fact, in the mass interval under our consideration, $0.25\le\mu\le10$, the spectrum of $T_{reh}$ does not change if we continue to reduce $\Delta\tau$ to infinitesimal values. So, as a result of the third part of our numerical calculations, the effect of the reduction in the transition time is more perceptible for the higher masses and is more intense for the ${\xi}_{max}$ of each mass. As a result, although the highest possible reheating temperature for the case of $\Delta\tau=1$ corresponded to $\mu=2.5$ and its ${\xi}_{max}=1.21$, by considering the effect of $\Delta\tau$, the highest possible reheating temperature is achieved at $\mu=10$ and its ${\xi}_{max}=16.83$, even though its highest temperature in the parameter space of this mass for $\Delta\tau=1$ corresponded to its ${\xi}_{min}$ with a negative value. 

Hence, \textbf{by reducing the transition time to $\Delta\tau=0.001$, the highest reheating temperature in the second scenario of reheating, becomes $4.34\times 10^{13} GeV$.} This temperature, in fact, is a quite sufficient reheating temperature that is enough for the universe to enter the hot Big Bang era. As a result, the non-perturbative approach to gravitational particle production suggests that this scenario is a relevant scenario of reheating that should be considered as a naturally possible channel of reheating, especially in NO models of inflation. In this study, we tried to survey a wide range of the parameter space of the gravitational channel of reheating. Such a survey consists of the free parameters of the model of the spectator scalar field, $\mu$ and $\xi$, as well as the characteristic feature of the inflationary scenario, $\Delta\tau$.

For the sake of completeness, we presented an independent model that is analytically solvable as a mathematical tool parallel to the numerical ones. Then, using the properties of the Bessel's functions and introducing another piecewise scale factor, we tried to find an analytical formula for the gravitational particle production rate. Such a formula provides us with a better understanding of the role of each free parameter of the scalar field model, $\xi$ and $\mu$, in the GPP rate and $T_{reh}$. This analysis can strengthen our physical intuition and help us to find a quick and rough estimation of $T_{reh}$. There are differences in the results of our numerical and analytical calculations that arise from the difference between the two scale factors. But in some special cases, the general behavior of the analytical plots was in accordance with the numerical ones, which demonstrates that these analytical calculations can be considered as not-too-bad approximations for the numerical results. But on the other hand, these differences indicate the necessity of the numerical calculations if one tries to find a better estimation of $T_{reh}$ in the gravitational particle production scenario in a more realistic model for the evolution of the background. 

Hence, we have studied the gravitational reheating in both ways of analytical and numerical, and tried to provide the reader with a complete picture of this channel of reheating. 

\section*{Acknowledgment}

The authors sincerely thank L. H. Ford and Edward Kolb for their inspiring scientific studies and publications on the issue of gravitational particle production. Amirsalar Nikandish considers himself deeply indebted to the pioneering experts in this field through their deep papers. The authors also want to thank Soichiro Hashiba and Jun'ichi Yokoyama for their wonderful paper, which was our primary reference in this work. He also wants to acknowledge Mohammad Kahfi, Ali Salehi, Dr. Reza Naderi, and Soheil Hassankhanpour, for their technical help and support.    


\appendix 

\section{ The Adiabatic Vacuum Based on the WKB Approximation}\label{wkbadi}
The adiabatic vacuum prescription provides us with a natural definition of the vacuum state in curved space-time. In general, this prescription holds when the effect of gravity becomes negligible. \textbf{Hence, at first look, it seems that this procedure of defining the physical vacuum is an approximate way which becomes precise, in backgrounds where the changes of the metric with time are slow.} Turning back to the EoM, Eq. \ref{main}, and rewriting it in terms of the effective time-dependent frequency, we have:
\begin{equation}\label{a1}
\frac{d^2 \Delta({\kappa})}{d\tau^2}+\omega^{2}(\tau)\,\Delta_{\kappa}(\tau)=0
\end{equation}
As discussed in the paper, the procedure of adiabatic vacuum is usually done in cases when the adiabaticity condition holds as:
\begin{equation}\label{a2}
\left|
\frac{\omega_{\kappa}(\tau+\Delta\tau)-\omega_{\kappa}(\tau)}
     {\omega_{\kappa}(\tau)}
\right|
\simeq
\left|\frac{\omega'_{\kappa}(\tau)}{\omega_{\kappa}(\tau)}\Delta\tau\right|
=
2\pi\,\left|\frac{\omega'_{\kappa}(\tau)}{\omega_{\kappa}^{2}(\tau)}\right|
\ll 1
\end{equation}
This condition, along with the essential condition that during the adiabatic regime, we must have ${{{\omega}_{\kappa}}^2}(\tau)>0$, provides the conditions of the adiabatic prescription of the physical vacuum. These conditions hold in two important cases: first, when the gravitational background is slowly changing with time, and second, when the mode under consideration is a high-frequency mode. In our study, the first situation is considered as the adiabatic limit for the inflationary era, and the second one is considered as the adiabatic limit in the kination epoch. In the first studies of the adiabatic vacuum prescription, the adiabatic vacuum was defined through the well-known WKB approximate solution as: 
\begin{equation}\label{a3}
\Delta_{\kappa}(\tau)=\frac{1}{\sqrt{\omega_{\kappa}(\tau)}}
\exp\!\left(
i\!\int_{\tau_\star}^{\tau}\!\omega_{\kappa}({\tau})\,d{\tau}
\right)
\end{equation}
This adiabatic solution can be written in the functional form of an asymptotic series. By finding the n-th order WKB solution, we can define the n-th order adiabatic vacuum state. But it is essential to note that \textbf{the WKB solution is in fact a divergent asymptotic series. Hence, the GPP rate obtained through this procedure is defined with a certain fundamentally limited accuracy. In fact, the WKB solution is just an approximation to the exact solution of the EoM for a certain cosmological epoch.} But in this appendix, we are going to describe in more detail the procedure that one has to follow to compute the GPP rate using the adiabatic vacuum notion. At the end, we will highlight how our procedure in this paper (non-perturbative approach) is built on the basis of the notion of the adiabatic vacuum, and on the other hand, what are the main differences between the WKB adiabatic vacuum and our chosen vacua.

To explain the WKB series, let us rewrite the EoM in the form usually written in mathematical references. If the EoM is written as:
\begin{equation}\label{a4}
\varepsilon^{2}\,\frac{d^2 \Delta_{\kappa}(\tau)}{d\tau^2}
+\omega_{\kappa}^{2}(\tau)\,\Delta_{\kappa}(\tau)=0
\end{equation}
Then, the adiabaticity condition becomes:
\begin{equation}\label{a5}
\varepsilon\,\left|\frac{d\omega_{\kappa}}{d\tau}\right|
\ll \omega_{\kappa}^{2}
\end{equation}
This condition guarantees that either the metric of the background is a slowly varying function of time or the mode function is a high-frequency mode with a short wavelength; the adiabatic regime occurs, and the adiabaticity condition is satisfied. Hence, the WKB solution can be written as:
\begin{equation}\label{a6}
\Delta_{\kappa}^{\mathrm{WKB}}(\tau)
=\frac{C}{\sqrt{\omega_{\kappa}(\tau)}}
\exp\!\left(\pm\frac{i}{\varepsilon}
\int^{\tau}\!\omega_{\kappa}(\tau')\,d\tau'\right)
\end{equation}
This approximate solution can be written in a generalized form as an n-th-order approximate solution:
\begin{equation}\label{a7}
\Delta_{\kappa}^{\mathrm{WKB}_n}(\tau)
=\frac{C}{\sqrt{W_{n}(\tau)}}
\exp\!\left(\pm\frac{i}{\varepsilon}
\int^{\tau}\!W_{n}(\tau')\,d\tau'\right)
\end{equation}
Here the function $W_n(\tau;\epsilon)$ is an approximate solution to the following equation:
\begin{equation}\label{a8}
\frac{1}{2}\frac{W''}{W}-\frac{3}{4}\frac{W'\,^2}{W\,^2}
=\frac{\omega_{\kappa}^{2}-W^{2}}{\varepsilon^{2}}
\end{equation}
This yields a power series solution in the parameter ${\epsilon}^2$. Using this approximate solution, we will try to compute the GPP rate for a mode function whose EoM is of the form of Eq. \ref{a4}. As we have explained throughout our paper many times, the choice of a specific mode function for a cosmological era is totally equivalent to the choice of a vacuum state. If one wants to choose the physical mode function, using the WKB solution in Eq. \ref{a7}, one should require that at an initial time, the mode function and its first time derivative match the WKB approximate solution as:
\begin{equation}\label{a9}
\frac{\Delta'(\tau)}{\Delta(\tau)}
=\frac{\Delta_{\mathrm{WKB}}^{'}(\tau)}
       {\Delta_{\mathrm{WKB}}(\tau)}
\end{equation}
By these requirements, the initial conditions of the mode function of the first epoch are obtained using the WKB solution. We denote this mode function as ${{\Delta}_{\kappa}}^i$. At a final time, in the second cosmological epoch, the mode function must satisfy the above conditions as well. \textbf{In many papers, the final mode function is considered to be the WKB solution at the final time.} We denote the final mode function as ${{\Delta}_{\kappa}}^f$. So, finally, we can write the Bogoliubov transformation between these two mode functions as:
\begin{equation}\label{a10}
\Delta_{\kappa}^{\mathrm{i}}(\tau)
=\alpha\,\Delta_{\kappa}^{\mathrm{f}}(\tau)
+\beta\,\Delta_{\kappa}^{\mathrm{f}\,*}(\tau)
\end{equation}
Hence, the GPP rate can be computed using the time-independent Bogoliubov coefficient $\beta$, as:
\begin{equation}\label{a11}
\beta
=\frac{\Delta_{\kappa}^{\mathrm{i}'}(\tau)\,\Delta_{\kappa}^{\mathrm{f}}(\tau)
-\Delta_{\kappa}^{\mathrm{i}}(\tau)\,\Delta_{\kappa}^{\mathrm{f}'}(\tau)}{i}
\end{equation}
In other studies that used the notion of the adiabatic vacuum, the values of the final mode function can be evaluated at an arbitrary time in the second cosmological epoch, $\tau$, just by using Eq. \ref{a9}, and substituting $\tau$ instead of ${\tau}_0$ (\textbf{Of course, it is one of the most important differences between the WKB method and the non-perturbative method that we have used in this paper.}). Hence, it is just the values of ${{\Delta}_{\kappa}}^i({\tau}_0)$ and ${{{{\Delta}_{\kappa}}^i}^{\prime}}({\tau}_0)$, that remained to calculate. \textbf{It is the essential key point in evaluating the GPP rate using the Bogoliubov transformation, which we have also used in our numerical calculation: To compute the value of these functions at time $\tau$, we have to solve the EoM from the initial time to the final time, using the initial conditions obtained from Eq. \ref{a9}. Then, with this solution in hand, and using the value of the final mode function that was obtained just by substituting the value of the final time in the WKB condition, Eq. \ref{a9}, we must calculate the GPP rate using Eq. \ref{a11}. Although the WKB approximation provides us with the appropriate initial conditions to select an initial physical mode function for a cosmological era, it can not be used to compute the values of the ${{\Delta}_{\kappa}}^i({\tau}_0)$ and ${{{{\Delta}_{\kappa}}^i}^{\prime}}({\tau}_0)$ in the final time. This is due to the fact that the WKB solution satisfies the condition in Eq. \ref{a9} at any time and hence can not provide us with the GPP rate. The mode function of the first era must be computed either using the numerical or analytical method from the initial time to the final time.} 

It is essential to mention that we can use the WKB approximation in different orders to select a mode function as the physical one, either at the initial or the final time. The difference between these mode functions is in their different approximation order, and hence, there is only an approximate agreement between them. So, \textbf{at first, there is a fundamental approximation in defining the adiabatic vacuum or mode function using the WKB approximation. As we have mentioned, in many studies, although the WKB approximation at the initial time is considered only as the initial conditions to obtain the mode function of the first era numerically from the initial time to the final time, the final mode function is simply considered to be the WKB solution at the final time. As we have seen, this solution is just an approximation of the exact mode function, and also, there is a fundamental limit in its accuracy. In addition, the WKB solutions with a different order are in agreement only in an approximate manner. But in the non-perturbative approach, we have found an exact mode function for the kination era, and hence, these challenges of the WKB accuracy are not mentioned in our approach.}

On the other hand, if one uses the WKB solution to compute the ${{\Delta}_{\kappa}}^i({\tau}_0)$ at the final time, the resultant GPP rate is not obtained correctly. \textbf{In the non-perturbative approach, we have solved the EoM from the initial time to the final time in the kination era, using the numerical method with sufficient accuracy.}

Hence, although the WKB solution is very useful to define the initial conditions for the mode function of the first era using the notion of the adiabatic vacuum, its accuracy is, in fact, insufficient to be used as the final mode function and also for the calculation of the GPP rate.

As a result, \textbf{although we have used the notion of the adiabatic vacuum to impose physically meaningful initial conditions on our solutions, in the non-perturbative approach, the final mode function, KV in Sec. \ref{theo}, was calculated in an exact manner, and we have not simply used the WKB solution as the final vacuum. In addition, to calculate the initial mode function at the final time, we have solved the EoM analytically, and with appropriate accuracy, by invoking numerical calculations. These essential differences can provide this path of study with a more accurate approach for calculations of the GPP rate and the resultant reheating temperature.} 

\section{Derivation of the EoM}\label{EoMder}
In this appendix, we are going to explain the main key steps that we have taken to derive the EoM of the spectator field, $\phi$, as well as the auxiliary field, $\chi$. To this end, starting with the action of the spectator field, Eq. \ref{action}, and varying the Lagrangian density of the free $\phi$ as described in Eq. \ref{freephi} with respect to $\phi$, we obtain:
\begin{equation}
\delta\mathcal{L}^{\text{free}}_{\phi}
= \sqrt{-g}\left(
-\tfrac{1}{2} g^{\mu\nu}\,\right)\partial_{\mu}\!\left(\delta\phi\right)\,\partial_{\nu}\phi
+\sqrt{-g}\left(
-\tfrac{1}{2} g^{\mu\nu}\right)\partial_{\mu}\phi\,\partial_{\nu}\left(\delta\phi\right)
-\xi R\sqrt{-g}\,\phi\,\delta\phi
- M_{\phi}^{2}\,\phi\,\delta\phi\sqrt{-g}
\label{var1}
\end{equation}
Now, using the technique of integration by parts, the above equation can be written in the following form:
\begin{align}
\delta\mathcal{L}^{\text{free}}_{\phi}
= \partial_{\mu}\!\left(\sqrt{-g}\left(-\tfrac{1}{2}g^{\mu\nu}\right)\partial_{\nu}\phi\,\delta\phi\right)\,\\
- \partial_{\mu}\!\left(\sqrt{-g}\left(-\tfrac{1}{2}g^{\mu\nu}\right)\partial_{\nu}\phi\right)\delta\phi\,\\
+ \partial_{\nu}\!\left(\sqrt{-g}\left(-\tfrac{1}{2}g^{\mu\nu}\right)\partial_{\mu}\phi\,\delta\phi\right)\,\\
- \partial_{\nu}\!\left(\sqrt{-g}\left(-\tfrac{1}{2}g^{\mu\nu}\right)\partial_{\mu}\phi\right)\delta\phi\,\\
- \xi R\sqrt{-g}\,\phi\,\left(\delta\phi\right)
- \sqrt{-g}\,M_{\phi}^{2}\,\phi\,\delta\phi
\label{var2}
\end{align}
In the above equation, the first and third terms can be eliminated. This is due to the fact that these terms are total derivatives and are equal to zero on the boundary. In the fourth term, we replace the indices $\mu$ and $\nu$ with each other, using the fact that these indices are dummy indices and the metric is a symmetric tensor. The EoM is obtained is one demands $\delta\mathcal{L}=0$ as:
\begin{equation}
\square \phi - \left(\xi R + M_\phi^{2}\right)\phi = 0
\label{var3}
\end{equation}
where we have used the definition of the physical D’Alembertian in curved space-time as introduced in Sec. \ref{theo}. By invoking the FLRW metric in Eq. \ref{metric}, using the fact that the auxiliary field $\chi$ is related to $\phi$ through $\phi=\frac{\chi}{a}$, and performing the derivatives in the D’Alembertian operator, the main EoM, Eq. \ref{main}, is obtained. It is worth noting that the determinant of the metric in terms of the cosmic time must be considered to be $\sqrt{-g}=a^3$. Also, the following relations are used in this derivation:

First of all, the derivatives of the spectator field with respect to the cosmic and conformal times are related to each other using the equations:
\begin{align}
\dot\phi &= \frac{1}{a}\,\phi',\\
\ddot\phi &= \frac{1}{a^{2}}\,\phi''-\frac{a'}{a^{3}}\,\phi'.
\end{align}
Where the dot is for the time derivative with respect to the cosmic time and the prime is for the derivative with respect to the conformal time. Secondly, the first time derivative of the spectator field, $\phi$, is related to the time derivative of the auxiliary field, $\chi$, as:
\begin{equation}
\phi'=\frac{d}{d\eta}\!\left(\frac{\chi}{a}\right)
=\,\frac{-a'}{a^{2}}\,\chi+\frac{1}{a}\,\chi'.
\end{equation}
Such an equation can also be found for the second derivative, which is useful for the derivation of the main EoM:
\begin{align}
\phi''&=\frac{d}{d\eta}\!\left(-\frac{a'}{a^{2}}\chi+\frac{1}{a}\chi'\right) \nonumber\\
&=\frac{-a''a^{2}+2aa'^2}{a^{4}}\,\chi\,-\frac{a'}{a^{2}}\,\chi'-\frac{a'}{a^{2}}\,\chi'+\frac{1}{a}\,\chi''.
\end{align}
At the end, it is also important to note that the curvature scalar can be written in terms of the conformal time, as:
\begin{equation}
R(\eta)=6\,\frac{a''}{a^{3}}.
\end{equation}
With this consideration and the above equations, Eq. \ref{main} is obtained in a straightforward manner.    

\section{Some Relevant Mathematical Properties of the Bessel Functions }\label{bes}
In this appendix, we are going to list some of the mathematical properties of the Bessel functions. These properties are going to be used in the derivation path of the mode function discussed in this paper, for instance, KV, BDV, and AKV. Due to their importance, in some of the mathematical textbooks, they have been named as Bessel function mathematical identities. 

First of all, it is very important to note that the Hankel functions of the first and second kinds can be written in terms of the $J$ and $Y$ Bessel functions as:
\begin{align}
H_{\alpha}^{(1)}(z) &= J_{\alpha}(z) + i\,Y_{\alpha}(z)\\
H_{\alpha}^{(2)}(z) &= J_{\alpha}(z) - i\,Y_{\alpha}(z)
\label{c1}
\end{align}
According to these equations, the inverse of these transformations can be written and is also very important for our calculations. This means that the $J$ and $Y$ functions can be written in terms of the Hankel functions as:
\begin{align}
J_{\alpha}(z) &= \tfrac{1}{2}\!\left(H_{\alpha}^{(1)}(z)+H_{\alpha}^{(2)}(z)\right)\\
Y_{\alpha}(z) &= \tfrac{1}{2i}\!\left(H_{\alpha}^{(1)}(z)-H_{\alpha}^{(2)}(z)\right)
\label{c2}
\end{align}
In our procedure regarding the issue of GPP, we have used the notion of adiabatic vacuum. This notion was used for both inflationary and kination eras. For the inflationary era, the notion of the adiabatic vacuum was used to set the relevant initial conditions in $\tau\rightarrow-\infty$. In this limit, all modes lie well inside the cosmological horizon and hence must behave like the modes in flat spacetime. On the other hand, due to the fact that the conformal time can not tend to infinity for the kination era, the notion of adiabatic vacuum was invoked for the high-frequency modes, where the quantity $\frac{{\kappa}^2}{a^2{\mu}^2}$ was the quantity that tends to infinity. In both cases, the arguments of Bessel functions must tend to infinity to impose the initial conditions. \textbf{Due to these reasons, it is essential to write the general solutions for both eras in terms of the Hankel functions instead of Bessel functions. Since it is a well-known feature of Bessel functions that if one wants to find the limiting behavior of a solution that contains Bessel functions at infinity, this limit is only meaningful if one applies Hankel functions, and not Bessel functions.} It is exactly the case in our study, and hence, the expansion of the Bessel function in terms of Hankel functions in Eq. \ref{c2} is of particular interest. 

According to the above discussion, it is useful to write the limiting behavior of the Hankel functions for large arguments, which is:  
\begin{align}
H_{z}^{(1)} &\sim \sqrt{\frac{2}{\pi z}}\;
e^{i\,\!\Big(z-\tfrac{1}{2}\nu\pi-\tfrac{1}{4}\pi\Big)}\\
H_{z}^{(2)} &\sim \sqrt{\frac{2}{\pi z}}\;
e^{-i\,\!\Big(z-\tfrac{1}{2}\nu\pi-\tfrac{1}{4}\pi\Big)}
\label{c3}
\end{align}

Based on the discussion above regarding the limiting procedure for using the notion of the adiabatic vacuum, it is also useful to express the Airy function in terms of the Hankel functions. If $\zeta = \frac{2}{3}z^{\frac{3}{2}}$, then the transformation equations between Airy and Hankel functions are as follows:
\begin{align}
\mathrm{Airy}(z)
&= \frac{1}{2}\sqrt{\frac{z}{3}}\; e^{\frac{2\pi i}{3}}\,H^{(1)}_{1/3}(\zeta) \\
&= \frac{1}{2}\sqrt{\frac{z}{3}}\; e^{-\frac{2\pi i}{3}}\,H^{(2)}_{1/3}(-\zeta)
\label{c4}
\end{align}
and also:
\begin{align}
\mathrm{Ai}(-z)
&= \frac{1}{2}\sqrt{\frac{z}{3}}
\left( e^{+\frac{\pi i}{6}} H^{(1)}_{1/3}(\zeta)
     + e^{-\frac{i \pi}{6}} H^{(2)}_{1/3}(\zeta) \right), \\
\mathrm{Bi}(-z)
&= \frac{1}{2}\sqrt{\frac{z}{3}}
\left( e^{\frac{2\pi i}{3}} H^{(1)}_{1/3}(\zeta)
     + e^{-\frac{2\pi i}{3}} H^{(2)}_{1/3}(\zeta) \right)
\label{c5}
\end{align}
During the course of our calculations, as discussed in Sec. \ref{theo} and also seen in Sections \ref{theo} and \ref{anaresults}, we need to compute the Wronskian between the mode functions of the kination and inflation eras. To this end, we need to calculate the derivative of the Hankel functions, too. So, it is worth mentioning that if the function $l$ is supposed to be one of the functions of the Bessel functions family, which means:
\begin{equation}
\ell_\nu = J_\nu,\; Y_\nu,\; H^{(1)}_\nu,\; H^{(2)}_\nu
\label{c6}
\end{equation}
Then we have:
\begin{equation}
\ell_\nu'(z) = \frac{1}{2}\Big(\ell_{\nu-1}(z) - \ell_{\nu+1}(z)\Big)
\label{c7}
\end{equation}
Using these properties and equations for the Bessel functions, we will describe the derivation path of the GPP rate in the following appendices.

\section{Derivation of the KV for the Numerical Calculations}\label{kvder}

In this appendix, we are going to explain the essential key steps in the derivation of the KV, needed for our numerical calculations. The final result of this calculation is written in Eq. \ref{kv}. We start from the EoM for the kination era, Eq. \ref{keom}, which is obtained by substituting the limiting behavior of the numerical scale factor function in the EoM of the auxiliary field, Eq. \ref{main}. This equation is in the form of the well-known Bessel equation and hence yields an analytical general solution as:
\begin{equation}
\Delta_{\kappa}(\tau)
= C_{1}\,\mathrm{Airy Ai}\!\left(\,\frac{- \kappa^{2}-\mu^{2}-\mu^{2}\tau}{(\mu^{2})^{2/3}}\right)
+ C_{2}\,\mathrm{Airy Bi}\!\left(\,\frac{-\kappa^{2}-\mu^{2}-\mu^{2}\tau}{(\mu^{2})^{2/3}}\right)
\label{d1}
\end{equation}
Using the transformation between the Airy functions and the Hankel functions, Eq. \ref{c4} and Eq. \ref{c5}, and also by defining a new function, $y$ as:
\begin{equation}
y \equiv \frac{\kappa^{2}+\mu^{2}+\mu^{2}\tau}{(\mu^{2})^{2/3}}
\label{d2}
\end{equation}
The general solution of the EoM can be rewritten as:
\begin{align}
\Delta_{\kappa}(\tau)
&= C_{1}\,\mathrm{Ai}(-y) + C_{2}\,\mathrm{Bi}(-y) \\
&= \sqrt{y}\left(\frac{C_{1}e^{i\pi/6}+C_{2}e^{2\pi i/3}}{2\sqrt{3}}\right)
   H^{(1)}_{1/3}\!\left(\tfrac{2}{3}\,y^{3/2}\right)
 + \sqrt{y}\left(\frac{C_{1}e^{-i\pi/6}+C_{2}e^{-2\pi i/3}}{2\sqrt{3}}\right)
   H^{(2)}_{1/3}\!\left(\tfrac{2}{3}\,y^{3/2}\right)
\label{d3}
\end{align}
As is obvious from the new form of the general solution, the coefficients in the parentheses are constant coefficients, and hence, for the sake of simplicity, we define two new constant coefficients as:
\begin{equation}
Q_{1}\equiv \frac{C_{1}e^{i\pi/6}+C_{2}e^{2\pi i/3}}{2\sqrt{3}},
\qquad
Q_{2}\equiv \frac{C_{1}e^{-i\pi/6}+C_{2}e^{-2\pi i/3}}{2\sqrt{3}}
\label{d4}
\end{equation}
It is obvious that the issue of finding the constants $C_1$ and $C_2$ by imposing the adiabatic initial conditions on the solution is, in fact, equivalent to finding the constant coefficients $Q_1$ and $Q_2$. Hence, the general solution can be written in the following new form:
\begin{equation}
\Delta_{\kappa}(\tau)
= Q_{1}\,\sqrt{y}\;
H^{(1)}_{1/3}\!\left(\tfrac{2}{3}\,y^{3/2}\right)
+ Q_{2}\,\sqrt{y}\;
H^{(2)}_{1/3}\!\left(\tfrac{2}{3}\,y^{3/2}\right)
\label{d5}
\end{equation}
To determine the constant coefficients, we have to obtain the adiabatic limit of this solution. As discussed earlier, the adiabatic limit in the kination era is obtained from the high-frequency modes, which means for ${\kappa}^2 \gg a^2{\mu}^2$. This limit corresponds to the notion of vacuum at short distances, or equivalently, modes deep inside the horizon for which we have $\kappa\tau\gg1$. Starting with the function $y$, we have:
\begin{equation}
y=\frac{\kappa^{2}+\mu^{2}+\mu^{2}\tau}{(\mu^{2})^{2/3}}
=\frac{\kappa^{2}}{(\mu^{2})^{2/3}}
\left(1+\frac{\mu^{2}}{\kappa^{2}}+\frac{\mu^{2}}{\kappa^{2}}\tau\right)
\label{d6}
\end{equation}
Hence the $\sqrt{y}$ becomes:
\begin{equation}
\sqrt{y}\;\sim\;
\frac{\kappa}{(\mu^{2})^{1/3}}
+\mathcal{O}\!\left(\frac{\mu^{4/3}}{\kappa}\right)
\label{d7}
\end{equation}
Also, the quantity $\frac{2}{3}y^{\frac{2}{3}}$, needed for obtaining the adiabatic limit of the solution, can be written as:
\begin{equation}
\frac{2}{3}\,y^{3/2}\;\sim\;\frac{2}{3}\frac{\kappa^{3}}{\mu^{2}}+\kappa+\kappa\tau
\label{d8}
\end{equation}
In both of the above limiting forms, the first terms are the dominant terms. In addition, in the limiting form of $\frac{2}{3}y^{\frac{2}{3}}$, there is a time-dependent term, $\kappa\tau$, which demonstrates the time-dependent behavior of the mode function in its limiting form. We have to keep this term also, but the other terms can be eliminated compared with the dominant terms. By substituting these dominant terms and using the asymptotic form of the Hankel functions for the large argument, Eq. \ref{c3}, we obtain:
\begin{equation}
\Delta_{\kappa}(\tau)\xrightarrow[\kappa\to\infty]{}\;
Q_{1}\frac{\kappa}{(\mu^{2})^{1/3}}
\sqrt{\frac{2}{\pi \,\tfrac{2}{3}\,y^{3/2}}}\,
\,e^{\,i\frac{2}{3}y^{3/2}}\,e^{-i\frac{5\pi}{12}}
+Q_{2}\frac{\kappa}{(\mu^{2})^{1/3}}
\sqrt{\frac{2}{\pi \,\tfrac{2}{3}\,y^{3/2}}}\,
\,\,e^{-i\frac{2}{3}y^{3/2}}\,e^{\frac{i5\pi}{12}}
\label{d9}
\end{equation}
As we have seen in Eq. \ref{d8}, the time-dependency of $\frac{2}{3}y^{\frac{2}{3}}$ is of the form of $\kappa\tau$. On the other hand, the adiabatic condition forces the behavior of the mode function in the adiabatic limit to be like the behavior of the mode function in Minkowski space-time, $\frac{1}{\sqrt{2\kappa}}e^{-i\kappa\tau}$. As a result, the coefficient $Q_1$ must be set equal to zero, and there remains only $Q_2$ to determine. This can be done using the limiting behavior of $y$ in Eq. \ref{d8}, and we obtain:  
\begin{equation}
Q_{2}
=\sqrt{\frac{\pi}{6}}\;\mu^{-1/3}\;
e^{i\,\left\{\frac{2}{3}\frac{\kappa^{3}}{\mu^{2}}+\kappa-\frac{5\pi}{12}\right\}}
\label{d10}
\end{equation}
This provides us with the special form of the physical mode function or the physical vacuum of the kination era in Eq. \ref{kv}.

\section{Derivation of the $T_{reh}$ for the Two Reheating Scenarios in Our Study}\label{trehder}

In this appendix, we are going to provide our readers with the key steps in the derivation path of the reheating temperature resulting from the GPP. As discussed in Sec. \ref{theo}, there are two possible reheating scenarios in our case, and we are going to present the derivation path for both scenarios. 

In the first scenario, the gravitationally produced superheavy scalar particles decay and then thermalize instantly after their production. As a result, the energy density of the produced thermal bath, ${\rho}_r$, should be considered equal to the energy density of the scalar field, ${\rho}_{\phi}$, just after the transition time. We have:
\begin{equation}
\rho_{\mathrm{rad}}^{\,I}=\rho_{\phi}^{\,I}
\label{e1}
\end{equation}
Since the scalar particle decays and thermalizes instantly, its energy density, which is equal to the energy density of radiation, will decay as $a^{-4}$ with the expansion of the universe. Hence:
\begin{equation}
\rho_{\rm rad}(t)=\rho_{\rm rad}^{I}\,a^{-4}
=\rho_{\phi}^{I}\,a^{-4}
\label{e2}
\end{equation}
In the above equations, the ${{\rho}_{rad}}^I$ and the ${{\rho}_{\phi}}^I$ denote the energy density of created radiation and scalar particles just after the inflation-kination transition, respectively. After the transition time, the dominant energy component of the universe is still the energy density of the inflaton. Since our inflationary model is considered to be a NO model, after kination, this energy density is due to the energy density stored in the kinetic part of the inflaton. At a certain moment of time, or equivalently, a specific value of the scale factor, the energy density of the radiation becomes equal to that of the inflaton, and after that, the radiation becomes the dominant component of the universe. We denote this moment of time as the reheating time or the time of equality, and after this time, the universe enters the radiation domination (RD) era. The temperature of the thermal bath at this moment is considered to be the reheating temperature of the first scenario. To obtain the scale factor at the equality time, we note that:
\begin{equation}
H_{\rm inf}^{2}
=\frac{1}{3\,M_{\rm Pl}^{*2}}\;\rho_{\rm inf}^{I}
\label{e3}
\end{equation}
This is because the Hubble parameter is considered to be almost constant during inflation. Just after the transition time, we use the approximation that the value of the Hubble parameter does not change significantly, and hence, the above equation is valid. Since during the kination era, the energy density of the inflaton scales with $a^{-6}$, we have:   
\begin{equation}
\rho_{\rm inf}(t)=\rho_{\rm inf}^{I}\,a^{-6}
=3\,M_{\rm Pl}^{*2}\,H_{\rm inf}^{2}\,a^{-6}
\label{e4}
\end{equation}
Where the last equality is written by substituting Eq. \ref{e3}. At the time of equality, we have ${\rho}_{rad}={\rho}_{Inf}$. Hence:
\begin{equation}
\rho_{\rm rad}^{I}\,a^{-4}=\rho_{\rm inf}^{I}\,a^{-6}
\label{e5}
\end{equation}
And as a consequence:
\begin{equation}
3\,M_{\rm Pl}^{*2}\,H_{\rm inf}^{2}\,a^{-6}
=\rho_{\phi}^{I}\,a^{-4}
\label{e6}
\end{equation}
So the scale factor at the time of equality is obtained by solving the above equation as:
\begin{equation}
a_{\rm equ}^{2}
=\frac{3\,M_{\rm Pl}^{*2}\,H_{\rm inf}^{2}}{\rho_{\phi}^{I}}
\label{e7}
\end{equation}
Using this scale factor and also Eq. \ref{e1}, the energy density of the radiation at the time of equality (reheating time) is obtained as:
\begin{equation}
\rho_{\rm rad}^{\rm equ}
=\rho_{\phi}^{I}\,a_{\rm equ}^{-4}
=\frac{(\rho_{\phi}^{I})^{3}}{9\,M_{\rm Pl}^{*4}\,H_{\rm inf}^{4}}
\label{e8}
\end{equation}
As we have mentioned earlier, the produced thermal bath is supposed to be thermalized instantly, and hence, the relation between its energy density and its temperature is expressed through the Black-Body equation:
\begin{equation}
\rho_{\rm rad}^{\rm BB}
=\frac{\pi^{2}}{30}\,g_{*}\,T^{4}
\label{e9}
\end{equation}
Substituting the energy density of the radiation at the reheating time from Eq. \ref{e8} in the Black-Body equation, the reheating temperature of the first reheating scenario is obtained as:
\begin{equation}
T_{\rm reh}^{(1)}
=\left(\frac{10}{3\pi^{2}g_{*}}\right)^{\!1/4}
\frac{1}{M_{\rm Pl}^{*} H_{\rm inf}}
\left(\rho_{\phi}^{I}\right)^{3/4}
\label{e10}
\end{equation}
Using this formula, the reheating temperature in the first scenario can be obtained by substituting the resultant energy density of the gravitationally produced particles from both analytical and numerical calculations. To obtain the reheating temperature for the second scenario, we note that in that case, the scalar particles decay at the time of equality. In fact, they decay just at the time that their energy density dominates the energy density of the inflaton. Due to the discussion presented in Sec. \ref{theo}, the gravitationally produced particles scale like the pressureless matter with $\omega = 0$. Hence, from the transition time till the time of equality, their energy density is scaled as $a^{-3}$. Comparing this situation with the previous case, we conclude that in this scenario, the created particles experience less dilution due to the expansion of the universe and hence, we expect that the resultant reheating temperature should be higher than the first scenario. In this case, we have:
\begin{equation}
\rho_{\phi}(t)=\rho_{\phi}^{I}\,a^{-3}
\label{e11}
\end{equation}
This is while the energy density of the kinetic part of the inflaton scales as:
\begin{equation}
\rho_{\rm Inf}(t)=\rho_{\rm Inf}^{I}\,a^{-6}
=3\,M_{\rm Pl}^{*2}\,H_{\rm inf}^{2}\,a^{-6}
\label{e12}
\end{equation}
As a result, at the time of equality, we should have:
\begin{equation}
\rho_{\phi}^{I}\,a^{-3}
=3\,M_{\rm Pl}^{*2}\,H_{\rm inf}^{2}\,a^{-6}
\label{e13}
\end{equation}
By solving the above equation, the scale factor of the equality time (reheating time) is obtained as:
\begin{equation}
a_{\rm equ}^{-3}
=\frac{\rho_{\phi}^{I}}{3\,M_{\rm Pl}^{*2}\,H_{\rm inf}^{2}}
\label{e14}
\end{equation}
Hence, the energy density of the gravitationally produced particles is obtained as written in the following equation. This energy density is transferred to the thermal bath due to the decay of the scalar particles to the SM fields. Hence, at the time of equality, the energy density of the scalar particles should be considered equal to the energy density of the radiation:
\begin{equation}
\rho_{\phi}^{\rm equ}
=\rho_{\phi}^{I}\,a_{\rm equ}^{-3}
=\frac{(\rho_{\phi}^{I})^{2}}{3\,M_{\rm Pl}^{*2}\,H_{\rm inf}^{2}}
=\rho_{\rm rad}
\label{e15}
\end{equation}
Just like the first scenario, assuming that the thermalization occurs instantly and invoking the Black-Body relation, we have:
\begin{equation}
\frac{\pi^{2}}{30}\,g_{*}\,T_{\rm reh}^{4}
=\frac{\left(\rho_{\phi}^{I}\right)^{2}}{3\,M_{\rm Pl}^{*2}\,H_{\rm inf}^{2}}
\label{e16}
\end{equation}
And hence, the reheating temperature in the second scenario is obtained as:
\begin{equation}
T_{\rm reh}^{(2)}
=\left(\frac{10}{\pi^{2}g_{*}}\right)^{\!1/4}
\frac{1}{\left(M_{\rm Pl}^{*}H_{\rm inf}\right)^{1/2}}
\left(\rho_{\phi}^{I}\right)^{1/2}
\label{e17}
\end{equation}
which is obviously greater than the ${T_{reh}}^{(1)}$. 

\section{Tables of the Numerical Calculations}\label{tab}

In this appendix, we provide a table that consists of the details of the numerical study of the effect of the non-conformal values of the gravitational coupling. For each mass, the values of ${\xi}_{L}$ and ${\xi}_{R}$ are obtained using the stability equation, Eq. \ref{p5}. To avoid tachyonic instability, we survey the parameter space from ${\xi}_{min}$ to ${\xi}_{max}$ that are chosen so close to ${\xi}_{L}$ and ${\xi}_{R}$ that we do not lose any part of the parameter space. ${\xi}_{best}$ corresponds to the non-conformal coupling in the parameter space that can result in the highest possible reheating temperature for each value of mass. The highest reheating temperature of both scenarios for each mass and its ${\xi}_{best}$ is also obtained. All of these parameters and their values are tabulated in Table. \ref{table} for better understanding. The values in this table are used to plot the important Fig. \ref{fig6}.    

\begin{table}[t!]
  \caption{Non-conformal coupling intervals and the highest reheating temperatures}
  \label{table}
  
  \begin{center}\scriptsize{
  \begin{tabular}{|c||c|c|c|c|c|c|c|}
    \hline
    $\mu$ & $\xi_{L}$ & $\xi_{R}$ & $\xi_{\min}$ & $\xi_{\max}$ & $\xi_{\text{best}}$ &
    $T_{\mathrm{reh}}^{(1)\,\text{best}}~(\mathrm{GeV})$ &
    $T_{\mathrm{reh}}^{(2)\,\text{best}}~(\mathrm{GeV})$ \\ \hline

    0.25 &  0.16140 &  0.17710 &  0.16160 &  0.17660 &  0.16160 & $1.42\times 10^{5}$  & $5.18\times 10^{8}$  \\ \hline
    0.50 &  0.14580 &  0.20830 &  0.14700 &  0.20700 &  0.20700 & $6.89\times 10^{5}$  & $1.48\times 10^{9}$  \\ \hline
    0.75 &  0.11980 &  0.26040 &  0.12080 &  0.25880 &  0.25880 & $2.28\times 10^{6}$  & $3.29\times 10^{9}$  \\ \hline
    1.00 &  0.08333 &  0.33333 &  0.08830 &  0.32830 &  0.32830 & $4.92\times 10^{6}$  & $5.50\times 10^{9}$  \\ \hline
    1.50 & -0.02100 &  0.54170 & -0.02000 &  0.54100 &  0.54100 & $1.26\times 10^{7}$  & $1.03\times 10^{10}$ \\ \hline
    2.00 & -0.16670 &  0.83330 & -0.16620 &  0.83280 &  0.83280 & $1.73\times 10^{7}$  & $1.27\times 10^{10}$ \\ \hline
    2.50 & -0.35420 &  1.20830 & -0.35280 &  1.20720 &  1.20720 & $1.84\times 10^{7}$  & $1.32\times 10^{10}$ \\ \hline
    3.00 & -0.58330 &  1.66670 & -0.58000 &  1.66400 &  1.66400 & $1.76\times 10^{7}$  & $1.29\times 10^{10}$ \\ \hline
    3.50 & -0.85420 &  2.20830 & -0.85330 &  2.20670 &  2.20670 & $1.62\times 10^{7}$  & $1.22\times 10^{10}$ \\ \hline
    4.00 & -1.16670 &  2.83330 & -1.16470 &  2.83130 &  2.83130 & $1.47\times 10^{7}$  & $1.14\times 10^{10}$ \\ \hline
    4.50 & -1.52080 &  3.54170 & -1.51870 &  3.53930 &  3.53930 & $1.34\times 10^{7}$  & $1.07\times 10^{10}$ \\ \hline
    5.00 & -1.91670 &  4.33300 & -1.91500 &  4.33100 &  4.33100 & $1.18\times 10^{7}$  & $9.87\times 10^{9}$  \\ \hline
    5.50 & -2.35420 &  5.20830 & -2.35330 &  5.20670 &  5.20670 & $1.72\times 10^{7}$  & $1.27\times 10^{10}$ \\ \hline
    6.00 & -2.83330 &  6.16670 & -2.83030 &  6.16370 & -2.83330 & $1.06\times 10^{7}$  & $9.20\times 10^{9}$  \\ \hline
    6.50 & -3.35420 &  7.20830 & -3.35330 &  7.20670 & -3.35330 & $1.05\times 10^{7}$  & $9.13\times 10^{9}$  \\ \hline
    7.00 & -3.91670 &  8.33330 & -3.91170 &  8.32830 & -3.91170 & $8.91\times 10^{6}$  & $8.17\times 10^{9}$  \\ \hline
    7.50 & -4.52080 &  9.54170 & -4.50670 &  9.53330 & -4.50670 & $7.27\times 10^{6}$  & $7.13\times 10^{9}$  \\ \hline
    8.00 & -5.16670 & 10.83330 & -5.16170 & 10.82830 & -5.16170 & $6.71\times 10^{6}$  & $6.76\times 10^{9}$  \\ \hline
    8.50 & -5.85420 & 12.20830 & -5.85330 & 12.20670 & -5.85330 & $6.43\times 10^{6}$  & $6.57\times 10^{9}$  \\ \hline
    9.00 & -6.58330 & 13.66670 & -6.56830 & 13.65170 & -6.56830 & $6.05\times 10^{6}$  & $6.31\times 10^{9}$  \\ \hline
    9.50 & -7.35420 & 15.20830 & -7.35330 & 15.20670 & -7.35330 & $6.08\times 10^{6}$  & $6.33\times 10^{9}$  \\ \hline
   10.00 & -8.16670 & 16.83330 & -8.16170 & 16.82830 & -8.16170 & $6.10\times 10^{6}$  & $6.34\times 10^{9}$  \\ \hline
  \end{tabular}}
  \end{center}
\end{table}

\section{Derivation of the Scale Factor for the Analytical Calculation}\label{anscder}

In this appendix, we are going to provide the reader with the steps that we have taken to choose the piecewise scale factor needed for our analytical calculations, as in Eq. \ref{scalean}. To this end, we note that it is a well-known fact that the cosmic scale factor depends on cosmic time as:
\begin{equation}
a(t)\;\propto\; t^{\frac{2}{3(1+W)}}
\label{g1}
\end{equation}
As a result, for the kination era with $W=1$, we have $a \propto t^{\frac{1}{3}}$. According to the definition of the conformal time, $d\eta=\frac{dt}{a}$ or $\eta=\int\frac{dt}{a}$, we conclude that for the kination era $a^2\propto\eta$. Considering the constants of integration, and in terms of the dimensionless conformal time,$\tau$, we obtain for the scale factor of the kination era that:
\begin{equation}
a^{2}=p\,\tau+q,
\qquad
a=(p\,\tau+q)^{1/2}
\label{g2}
\end{equation}
In this functional form, the $p$ and $q$ are constants of integration that should be found by imposing relevant constraints. For further considerations, which we will explain in what follows, we have to calculate the first and second time derivatives of this scale factor. We have:
\begin{equation}
\frac{da}{d\tau}=\frac{p}{2}\,\frac{1}{(p\tau+q)^{1/2}},
\qquad
\frac{d^{2}a}{d\tau^{2}}=\frac{-p^{2}}{4}\,\frac{1}{(p\tau+q)^{3/2}}
\label{g3}
\end{equation}
On the other hand, for the inflationary era, we know that the Hubble parameter, $H$, is almost constant. As a result, due to its definition, $\frac{\dot{a}}{a}=H$, we conclude that $a\propto e^{Ht}$. Hence, in terms of the dimensionless conformal time, we have to consider the behavior of the scale factor as:
\begin{equation}
a=-\frac{1}{\tau},
\qquad
a^{2}=\frac{1}{\tau^{2}}
\label{g4}
\end{equation}
This is exactly what we have considered in Eq. \ref{scalean} for the inflationary era. Comparing this scale factor with the one in Eq. \ref{scalenum}, we see that for the inflationary era, the numerical scale factor is of the form $\frac{a}{1+{\tau}^2}$, but the chosen scale factor above behaves as $\frac{1}{{\tau}^2}$. This is due to the fact that if we substitute the first functional form into the EoM, the analytical solution can not be obtained. In other words, the analytical general solution to the EoM with this form of the scale factor can not be found. Hence, we choose the form of $\frac{1}{{\tau}^2}$for which the EoM can yield an analytical solution. This approximation does not provide us with any challenge. Since during the inflationary era, the conformal time tends to infinity, the number $1$ in the Denominator of $\frac{a}{1+{\tau}^2}$ can be ignored without any challenge.

With these scale factors for the kination and inflation in hand, we can build the piecewise function for the scale factor of analytical calculations. To this end, we have to determine the constants $p$ and $q$. It is essential to keep in mind that the scale factor and its first time derivative must be continuous at the transition time between inflation and kination, ${\tau}_f$. Hence, we have two constraints essential to determine $p$ and $q$. Although the first one guarantees the continuity of the scale factor, the second constraint will provide us with the continuity of the Hubble parameter as well. Both of these conditions are essential for building a consistent cosmological model. Using these conditions, we set equal the scale factor of the kination and inflation and their first derivative at the transition time. Hence, we find that:
\begin{equation}
p=-\frac{2}{\tau_{f}},
\qquad
q=2+\frac{1}{\tau_{f}^{2}}
\label{g5}
\end{equation}
As a result, the scale factor for the kination era becomes:
\begin{equation}
a(\tau)=\left(-\frac{2}{\tau_f}\,\tau+\frac{1}{\tau_f^{2}}+2\right)^{1/2},
\qquad
a^{2}(\tau)=-\frac{2}{\tau_f}\,\tau+\frac{1}{\tau_f^{2}}+2
\label{g6}
\end{equation}
This scale factor and also the scale factor of the inflationary era, Eq. \ref{g4}, and their time derivatives are continuous and together will build the piecewise function for our analytical calculation in Sec. \ref{anaresults}. 

There remains only another important note that is worth attention. \textbf{This scale factor is built so that if one sets the transition time, ${\tau}_f$, equal to $-1$, its value becomes normalized to unity. This is because the functional form of the scale factor that we have chosen for our numerical analysis becomes unity at the end of the inflation. In Sec. \ref{anaresults}, one of our goals is to compare our numerical results with analytical results. This is essential for highlighting the strengths and weaknesses of the analysis. Hence, it is necessary for this comparison that the normalization of the scale factor in both analyses be the same.}

\section{Derivation of the BDV and AKV for the Analytical Calculations and the resultant ${\lvert{\beta}_{\kappa}\lvert}^2$}\label{ander}
 In this appendix, we are going to explain in detail the key steps in the derivation of the BDV and AKV needed for the analytical calculations in Sec. \ref{anaresults}. To this end, we invoke the analytical piecewise scale factor, presented in Eq. \ref{scalean}. By substituting the function corresponding to the inflationary era in the EoM of the auxiliary field, Eq. \ref{main}, the EoM for the inflationary era is obtained as Eq. \ref{eominf}, where the effective mass is defined in Eq. \ref{M}.

The general analytical solution to this equation is obtained as:
\begin{equation}
\Delta_{\kappa}(\tau)
=C_{1}\sqrt{\tau}\,J_{\frac{1}{2}\sqrt{1-4M^2}}(\kappa\tau)
+C_{2}\sqrt{\tau}\,Y_{\frac{1}{2}\sqrt{1-4M^2}}(\kappa\tau)
\label{f1}
\end{equation}
It is worth noting that the argument of the Bessel function can not be a negative-valued number. But the conformal time itself is negative-valued during the inflationary era. Hence, this solution might seem problematic to be considered as the analytical solution. There is, in fact, a systematic way to overcome this challenge, and it lies in a symmetry which is manifest in the EoM of the inflationary era. The EoM of the inflationary era, Eq. \ref{eominf}, is an even EoM for the parameter $\tau$, which means that if one replaces the parameter $\tau$ with $-\tau$, the EoM will not be changed at all. Hence, we can only replace the $\tau$ in the general solution with $-\tau$ and then the general solution for the inflationary era becomes:
\begin{equation}
\Delta_{\kappa}(\tau)
=C_{1}\sqrt{-\tau}\,J_{{\frac{1}{2}\sqrt{1-4M^2}}}(-\kappa\tau)
+C_{2}\sqrt{-\tau}\,Y_{{\frac{1}{2}\sqrt{1-4M^2}}}(-\kappa\tau)
\label{f2}
\end{equation}
For the sake of simplicity in writing the equations, we define a new variable, $\nu$, as:
\begin{equation}
\nu\equiv\frac{1}{2}\sqrt{\,1-4M^{2}\,}
=\frac{1}{2}\sqrt{\,1-4\!\left(\mu^{2}+12\xi-2\right)}
\label{f3}
\end{equation}
To determine the constant coefficients, $C_1$ and $C_2$, we have to impose the relevant adiabatic initial conditions on this general solution. These initial conditions for the inflationary era, as discussed before, are obtained by requiring that $\tau \rightarrow -\infty$. In this limit, the mode function has to tend toward the behavior of the mode function corresponding to the vacuum in flat spacetime. The mode function, defining the vacuum state obtained by imposing this initial condition, is exactly the well-known Bunch-Davies vacuum. Hence, since we have to find the limit of the general solution at infinity, it is essential to rewrite the solution in terms of the Hankel functions. This is due to the fact that to find the limit of the functions of the family of Bessel functions at infinity, it is better to rewrite them in terms of the Hankel functions. Using the transformations in equations \ref{c1} and \ref{c2}, the general solution can be written as:
\begin{equation}
\Delta_{\kappa}(\tau)=
\frac{1}{2}\!\left(C_{1}+\frac{C_{2}}{i}\right)\sqrt{-\tau}\,H_{\nu}^{(1)}(-\kappa\tau)
+\frac{1}{2}\!\left(C_{1}-\frac{C_{2}}{i}\right)\sqrt{-\tau}\,H_{\nu}^{(2)}(-\kappa\tau)
\label{f4}
\end{equation}
Now, we impose the adiabatic or Bunch-Davis initial condition on this solution as:
\begin{equation}
\Delta_{\kappa}(\tau)\xrightarrow[\tau\to-\infty]{} \frac{1}{\sqrt{2\kappa}}\,e^{-i\kappa\tau}
\label{f5}
\end{equation}
To use this initial condition, we have to invoke the limiting form of the Hankel functions for the large argument, as is written in Eq. \ref{c3}. Since we desire that the behavior of the mode function becomes the form of the mode function which defines the vacuum in flat spacetime at past infinity, the coefficient $C_2$ must be $C_2=iC_1$ and hence, the general solution becomes:
\begin{equation}
\Delta_{\kappa}(\tau)=C_{1}\sqrt{-\tau}\,H_{\nu}^{(1)}(-\kappa\tau)
\label{f6}
\end{equation}
And it is only the coefficient $C_1$ that needs to be determined. If we equate the general solution to the mode function of the flat space-time, as in Eq. \ref{f5}, the coefficient $C_1$ is determined as:
\begin{equation}
C_{1}=\frac{\sqrt{\pi}}{2}\;e^{\,i\frac{(2\nu+1)\pi}{4}}
\label{f7}
\end{equation}
Hence, the mode function, which defines the physical vacuum corresponding to the inflationary era, is obtained as:
\begin{equation}
\Delta_{\kappa}^{\inf}(\tau)=
\frac{1}{2}\sqrt{-\pi\tau}\;e^{\,i\frac{(2\nu+1)\pi}{4}}\,H_{\nu}^{(1)}(-\kappa\tau)
\label{f8}
\end{equation}
With this mode function in hand, it is time to determine the physical mode function of the kination era. Substituting the second function in the piecewise scale factor and ignoring the term containing the curvature scalar in the EoM, the final form of the EoM for the kination era becomes the equation in Eq. \ref{eomkin}. The general analytical solution for this equation becomes:
\begin{equation}
\Delta_{\kappa}(\tau)=
C_{1}\,\operatorname{Airy Ai}\!\left(
\frac{-\kappa^{2}-2\mu^{2}-\dfrac{\mu^{2}}{\tau_{f}^{2}}+2\mu^{2}\dfrac{\tau}{\tau_{f}}}
{2^{2/3}\!\left(\dfrac{\mu^{2}}{\tau_{f}}\right)^{2/3}}
\right)
+ 
C_{2}\,\operatorname{Airy Bi}\!\left(
\frac{-\kappa^{2}-2\mu^{2}-\dfrac{\mu^{2}}{\tau_{f}^{2}}+2\mu^{2}\dfrac{\tau}{\tau_{f}}}
{2^{2/3}\!\left(\dfrac{\mu^{2}}{\tau_{f}}\right)^{2/3}}
\right)
\label{f9}
\end{equation}
Again, to impose the relevant adiabatic initial conditions, we have to find the limit of the Airy functions at infinity. But we have to keep in mind that in the case of the kination era, the adiabatic initial conditions are defined by requiring that $\kappa\rightarrow\infty$, or more precisely, $\frac{{\kappa}^2}{a^2{\mu}^2}\rightarrow\infty$. Hence, it is essential to rewrite the general solution in terms of the Hankel functions. Using the transformation between the Airy functions and the Hankel functions in Eq. \ref{c4} and Eq. \ref{c5}, and defining a new variable, $x$, as:
\begin{equation}
x(\tau)\equiv
\frac{\kappa^{2}+2\mu^{2}+\dfrac{\mu^{2}}{\tau_{f}^{2}}
-2\mu^{2}\dfrac{\tau}{\tau_{f}}}
{2^{2/3}\!\left(\dfrac{\mu^{2}}{\tau_{f}}\right)^{2/3}}
\label{f10}
\end{equation}
The general solution can be written in terms of the Hankel functions as:
\begin{equation}
\Delta_{\kappa}(\tau)=
\sqrt{x}\!\left(
\frac{C_{1}\,e^{\,i\pi/6}+C_{2}\,e^{\,2\pi i/3}}{2\sqrt{3}}
\right) H^{(1)}_{1/3}\!\!\left(\tfrac{2}{3}x^{3/2}\right)
+\sqrt{x}\!\left(
\frac{C_{1}\,e^{-i\pi/6}+C_{2}\,e^{-2\pi i/3}}{2\sqrt{3}}
\right) H^{(2)}_{1/3}\!\!\left(\tfrac{2}{3}x^{3/2}\right)
\label{f11}
\end{equation}
The two parentheses in the above equations are totally constant. Hence, for the sake of simplicity in writing and also calculations, we define two new constants as:
\begin{equation}
T_{1}\equiv \frac{C_{1}\,e^{\,i\pi/6}+C_{2}\,e^{\,2\pi i/3}}{2\sqrt{3}},
\qquad
T_{2}\equiv \frac{C_{1}\,e^{-i\pi/6}+C_{2}\,e^{-2\pi i/3}}{2\sqrt{3}}
\label{f12}
\end{equation}
Determining the constant coefficients $C_1$ and $C_2$ is now equivalent to determining the constants $T_1$ and $T_2$. As a result, the general solution for the kination era becomes:
\begin{equation}
\Delta_{\kappa}(\tau)=
T_{1}\,\sqrt{x}\; H^{(1)}_{1/3}\!\left(\tfrac{2}{3}x^{3/2}\right)
+T_{2}\,\sqrt{x}\; H^{(2)}_{1/3}\!\left(\tfrac{2}{3}x^{3/2}\right)
\label{f13}
\end{equation}
To determine the constant coefficients, we have to invoke the adiabatic initial conditions for the kination era. To this end, first of all, we will try to find the limiting form of $\sqrt{x}$ and $\frac{2}{3}x^{\frac{3}{2}}$ in the adiabatic limit, $\frac{{\kappa}^2}{a^2{\mu}^2}\rightarrow\infty$. Starting with the function $x$, we have:
\begin{equation}
x(\tau)=
\frac{\kappa^{2}+2\mu^{2}+\dfrac{\mu^{2}}{\tau_{f}^{2}}-2\mu^{2}\dfrac{\tau}{\tau_{f}}}
{2^{2/3}\!\left(\dfrac{\mu^{2}}{\tau_{f}}\right)^{2/3}}
=\frac{\kappa^{2}}{\left(\dfrac{2\mu^{2}}{\tau_{f}}\right)^{2/3}}
\left(
\frac{2\mu^{2}}{\kappa^{2}}
+\frac{\mu^{2}}{\kappa^{2}\tau_{f}^{2}}
-\frac{2\mu^{2}}{\kappa^{2}}\frac{\tau}{\tau_{f}}+1
\right)
\label{f14}
\end{equation}
Hence, using Newton's binomial expansion, the limiting form of the $\sqrt{x}$ becomes:
\begin{equation}
\sqrt{x}\;\sim\;
\frac{\kappa}{\bigl(2 \frac{\mu^{2}}{|\tau_{f}|}\bigr)^{1/3}}
+\mathcal{O}\!\left(\frac{\mu^{4/3}}{\kappa}\right)
\label{f15}
\end{equation}
Where the first term is the dominant one. Following the same procedure, the limiting form of the $\frac{2}{3}x^{\frac{3}{2}}$ becomes:
\begin{equation}
\frac{2}{3}\,x^{3/2}\;\sim\;
(-\tau_{f})\,\frac{\kappa^{3}}{3\mu^{2}}
+(-\tau_{f}){3/2}\,\kappa
+\kappa\,\tau
\label{f16}
\end{equation}
Substituting these limiting forms and invoking the asymptotic behavior of the Hankel functions in Eq. \ref{c3}, the limiting form of the general solution is obtained. Now we impose the adiabatic initial condition on this solution. We know that in the limit of $\frac{{\kappa}^2}{a^2{\mu}^2}\rightarrow\infty$, the mode function must behave as the mode function of the flat space-time. Hence, the coefficient $T_1$ must be equal to zero, and the coefficient $T_2$ is obtained as:
\begin{equation}
T_{2}
=\sqrt{\frac{\pi}{6}}
\left(\frac{2\mu^{2}}{-\tau_{f}}\right)^{-1/6}
e^{i\,\left\{
(-\tau_{f})\frac{\kappa^{3}}{3\mu^{2}}
+(-\tau_{f})^{3/2}\kappa
-\frac{5\pi}{12}
\right\}}
\label{f17}
\end{equation}
As a result, the physical mode function for the kination era is obtained as:
\begin{equation}
\Delta_{\kappa}^{\mathrm{kin}}(\tau)=
\sqrt{\frac{\pi}{6}}
\left(\frac{2\mu^{2}}{-\tau_{f}}\right)^{-1/6}
e^{i\,\left\{
(-\tau_{f})\frac{\kappa^{3}}{3\mu^{2}}
+(-\tau_{f})^{3/2}\kappa
-\frac{5\pi}{12}
\right\}}
\;\sqrt{x}\;
H^{(2)}_{1/3}\!\left(\tfrac{2}{3}x^{3/2}\right)
\label{f18}
\end{equation}
With the physical mode functions of the inflation (Eq. \ref{f8}) and the kination (Eq. \ref{f18}) in hand, now it is time to calculate the Bogoliubov coefficient ${\lvert{\beta}_{\kappa}\lvert}^2$ as the Wronskian of these two functions. To this end, we have to compute the time derivative of these mode functions. Invoking the derivative equation of the Bessel functions in Eq. \ref{c7}, the time derivative of the mode function of the inflationary era becomes:
\begin{equation}
\big(\Delta_{\kappa}^{\inf}\big)'(\tau)=
e^{\,i(\frac{2\nu+1}{4})\pi}\,
\Bigg\{
-\frac{1}{4}\sqrt{\frac{\pi}{-\tau}}\;
H^{(1)}_{\nu}\!\left(-\kappa\tau\right)
-\frac{\kappa}{4}\sqrt{-\pi\tau}\;
\Big(
H^{(1)}_{\nu-1}\!\left(-\kappa\tau\right)
- H^{(1)}_{\nu+1}\!\left(-\kappa\tau\right)
\Big)
\Bigg\}
\label{f19}
\end{equation}
Similarly, the time derivative for the kination physical mode function can be calculated as:
\begin{align}
({{{\Delta}_{\kappa}}^{kin}})^{\prime}(\tau)=\Bigg\{
\sqrt{\frac{\pi}{6}}\,(2\mu^{2})^{1/6}
\frac{(\tau_{f})^{2/3}}{(-\tau_{f})^{5/6}}\,
\frac{1}{2\sqrt{x}}\;
H^{(2)}_{1/3}\!\!\left(\tfrac{2}{3}\,x^{3/2}\right)+...\\
...+\sqrt{\frac{\pi}{6}}\,(2\mu^{2})^{1/6}
\frac{\tau_{f}^{2/3}}{(-\tau_{f})^{5/6}}\,
\frac{x}{2}\,
\Big(
H^{(2)}_{-2/3}\!\!\left(\tfrac{2}{3}\,x^{3/2}\right)
- H^{(2)}_{4/3}\!\!\left(\tfrac{2}{3}\,x^{3/2}\right)
\Big)
\Bigg\}
e^{ i\,\left\{
(-\tau_{f})\frac{\kappa^{3}}{3\mu^{2}}
+(-\tau_{f})^{3/2}\kappa
-\frac{5\pi}{12}
\right\}}
\label{f20}
\end{align}
Now, the parameter ${\lvert{\beta}_{\kappa}\lvert}^2$, responsible for the GPP rate at the transition time, $\tau={\tau}_f$, is calculated as:
\begin{align}
|\beta_{\kappa}|=
\left|
\frac{
W\!\left[\Delta_{\kappa}^{\mathrm{kin}},\,\Delta_{\kappa}^{\inf}\right]
}{
W\!\left[\Delta_{\kappa}^{\mathrm{kin}},\,\Delta_{\kappa}^{\mathrm{kin}*}\right]
}
\right|_{\tau=\tau_f}
&=\Bigg|
e^{\tfrac{1}{2} i \nu \pi}\,
\Bigg|\Bigg|
\sqrt{\tfrac{\pi}{6}}
\left(\tfrac{2\mu^{2}}{-\tau_{f}}\right)^{-1/6}
x_{f}^{1/2}\,
H^{(2)}_{1/3}\!\Big(\tfrac{2}{3}x_{f}^{3/2}\Big)
\\[2pt]
&\hspace{1.1cm}
\left\{
-\tfrac{1}{4}\sqrt{\tfrac{\pi}{-\tau_{f}}}\,H^{(1)}_{\nu}(-\kappa\tau_{f})
-\tfrac{\kappa}{4}\sqrt{-\pi\tau_{f}}\,
\big[H^{(1)}_{\nu-1}(-\kappa\tau_{f})-H^{(1)}_{\nu+1}(-\kappa\tau_{f})\big]
\right\}
\\[2pt]
&\hspace{1.1cm}
-\sqrt{\tfrac{\pi}{6}}\,(2\mu^{2})^{1/6}
\left(\frac{\tau_{f}^{2/3}}{(-\tau_{f})^{5/6}}\right)
\\[2pt]
&\hspace{1.1cm}
\left\{
\tfrac{1}{4}x_{f}^{-1/2}\,
H^{(2)}_{1/3}\!\Big(\tfrac{2}{3}x_{f}^{3/2}\Big)
+\tfrac{1}{4}x_{f}\,
\Big[H^{(2)}_{-2/3}\!\Big(\tfrac{2}{3}x_{f}^{3/2}\Big)
- H^{(2)}_{4/3}\!\Big(\tfrac{2}{3}x_{f}^{3/2}\Big)\Big]
\right\}
\\[2pt]
&\hspace{1.1cm}
\sqrt{-\pi\tau_{f}}\,H^{(1)}_{\nu}(-\kappa\tau_{f})
\Bigg|
\label{f21}
\end{align}
By setting ${\tau}_f=-1$, which guarantees that our analytical scale factor normalization is in accordance with our numerical one,\ref{scalenum}, the GPP rate is obtained as Eq. \ref{bettafin}. In this equation, we have:
\begin{equation}
x_{f}\equiv
\frac{\kappa^{2}+\dfrac{\mu^{2}}{\tau_{f}^{2}}}
{\left(\dfrac{2\mu^{2}}{\tau_{f}}\right)^{2/3}},
\qquad
\nu\equiv \frac{1}{2}\sqrt{1-4M^{2}},
\qquad
M^{2}\equiv \mu^{2}+12\,\xi-2
\label{f22}
\end{equation}
\bibliography{Main.bib}
\bibliographystyle{JHEP}
\end{document}